\documentclass{llncs}

\usepackage[round]{natbib} 

\usepackage{graphicx} %to inset images
\usepackage{amsmath,amssymb,amsfonts}%it is good to import first two
\usepackage{float}
\usepackage{enumerate}%personalized 
\usepackage[margin=1in]{geometry}
\usepackage{amsfonts}%for real numbers mark in mathematics ,symbols
\usepackage{xcolor}
\usepackage[colorlinks=true, linkcolor=blue, citecolor=blue]{hyperref}

\definecolor{editcolor}{RGB}{0, 0, 200} % Red color for added text

\usepackage{tikz}
\usepackage{pgfplots}
\usepackage{comment}
\usetikzlibrary{calc}
\usepackage{subfig}
\usepackage{algorithm}
\usepackage{algorithmic}
\usepackage{booktabs}
\usepackage{multirow, tabularx, booktabs}
\begin{document}

%\begin{frontmatter}

\title{
Large-Scale Portfolio Optimization Problem Under Cardinality Constraint With Enhanced Multi-Objective Evolutionary Algorithms
}
\author{    Danial Ramezani $^{\text{a}}$, Mostafa Abouei Ardakan$^{\text{a}}$}
\institute{   $^{\text{a}}$ Department of Industrial Engineering, Faculty of Engineering, Kharazmi University, Tehran, Iran} 

\maketitle


\begin{abstract}
Decision-making is posing an increasingly formidable challenge to investors because of the growing number of alternatives available in financial markets. A hot area of research over the past few decades has been portfolio optimization that seeks to determine how much an investor should invest in which asset. Introducing real-world conditions to the optimization model turns the problem into an NP-hard one for whose solution exact methods become inefficient; hence, researchers have turned to evolutionary algorithms to approximate solutions. In this paper, strengthening strategies are presented for multi-objective evolutionary algorithms that can provide a faster convergence rate and extensive search ability in the portfolio optimization problem under the cardinality constraint. To implement those features, a unique solution representation, a novel operator, and new repair mechanisms are introduced for solving the aforementioned problem in which lower and upper limits are set on the number of assets in the portfolio. For this purpose, new mating strategies along with the aforesaid package are implemented in well-known multi-objective evolutionary algorithms to solve the problem. The customized algorithms are subsequently tested against traditional ones using well-known market indices as benchmarks. Results indicate that the proposed strategy not only provides better approximations but also converges faster as well at no loss of performance with increasing number of assets in the market.
\end{abstract}


%\end{frontmatter}

%\pagebreak
\section{Introdution:}

Portfolio selection is well-known to investors and fund managers and is a widely used
strategy in financial markets. It, indeed, involves the selection of capital and its proper 
allocation to various potential assets aimed at maximizing return but minimizing risk.

Investors and portfolio managers seek to identify the best investment opportunities with the 
highest return and lowest risk  (\cite{pedersen2021enhanced}). Modern portfolio theory was pioneered by Harry Markowitz who put forth the concept of diversification as a 
solution to mitigating portfolio risk. He introduced variance as a risk measurement; his 
mean-variance (MV) model is based on minimizing risk for a certain level of return (\cite{RePEc:bla:jfinan:v:7:y:1952:i:1:p:77-91}). Although Markowitz’s revolutionary work altered quantitative finance, it 
suffered from certain shortcomings from a practical perspective; for instance, it ignored such 
real-world constraints (\cite{ertenlice2018survey}) as cardinality constraint (CC) that puts 
limits on the number of stocks in the portfolio, boundary constraint (BC) that imposes a limit 
on the portion of capital allowed for investment in an individual asset within a portfolio, 
transaction costs (\cite{thakkar2021comprehensive}), round lot constraint (\cite{almahdi2019constrained}), class constraint (\cite{almahdi2019constrained}), sector capitalization constraint (\cite{golmakani2011constrained}), and turnover constraint (\cite{clarke2002portfolio}). The portfolio 
optimization problem (POP) has attracted researchers’ attention and encouraged a 
significant amount of research work to extend the MV model or to incorporate real-world 
constraints.

\cite{konno1991mean} considered mean absolute error as a risk measure for addressing nonlinearity in the MV model. Later, value at risk (VaR) (\cite{jorion1997value}) gained attention in modeling
portfolio risks. Due to its shortcomings, namely subadditivity and convexity,
\cite{rockafellar2000optimization} put forth the concept of minimizing conditional 
value at risk (CVaR) instead of VaR. Using S\&P 500 data, \cite{shalu2024computational} demonstrated that expectile-based portfolio optimization models outperform the CVaR objective function.

The selection of a method for solving an optimization problem largely depends on the researcher’s objectives regarding desired outcomes. When models are complex, non-exact methods, which offer faster, approximate solutions that can still be considered feasible in many cases, are generally preferred due to their ability to handle higher complexity more efficiently. On the other hand, exact methods are more time-intensive but guarantee optimal solutions (\cite{milhomem2020analysis}). 
On account of the fact that adding the above constraints would make the problem an NP-hard one (\cite{erwin2023meta}), almost one fifth of the studies reported in the literature made use of exact 
approaches and machine learning (ML) algorithms to solve POPs (\cite{kalayci2019comprehensive}).

Portfolio optimization can be addressed using various exact algorithms and techniques. \cite{xu2024efficient} formulated the POP, where they considered only a limited number of assets for investment, as a max-min optimization problem using dual theory, and employed the depth-first branch and bound method to identify the globally optimal investment selection strategy. \cite{kobayashi2021bilevel} reformulated the POP with CVaR as a risk measure into a bilevel optimization problem and presented a cutting-plane algorithm to solve it exactly. \cite{cui2020hybrid} also considered CVaR as a risk measure in their proposed two-stage stochastic POP. To solve this model, they employed a hybrid combinatorial approach that integrates a hybrid algorithm with a linear programming solver. \cite{vielma2008lifted} utilized a branch-and-bound based algorithm for a multiple POP with different 
objective functions such as classical MV and shortfall. Results demonstrated that their 
algorithm outperformed common solvers. \cite{akbay2020parallel} presented an algorithm that
divided POP into two parts: choosing stocks via the variable neighborhood search (VNS) 
algorithm, and assigning weights using the quadratic programming.

Instances of using ML and deep learning techniques, either to solve the problem directly, to predict parameters, or alongside other methods, can also be found in the literature. \cite{fernandez2007portfolio} applied Hopfield networks, a type of neural network designed to solve optimization problems, to solve the POP. \cite{chen2021mean} applied XGBoost with improved firefly algorithm for predicting stock 
prices and utilized Monte-Carlo simulation to construct optimal portfolio out of good-quality 
stocks. \cite{paiva2019decision} proposed the support vector machine for identifying eligible 
stocks and the MV model for assigning weights to the portfolio. \cite{behera2023prediction} applied 
various ML regression algorithms along with the mean–VaR model to solve the POP; results 
indicated that the mean–VaR model accompanied by adaptive boosting outperformed the 
other models. \cite{wang2020portfolio} deployed various ML techniques to forecast stock returns and identify the most promising ones out of a pool of assets and, further, used MV to 
optimize allocations. \cite{wu2022construction} argued that conventional methods were unable to 
consider the market condition impacts and that they would fail to select assets from a large 
pool of stocks. They, therefore, proposed an ML technique and the K-means algorithm to 
predict trends and select stocks for optimization, respectively. \cite{zhang2020cost} presented a novel cost-sensitive reward function that incorporates patterns in prices and correlations among assets. The goal was to maximize the accumulated return while adhering to constraints such as transaction costs and risk management. \cite{ma2021portfolio} integrated ML and deep learning models with portfolio optimization techniques to enhance stock selection and return prediction, showing that the MV method combined with random forest forecasting outperforms other approaches. \cite{sadik2023novel} proposed using regularization techniques to optimize predictability through computer simulations for mean-reverting portfolio selection.

Since exact techniques fail to render efficient solutions for POP, nearly four-fifths of researchers have turned to metaheuristic algorithms (\cite{kalayci2019comprehensive}). \cite{chang2000heuristics} included CC and BC in the MV model 
and showed that efficient frontier (EF) might become discontinuous and invisible to exact 
approaches when each of the aforementioned constraints are introduced into a model. They, 
therefore, employed genetic algorithm (GA), tabu search, and simulated annealing in their 
model and tested the results on 5 market indices for a stock number of up to 225. Additionally, a greater body of literature has been 
over the past few years devoted to multi-objective models compared to single-objective ones 
due to the nature of conflicting objectives of the MV model (\cite{mansour2019multi}). What's more, the multi-objective approach is found to be capable of providing sets of portfolios that appeal to all investors, as they have different preferences for portfolio types based on their desired risk levels (\cite{gunjan2022brief}).  \cite{kalayci2020efficient} proposed a 
hybrid metaheuristic algorithm based on the key features of GA, ant colony optimization, and 
artificial bee colony optimization to address the cardinality-constrained portfolio 
optimization problem (CCPOP). \cite{song2023enhanced}  proposed a co-evolutionary multi-swarm 
adaptive differential evolution algorithm that was capable of resolving problems of
premature convergence and search stagnation by distinguishing between population and 
multi-operator parallel search strategy. \cite{kizys2022simheuristic} merged VNS with Monte-Carlo 
simulation both to improve the performance of the algorithm and to deal with uncertainty 
in their model. \cite{morteza2023improved} proposed a multi-population whale optimization 
algorithm with three objective functions, namely, return, risk, and liquidity. Moreover, they 
updated parameters with reward and penalty using reinforcement learning. \cite{moehle2023portfolio} developed a heuristic algorithm based on the alternating direction method of multipliers alternating direction method of multipliers for separable and nonconvex terms in the objective function, solving the problem with moderate accuracy. \cite{corazza2021novel} presented a hybrid particle swarm optimization (PSO) algorithm. They reformulated
POP as an unconstrained problem with an exact penalty function, which dynamically changes 
its parameters in order to reduce possible infeasible solutions throughout the process. \cite{silva2019multi}  proposed an adaptive ranking PSO algorithm that is based on three mechanisms 
for ranking solutions. In this model, the non-dominated sorting and crowding distance are 
initially applied before solutions are ranked based on crowding distance and cost-benefit 
(CB) defined as the ratio of expected return to risk. The summation of scores assigned from 
each ranking method determines the final rank; if there is a tie, the particle with the higher 
CB will be chosen. \cite{sadeghi2024memetic} proposed a memetic quantum-inspired genetic algorithm integrating quantum rotation gate mutations with tabu search to balance exploration and exploitation, enhancing convergence speed and solution accuracy, and demonstrate its superiority over existing methods on benchmark optimization problems. \cite{gunjan2024quantum} proposed quantum-inspired versions of GA, differential evolution, and PSO with enhanced crossover, regularization, and dynamic parameter tuning to solve constrained portfolio optimization using over a decade of stock data from NASDAQ, BSE, and Dow Jones.

\cite{zheng2022novel} proposed a multi-population parallel non-dominated sorting 
genetic algorithm (NSGA-II) with distinct evolutionary strategies that both employed normal 
distribution crossover and simulated binary crossover for each population. Furthermore, the 
algorithm used the Fuzzy C-means and grey relational projection to yield compromised 
portfolios for various investors with different risk profiles.  \cite{wang2022multi} took market 
impact cost into account and appended a new local search operator to the NSGA-II algorithm 
along with features from global search derived from NSGA-II. \cite{solares2019handling} proposed 
a multi-objective evolutionary algorithm based on decomposition with confidence intervals 
that describes the solutions. \cite{sanjalawe2025recent} provided a comprehensive review of the secretary bird optimization algorithm, detailing its recent advancements, variants, and applications across multiple domains, including portfolio optimization, while addressing challenges like premature convergence and parameter sensitivity, and highlighting future research directions for hybridization and adaptive improvements.

Although some reports in the literature make use of penalty functions and infeasibility 
tolerance for handling constraints, the vast majority employ a repair mechanism to address 
this issue (\cite{kalayci2019comprehensive}). \cite{chang2000heuristics} proposed a method to handle CC in 
portfolio and to ensure that the number of assets remains within bounded limits. In their 
repair procedure for GA, if the number of stocks in portfolio are greater than a maximum K, 
the asset with the lowest weight would be eliminated from the portfolio. On the other hand, 
if it is lower than K, a random stock from parents will be selected. \cite{anagnostopoulos2010portfolio} presented a tri-objective portfolio optimization model such that the 
number of assets in the portfolio should be minimized. \cite{pai2009evolutionary} applied k-means clustering to tackle CC. The k clusters in k-means would be the maximum number of 
stocks that the investor would have to choose from among N investible assets; the authors 
addressed the problem by selecting only one stock out of each cluster. \cite{cura2009particle} proposed 
a probabilistic repair method for CC. According to this method, when an asset is to be added 
to the portfolio, either a random asset is chosen or one is added with equal probability that 
has the highest c-value defined as the mean return to mean risk with respect to the aversion 
parameter which is not present in the portfolio. For removing additional stocks, either a 
random stock is removed or that with the lowest c-value is with equal probability. For BC, if 
the weight of an asset is higher or lower than the maximum or the minimum bound, either 
the upper bound or the lower one, respectively, is assigned to the asset; otherwise, the weight of the asset is determined via a formula given in their paper. \cite{zhao2021multiple} used 
a similar approach to handling CC, but the risk aversion parameter was disregarded. \cite{kaucic2019portfolio} proposed the combined NSGA-II and strength Pareto evolutionary algorithm 
with a new reproduction process and compared it with the reproduction process used by \cite{anagnostopoulos2011mean}. Moreover, they used the method in \cite{liagkouras2015efficient} to repair solutions to repair solutions. \cite{de2022multiobjective} proposed operators within a 
multi-objective genetic algorithm whose constraints were handled within the algorithm. \cite{yang2024dynamic} presented a dynamic tri-population multi-objective evolutionary algorithm that balances convergence, feasibility, and diversity for constrained multi-objective optimization by evolving multiple populations with adaptive resource allocation and constraint violation handling, demonstrating competitive performance on 57 benchmarks and 21 real-world problems.

This study presents fast-converging and extensive search strategies in evolutionary algorithms for portfolio optimization problem under the cardinality constraint, with two conflicting objectives. One feature of the study is a new encoding procedure designed to facilitate more efficient handling of the CC, which also enables us to introduce and implement our strategies more easily. A second aspect involves a novel handling mechanism and a new operator, both introduced not only to ensure the feasibility of the generated solutions but also to make them feasible in the best possible way. Third, new mating strategies are proposed for producing solutions and optimizing the searching process. Aforesaid novelties are meant to achieve faster convergence rate and improve the process of exploration for faster and better approximations, especially in larger markets. To test the proposed package, it is implemented in multi-objective evolutionary algorithms, and the improved algorithms are tested and validated against the conventional versions applied to three well-known indices from the OR-library (\cite{beasley1990or}), namely S\&P 100, German DAX 100, Japanese Nikkei 225, and Tehran Stock Exchange (TSE).

The rest of the paper is organized as follows. \hyperref[Mathematical formulation]{{Section 2}}  presents the mathematical model of 
the cardinality constrained mean-variance portfolio optimization problem. \hyperref[Methodology]{{Section 3}} describes the proposed approach including mating strategies, the new operator, CC, and the 
BC repair techniques. \hyperref[Experiment results]{{Section 4}}  presents and discusses the experimental results. Finally, \hyperref[Conclusion and future research]{{Section 5}} concludes the work and suggests future directions.

\section{Mathematical formulation:}
\label{Mathematical formulation}
MV portfolio optimization consists of a pair of conflicting objectives: maximizing return and 
minimizing risk. The multi-objective approach results in a variety of non-dominated 
portfolios with varying levels of risk and return suitable for different classes of investors. 
Each solution represents a tradeoff between the objectives, forming a curve known as the EF. The EF is a graphical representation of the set of optimal portfolios in a feasible search space, where no objective can be improved without worsening another. The EF shows the best possible tradeoff between risk and return, highlighting the set of portfolios that offer the highest expected return for a given level of risk, or the lowest risk for a given return. The 
adoption of a multi-objective approach in this study arises from the knowledge that finding 
a single solution as a compromise between both objectives is arduous due to the divergent 
preferences of investors.

Additionally, incorporating a cardinality constraint in portfolio optimization is particularly important, especially in real-world scenarios. Generally, achieving diversification does not require investing in all assets in the market; selecting a subset of $K$ assets can provide similar benefits at a lower cost for the investor, although the optimal value of $K$ remains unclear (\cite{jimbo2017portfolio}). Considering the CC makes the problem more realistic by reducing management costs and complexity, which are often associated with the transactions required for large portfolios (\cite{esmaeily2023portfolio}). It simplifies the monitoring and control of unrealized profits and losses for investors (\cite{tadonki2003portfolio}). The concept of CC is widely used in portfolio optimization, from large firms to high-frequency traders (\cite{leung2022cardinality}).

The multi-objective CCPOP can be mathematically formulated as 
follows:

\begin{equation}\label{Eq:eqRisk}
\min\,   f_{1}= \sum_{i=1}^{N}\sum_{j=1}^{N} w_iw_j\sigma_{ij}   
\end{equation}
\begin{equation}\label{Eq:eqReturn}
\max\,   f_{2}= \sum_{i=1}^{N} w_i\mu_{i}
\end{equation}
\\Subject to:
\begin{equation}\label{Eq:eqSumWeight}
\sum_{i=1}^{N} w_{i} = 1
\end{equation}
\begin{equation}\label{Eq:eqCardinalityBound}
K_{\text{min}} \leq \sum_{i=1}^{N} z_{i} \leq K_{\text{max}}
\end{equation} 
\begin{equation}\label{Eq:eqWeightBound}
\epsilon_{i}z_{i} \leq w_{i} \leq \delta_{i}z_{i},\hspace{5mm} \forall i=1,...,N
\end{equation} 
\begin{equation}\label{Eq:eqWi}
w_{i} \geq 0 ,\hspace{5mm} \forall i=1,...,N
\end{equation} 
%\begin{equation}\label{Eq:eqWeightBoundValues}
%0 \leq \epsilon_{i} \leq \delta_{i} \leq 1 ,\hspace{5mm} i=1,...,N
%\end{equation} 
\begin{equation}\label{Eq:eqZi}
z_{i} \in \{0,1\} ,\hspace{5mm} \forall i=1,...,N
\end{equation}

Equations \hyperref[Eq:eqRisk]{(\ref{Eq:eqRisk})} and \hyperref[Eq:eqReturn]{(\ref{Eq:eqReturn})} are objective functions that seek to minimize risk and maximize return, 
respectively, where $N$ is the number of available assets, $w_{i}$ and $w_{j}$ are the decision variables that represent the proportion of budget allocated to assets $i$ and $j$, respectively, $\sigma_{ij}$ is the covariance between assets $i$ and $j$, and $\mu_{i}$ is the expected return of asset $i$. Equation \hyperref[Eq:eqSumWeight]{(\ref{Eq:eqSumWeight})} ensures that all the budget has to be invested. Equation \hyperref[Eq:eqCardinalityBound]{(\ref{Eq:eqCardinalityBound})} is the CC that limits the number 
of assets in the portfolio, which is delineated by $K_{\text{max}}$ and $K_{\text{min}}$ representing the maximum 
and minimum numbers of assets that can be included in the portfolio, respectively. 
Furthermore, $z_i$ is an integer variable which is 1 if asset $i$ is included in the portfolio, and 0,
otherwise. In this study, we impose upper and lower limits on CC instead of a fixed number 
of desirable assets known as $K$; clearly, this is more flexible and can cover other types of 
formulations, as well. It can be transposed either to $\sum_{i=1}^{N} z_{i} \leq K$ when ($K_{\text{min}}=1$, $K_{\text{max}}=K$, $K<N$), which means that the number of assets in the portfolio should not 
exceed an integer number $K$; or to its strict formulation $\sum_{i=1}^{N} z_{i} = K$ when ($K_{\text{min}},K_{\text{max}}=K, K<N$), which can lead to rigorous limits on the number of assets. Equation \hyperref[Eq:eqWeightBound]{(\ref{Eq:eqWeightBound})} guarantees that if asset $i$ is chosen, its dedicated proportion lies between $[\epsilon_{i}$,$\delta_{i}]$ where $0 \leq \epsilon_{i} \leq \delta_{i} \leq 1 ,\hspace{5mm} (i=1,...,N)$. Finally, equations \hyperref[Eq:eqWi]{(\ref{Eq:eqWi})} and \hyperref[Eq:eqZi]{(\ref{Eq:eqZi})} disallows short selling and specify domains of the decision variables.

\section{Methodology:}
\label{Methodology}
As mentioned before, approximation algorithms, compared to exact methods, are more 
effective and reasonable approaches for addressing the portfolio optimization problem as an 
NP-hard. Moreover, a multi-objective approach is preferable due to the opposing objectives 
sought in such problems. NSGA-II is a well-known algorithm that is widely used in the 
literature because of its merits, including its ability to explore the problem’s search space  (\cite{verma2021comprehensive}). It also has proven effective in multi-objective optimization, including portfolio optimization, and is well-recognized within the engineering community (\cite{ma2023comprehensive}). For the purposes of this study, the NSGA-II algorithm was selected to be modified by adopting a new solution encoding system with new handling mechanisms, and incorporating a new operator to address the CC. Additionally, its mating strategies were altered to enhance both exploration and exploitation capabilities. Below is a desctiption of the 
approach adopted to tackle the cardinality constrained portfolio optimization problem.

In \hyperref[Solution encoding and constraint handling]{{Subsection 3.1}}, the solution representation and the approach proposed for handling 
constraints are introduced. The overview of NSGA-II is provided in \hyperref[NSGA-II]{{Subsection 3.2}}. In \hyperref[Proposed NSGA-II]{{Subsection 3.3}}, we present our mating strategies and the new operator. \hyperref[Cardinality constraint handling techniques]{{Subsection 3.4}} describes the techniques proposed for handling CC. Finally, the mechanisms for handling BC 
are described in \hyperref[Boundary constraint handling technique]{{Subsection 3.5}}.

\subsection{Solution encoding:}
\label{Solution encoding and constraint handling}
Representing solutions and measuring objective(s) are crucial prerequisites to the 
application of an EA to a problem. Although measurement of the objective function for each 
optimization model is straightforward, encoding the solution might vary with each model. 
Encoding solutions has a tremendous impact on the performance of an EA. When applied to 
optimization problems, wide use is made of real-coding solution representation in EAs. In 
EAs, the size of the solution representation is fixed to the number of assets in the market and each 
cell represents the weight of each asset. For CCPOP, this approach requires mechanisms to 
remove or add assets if the portfolios are to be made feasible in terms of CC. From a practical 
perspective and due to the costs of monitoring and transactions, the most commonly 
imposed constraint involves a limit on the maximum, rather than the minimum, number of 
assets in the portfolio. Furthermore, CC handling techniques employed for removing 
additional assets from the portfolio in the encoding systems in question are typically more 
time consuming than adding stocks since the number of desirable assets are generally far
less than that of investable ones. Moreover, imposing a minimum of diversification is less 
common and less time-consuming since the gap between $K_{\min}$ and the number of assets in 
the infeasible portfolio is not generally large.

We restricted the decision variables from the universe of assets to the upper bound of CC. 
Our chromosome is composed of two rows of a fixed size, $K_{\text{max}}$, where each element in the 
first row is the asset selected for the portfolio and its corresponding element in the second
row indicates its weight in the portfolio.

For better illustration, imagine a POP with $K_{\min} = 3$ and $K_{\max} = 5$. It must be noted that it 
is impossible to have more than five assets in this approach because the size of each 
chromosome is restricted to five and any values less than that would cause duplicate 
stocks. However, duplication does not necessarily mean infeasibility; rather, the 
chromosome would be feasible as long as there are at least three unique assets in it. If the 
number of unique assets in the chromosome falls below three, then the proposed CC 
handling techniques come into play to repair the chromosome. \hyperref[fig:Chromosome_structure]{Figure 1} depicts the structure 
of a feasible chromosome in the above POP example where there are three unique assets ($i, j, k$) and asset $j$ is repeated three times in the portfolio. The proportional weights of assets $i$ and $k$ are their corresponding weights in the second vector while that of $j$  is the summation of those of the positions in the second row in which asset $j$ appears on the first vector. Hence, The capital allocated to asset $i$ is $0.20$ that to asset $j$ is ($0.15+0.25+0.10$) and one to $k$ is $0.30$ percent of the total budget.

\begin{figure}[htbp]
\centering
\begin{tikzpicture}
  \draw (0,0) rectangle (10,1); 
  \foreach \i in {1,...,5} {
  	\ifnum\i=5
  		\node at (\i*2-1,0.5) {$k$};
  	\else
    	\draw (\i*2,0) -- (\i*2,1);
    	\ifnum\i=1
    		\node at (\i*2-1,0.5) {$i$};
    	\else
    		\node at (\i*2-1,0.5) {$j$};
    	\fi
    \fi
  }

  \draw (0,-1) rectangle (10,0);
  \foreach \i in {1,...,5} {
  	\ifnum\i=5
  		\node at (\i*2-1,-0.5) {$0.30$};
  	\else
    	\draw (\i*2,-1) -- (\i*2,0);
    	\ifnum\i=1
    		\node at (\i*2-1,-0.5) {$0.20$};
    	\fi
    	\ifnum\i=2
    		\node at (\i*2-1,-0.5) {$0.15$};
    	\fi
    	\ifnum\i=3
    		\node at (\i*2-1,-0.5) {$0.25$};
    	\fi
    	\ifnum\i=4
    		\node at (\i*2-1,-0.5) {$0.10$};
    	\fi

    \fi
  }
\end{tikzpicture}
\label{fig:Chromosome_structure}
\caption{Structure of a chromosome in a POP with size 5}
\end{figure}

We believe this representation is superior to the standard ones along the following lines. 
Equation \hyperref[Eq:eqCardinalityBound]{(\ref{Eq:eqCardinalityBound})} forces the number of assets in each solution to fall within a predetermined
range. However, there could be violations in both directions that need to be fixed. As opposed 
to the repairing techniques reported in the literature that consider both situations, a 
violation in terms of exceeding $K_{\text{max}}$ has already been taken care of in this approach by the 
fact that the maximum size of each chromosome is equal to $K_{\text{max}}$.

\begin{comment}
In addition, this encoding system does not depend on the number of available assets, therefore it does not matter how large market data is, computation time should be identical for the same values in equation \hyperref[Eq:eqCardinalityBound]{(\ref{Eq:eqCardinalityBound})}.

In addition, by allowing repeated stocks in the first vector through the mating process the number of assets of offspring can be changed conveniently, and this representation should be able to find a wider range of solutions for the same values of CC faster especially where the gap between lower and upper bound is larger.
\end{comment}
\subsection{Non-Dominated Sorting Genetic Algorithm II:}
\label{NSGA-II}
It is straightforward in single-objective optimization algorithms to determine the fitness and 
the supremacy of solutions in the population via sorting them with regard to the objective(s); 
however, it is slightly different in multi-objective algorithms. Understanding the concept of 
dominance is a key point toward understanding multi-objective optimization problems in 
which solution (a) is considered superior to (b) if it were better in at least one objective and 
equal or better in other objectives. In this case, we would say solution (a) dominates solution 
(b) or, conversely, solution (b) is dominated by solution (a). Additionally, if there exist 
solutions (a) and (b) and if solution (a), in comparison with solution (b), is superior in one 
objective but underperforms in another, then both solutions (a) and (b) are non-dominated 
since none is superior to the other.

NSGA-II is a multi-objective algorithm based on GA that is commonly used to solve multi-objective optimization problems. Similar to other EAs, this algorithm is also based on 
Darwin’s theory of “survival of the fittest”. It preserves the better solutions in each 
generation, known as non-dominated solutions, to survive and reproduce similar solutions.

The algorithm initializes with a random population of solutions from which offspring 
solutions are subsequently generated via crossover and mutation operators to be added to 
the original population. Afterwards, the solutions thus generated are evaluated and sorted 
to select the better ones for the next generation. A non-dominated sorting is performed on 
chromosomes to determine the rank of each individual, with a lower rank implying better 
performance. All the individuals in the first rank are initially added, which will survive in the 
next generation and are thus eliminated from the sorting process. This process continues 
until the number of individuals added reaches a predetermined population size. If the 
number of individuals with rank (n) is greater than the remaining capacity of the population 
size, the crowding distance is applied to maintain solution diversity in each front. This 
approach ensures two advantages; first, solutions with better ranks, compared to others,
survive through each generation and, second, solutions close to each other are eliminated by choosing individuals with distances farther from the others in order to achieve a well-distributed Pareto front (PF). Once sorting is completed, the same 
processes are repeated until the algorithm reaches its stopping criteria. The pseudo code of 
NSGA-II is given in \hyperref[Nalgorithm]{{Algorithm 1}}.

\begin{algorithm}[H]
\label{Nalgorithm}
\caption{Non-Dominated Sorting Genetic Algorithm II}
\begin{algorithmic} 
\STATE Initialize a random population $p$;
\WHILE{Stopping criteria is not true}
%\FOR{k = 2 to 16}
\STATE Produce offspring $p^{'}$ with crossover and mutation;
%\STATE Add $p$ and $p^{'}$ to $A$;
\STATE $A = p \cup p^{'} $;
\STATE Evaluate individuals in $A$;
\STATE Pick the best individuals of size $|p|$ from set $A$ with respect to their rank and crowding distance values for next generation;
%\IF{eChange$<$improvementChange}
%\STATE optimalK$\leftarrow$k-1
%\STATE break
%\ENDIF
%\ENDFOR
\ENDWHILE
\end{algorithmic}
\end{algorithm}

\subsection{Our proposed strategy:}
\label{Proposed NSGA-II}
CCPOP can be considered as a two-step optimization problem. In the first step, a subset of stocks out of the universe of assets has to be determined
and in the next phase, capital allocation to selected assets has to be decided. We, therefore, modified NSGA-II to 
be more focused on exploring the search space in order to find high-quality combinations of 
assets (i.e., the first row of chromosomes) in the specified initial iterations. Next, the algorithm devotes more efforts to adjusting the proportional weights of portfolios 
to find their optimal capital allocations (i.e., the second row of chromosomes). To achieve 
these goal, new tournament selections and a modified mutation operator were developed
while a strategy was deployed to employ them throughout the optimization process. Finally, 
a novel operator referred to as the ‘explorer’ was developed.

\subsubsection{Tournament selection:}
\label{Crossover and tournament selection}
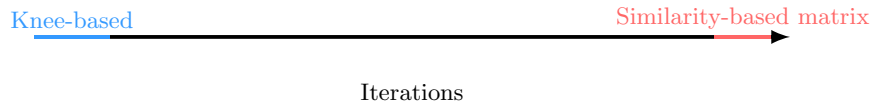
\begin{figure}[h]
\label{fig:Tornoumant}

\centering
\begin{tikzpicture}[>=latex]

    % Draw horizontal line
    \draw[->, line width=0.5mm] (0,0) -- (10,0);
    
    % Define colors
    \definecolor{colorA}{RGB}{51,153,255}
    \definecolor{colorB}{RGB}{255,102,102}
    
    % Calculate coordinates for color bands
    \coordinate (A) at ($(0,0)!0.1!(10,0)$);
    \coordinate (B) at ($(0,0)!0.9!(10,0)$);
    
    % Draw color bands and label them
    \fill[colorA] (0,-0.025) rectangle (A |- 0,+0.025) node[midway,above] {Knee-based};
    \fill[colorB] (B |- 0,+0.025) rectangle (9.75,-0.025) node[midway,above] {Similarity-based matrix};
    \node[below] at (5, -0.5) {Iterations};
\end{tikzpicture}
\caption{Range of activation of each tournament selection technique.
}
\end{figure}

Tournament selection in EAs is meant to manage the selection of individuals to generate 
offspring. It is characterized by certain benefits and challenges; while the benefit involves
awarding higher chances to elite solutions to generate offspring that would lead to more 
promising generations, the challenge lies in the resulting increased complexity of the 
algorithm. Two kinds of tournament selection are then used whose domains of activation are
presented in \hyperref[fig:Tornoumant]{{Figure 2}}.

The first method is based on the Euclidian distance of points from the utopia point (EP); this 
is represented by a knee-based index in \hyperref[fig:Tornoumant]{{Figure 2}} and is active for $\alpha$ \% of early iterations. In our case, EP is the imaginary point that entails a solution with the highest return and the lowest risk. It is obvious that if such a point ever existed, it would dominate 
all the points in PF. Moreover, points located in the knee area of PF represent a balance 
between the two objectives; in other words, they are not too risky and provide more returns
compared to the less risky assets.

If the algorithm finds a proper combination of assets in a shorter time, the savings on 
computational resources and time can then be dedicated to readjustment of
capital allocations in portfolios. Also, randomly mating early individual solutions might
waste computational resources, resulting in lower convergence rates since the majority of 
the solutions in early generations are far from PF. Considering the fact that the unconstrained  PF is 
oriented toward EP in each iteration and that ease of finding better solutions is sought in 
early iterations, the tournament selection as captured by \hyperref[Knee-based tournament selection]{{Algorithm 2}}.

\begin{algorithm}[h]
\label{Knee-based tournament selection}
\caption{Knee-based tournament selection}
\begin{algorithmic} 
\STATE Let $p_i$ be $i$th solution from population $p$; 
\STATE Let $f_i$ be the coordinate of $p_i$ in the objective space;
\STATE Let $c$ be the coordinate of the eutopia point in iteration $i$;
%\STATE Let $n$ be the fixed number of population size in each iteration;
%\STATE Let $r_{e}$ be an exponential random number between 0 and 1 with the parameter of $\displaystyle \frac{n}{\alpha}$;
\STATE Let $r_{e} \sim \text{Exp}(\frac{|p|}{\upsilon})\quad \upsilon \in \mathbb{Z}^+$;
\STATE Let $n_{ci}$ be the rate of crossover in iteration $i$;
%\STATE Let $S_{i}$ be the array containing individuals in generation $i$;
%\STATE $e = \{\}$;
\STATE $L = \displaystyle\{x \hspace{2mm}|\hspace{2mm} x \subset p , |x| = [|p|(n_{ci})], \text{x follows uniform distribution}\}$;
%Sample of solutions' positions of size$[n.n_{ci}]$ picked with uniform probability; 

%\COMMENT{e.g. [6,20,...,1] implies that the 6th and 20th solutions from S are individuals to mate with other individuals of same size from another set ($P_{e}$) to produce 1st and 2nd offspring.}

%\STATE $d_{i} =$ The Euclidian distance of each individual from $C_{i}$;
\STATE $d_i= \displaystyle \sqrt{(f_i-c)^2} ,\hspace{2mm} \forall i=1,...,|p|$;
%\STATE $d^{'}_{i} =$ Sorted values of $S_{i}$ in ascending order;
%\STATE $d^{''}_{i} =$ The corresponding position of $d^{'}_{i}$ in $S_{i}$;

\STATE $d^{'} =$ Sorted $p$ based on their value of $d$ in ascending order;
%\STATE $d^{''}_i= $ Corresponded individual ;
\FOR{$j=1$ to $[|p|*n_{ci}]$}
%\STATE add \{$\min(n,[r_{e}*n])$\}th position of $d^{''}_{i} =$ to $P_{e}$  ;
\STATE Mate $L_j$ with $\displaystyle d^{'}_{\min(|p|,[r_{e}*|p|])}$;
\STATE Normalize weights considering boundaries;
\ENDFOR

\end{algorithmic}
\end{algorithm}

By picking out individuals from two different sets in \hyperref[Knee-based tournament selection]{{Algorithm 2}} solutions closer to EP are 
selected that can outperform their rivals with respect to offspring generation. Moreover, 
equal chances are given to the remaining solutions so that the algorithm would not stuck in 
local optima by only mating points closer to EP. By applying minimal guidance, it was 
ensured that the algorithm would not mate two unsatisfactory solutions with each other;
this additionally led to faster convergence.

The second type of tournament selection is based on similarity of individuals and is activated 
during final iterations to mate closer solutions. The idea behind this tournament is that once 
a decent number of solutions are found, similar pairs of portfolios that share the same or 
almost the same combinations of assets are selected for generating offspring. This ensures 
that almost the same portfolios with slightly different weights are produced during the 
mating process, which in turn guarantees that the algorithm searches for better capital 
allocations for each solution. To achieve this goal, it is essential to measure the distance 
between each two portfolios only based on their asset combinations in order to mate the 
most similar solutions. This is captured by \hyperref[Similarity-based tournament selection]{{Algorithm 3}}, in which $D$ is a $|p|*|p|$ upper triangular matrix whose main diagonal is set to a large negative number such that no solution 
is picked out with itself as a pair.

\begin{algorithm}[H]
\label{Similarity-based tournament selection}
\caption{Similarity-based matrix tournament selection}
\begin{algorithmic}
\STATE Let $M$ be a big number; 
%\STATE Let $D,P_{d}$ be an empty 2D array and an empty array respectively;
%\STATE Let $n_{ci}$ be the rate of crossover in iteration $i$;
%\STATE Let $p_i$ be population in generation $i$; 
\STATE Let $s_i$ be set of assets incorporated in $p_i$;
%\STATE Let $s_{ij}$ be the number of shared elements in solutions $i$th and $j$th;
 
%\STATE Let $n$ be the fixed number of population size in each iteration;
%\STATE Let $S_{i}$ be the array containing individuals in generation $i$;
%\STATE $S^{'}_{i} =$ Sets of unique assets of corresponding individuals in $S_{i}$;
\FOR{$j=1$ to $|p|$}
	\FOR{$k=j$ to $|p|$}
		\IF {$j=k$}
			\STATE $D_{jk} = -M$;
		\ELSE 
			\STATE $D_{jk} = |\{s_j\}\cap\{s_k\}|$;
			%\COMMENT{Constructing distance matrix}
			%Calculate $s_{jk}$ and add it to $D$; 
		\ENDIF
	\ENDFOR
\ENDFOR
\FOR{$l=1$ to $[|p|*n_{ci}]$}
%\STATE Add the position of maximum value of $D$ to $P_{D}$ and change its position in $D$ to $-1$;
\STATE $U = \{(p_j,p_k) \hspace{2mm}|\hspace{2mm} D_{jk}=\max(D)	\}$;
\STATE $D_{jk} = -1$;
\STATE Mate solutions in $U$ with each other;
\STATE Normalize weights considering boundaries;
\ENDFOR
\end{algorithmic}
\end{algorithm}

\subsubsection{Mutation:}
\label{Mutation}
Mutation operators are used to explore the whole search space in order to guarantee that 
the algorithm will not get trapped in local optima. As already mentioned, the algorithm is 
designed to explore the whole search space globally in its initial iterations and search more locally in final iterations. Since mutations have important impacts on this goal, use 
is made of various types of problem-specific mutations that may be divided up into three 
kinds based on their features and exploration capacity for use at each stage of optimization. 
Unlike the crossover operator that chooses an asset and its corresponding weight from 
parents, the mutation one operates in a slightly different manner. This operator  for our solution representation has two steps: mutating the first vector holding a 
combination of stocks, followed by changing the weight of the incorporated assets. The 
representation provides the opportunity for various levels of mutation based on its exploration capabilities. It should be noted that, compared to merely changing the weight of an asset 
of a chromosome, swapping an asset in the first vector has a greater power for exploring the 
search space. The former most likely finds a solution near its parents and is capable of
searching more locally and accurately; it, however, has less capable of throwing a point 
farther away in the search space. Adversely, the latter type of mutation possibly generates 
offspring much farther away from the chosen parent. It has more power to throw a point 
away from its initial position but less accurately. \hyperref[Mutations]{{Algorithm 4}}  presents the mutation 
techniques used in the current study.

A dynamic combination of the mutations cited above was employed to achieve an 
equilibrium among the blindly created points far from the original ones in order to explore the search space and transfer points intelligently through the process. Excluding the initial 
and final iterations that are somewhat biased toward exploring purposes, the second type of
mutation won a greater chance of being selected for mutating solutions.

\begin{figure}[H]%
\label{fig:Mutations}
    \centering
\includegraphics[width= 0.49\textwidth]{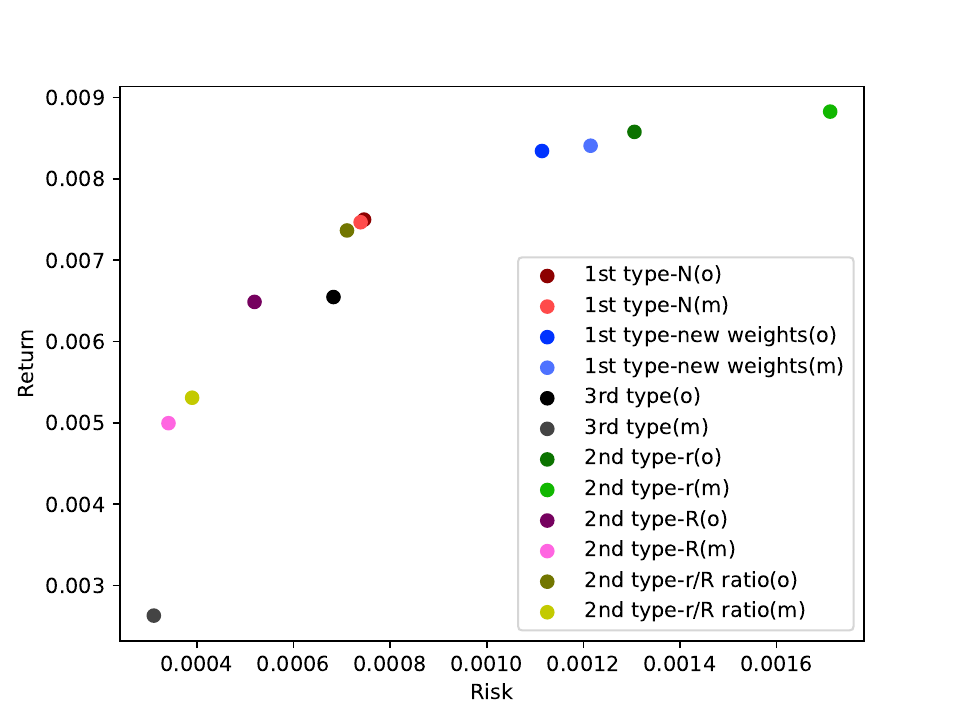}
\caption{Effect of each mutation on random individuals in random and different iterations.}
\end{figure}

Regarding the initial iterations where the objective is to find appropriate asset combinations, 
it is the third and second types of mutations that are mainly in charge with the set $P$. As is 
clear from \hyperref[fig:Mutations]{{Figure 3}}, creating new portfolios generally yields offspring far from the 
population, especially in the initial iteration that is meanwhile capable of preserving
diversity and introducing new mixes of assets to the population pool. Although these 
features are beneficial throughout the optimization process, they are more valuable in early 
iterations.

The principle that supports the use of various sets alongside exponential probability 
distribution in the second type of mutations is investors’ propensity to take risks. Assets are 
characterized by return and standard deviation (SD) of returns (also called volatility) as their
features. Since the historical return of stocks directly influences Equation \hyperref[Eq:eqReturn]{(\ref{Eq:eqReturn})}, assets with 
higher average returns exhibit a better performance in one objective. Even though risk 
objective determination depends on the correlation of assets within a portfolio, it is also 
affected by SD of returns; this is shown in Equation \hyperref[Eq:correlation]{(\ref{Eq:correlation})} where $\sigma_{i}$ is SD of returns of asset $i$ and $\rho$ is the correlation coefficient. 

\begin{equation} \label{Eq:correlation}
\begin{aligned}
\rho &=\displaystyle \frac{\sigma_{ij}}{\sigma_{i}*\sigma_{j}}\\
\sigma_{ij} &= (\sigma_{i}*\sigma_{j})*\rho
\end{aligned}
\end{equation}

$P^{s}$ is sorted by the ratio of returns to risks. Assets with a higher ratio are preferable for all 
investors with any risk profile since everybody prefers to invest in securities with higher 
returns and less volatility.

Risky assets are more attractive to risk-taking investors who seek to invest and build a 
diversified portfolio; hence, stocks with these attributes are more favored and more 
welcomed in half of the PF that contains portfolios with higher returns and volatility. The 
same logic justifies the employment of set $P^{R}$ for risk-averse investors. Although using an 
exponential distribution provides a chance of selection for all assets in the sets, we did not thoroughly disregard set $P$ with a uniform probability distribution so as to maintain 
diversity in the solutions. Using this technique, the algorithm was able to cover more of the 
search space.

Ultimately, the first type of mutations is more likely to be chosen in the final stage for better 
allocations of wealth to selected portfolios since the chance of modifying their asset 
combination is diminished. Moreover, according to \hyperref[fig:Mutations]{{Figure 3}}, this type is better capable of
searching more locally near selected individuals.

\hyperref[fig:Mutations]{{Figure 3}} demonstrates the fitness of random points before and after implementing the 
mutation techniques. The original point for the third type is a random sample from the 
population and the mutated point is a randomly generated portfolio. The fitness of points is 
recorded from different PFs and different iterations but only graphed together for better 
depiction of scalability. In this algorithm, $r$ and $R$ stand for return and risk, while $o$ and $m$ for original and mutated solutions, respectively. Also, $N$ stands for the normal random 
number method that is the first type.

\begin{algorithm}[H]
\label{Mutations}
\caption{Mutations}
\begin{algorithmic}
%\STATE Let $n$ be the fixed number of population size in each iteration;
%\STATE Let $p$ be population in generation $i$;
\STATE Let $n_{mi}$ be the rate of mutation in iteration $i$;
\STATE Let $P^{m}$ be the probability of each element being changed;
\STATE Let $P$ be the set holding universe of assets;% and $m$ be the  number of assets in it ; 
\STATE Let $P^{r}$,$P^{R}$,$P^{s}$ be the universe of assets sorted with respect to their returns, $\sigma$ of returns and ratio of $ \frac{\mu}{\sigma}$ in descending, ascending and descending orders, respectively;
%\STATE Let $N_{R}$ the universe of assets sorted with respect to their SD of returns in ascending order;
%\STATE Let $N_{s}$ be the universe of assets sorted with respect to their ratio of returns to SD of returns in descending order;
%\STATE Let $m$ be number of assets in $N$;

%\STATE Let $r_{e} \sim \mathcal{U}(0,1)$ be an exponential random number between 0 and 1 with the parameter of $\dfrac{m}{\beta}$;
\STATE Let $r_{u} \sim \mathcal{U}(0,1)$, $r_{e} \sim \text{Exp}(\frac{|N|}{\beta})$, $r_{n} \sim \mathcal{N}(0,\frac{1}{(K_{\text{max}})^{2\alpha}})$ , $\quad \alpha,\beta \in \mathbb{Z}^+$;
%\STATE Let $r_{n}$ be an Normal random number with mean 0 and standard deviation (SD) $\dfrac{1}{(K_{max})^\alpha}$;

\IF {the condition for the first type mutations is satisfied}
\STATE $P^{m} = \frac{1}{2}(P^{m})$;
%\STATE Use set $P$ to perform mutation on assets and give equal probability to all assets in the set to be selected each time replacement is required;
\STATE Perform mutation on assets and pick $\displaystyle P_{\{[(r_u)|P|]\}}$ if replacement is required;
\STATE $P^{m} = 2(P^{m})$;
\IF {the condition for normal search near weights is satisfied}
\STATE  $w_{i}= w_{i}+\max(0,w_{i}+r_{n})$ whenever the weight of the $i$th element has to be changed and leave it as $w_{i}$ otherwise;
\ELSE
\STATE Generate a new budget allocation for the subjected individuals;
\ENDIF
\ENDIF

\IF {the condition for the second type mutations is satisfied}
	\IF {the condition for set $P$ is satisfied}
	\STATE Perform mutation on assets and pick $\displaystyle P_{\{[(r_u)|P|]\}}$ if replacement is required;;
	\ENDIF
	\IF {the condition for set $P^{r}$ is satisfied}
	\STATE Pick random individuals with higher returns than the average in population ${i}$;
	\STATE Perform mutation on them and pick $P^r_{\{\min(|P|,[r_{e}*|P|])\}}$ if replacement is required;
	%\STATE Use set $N_{r}$ to perform mutation on assets and pick $\{\min(m,[r_{e}*|N|])\}$th assets from the set each time replacement is required;
	\ENDIF
	\IF {the condition for set $P^{R}$ is satisfied}
	\STATE Pick random individuals with less risk than the average in population ${i}$;
	\STATE Perform mutation on them and pick $P^R_{\{\min(|P|,[r_{e}*|P|])\}}$ if replacement is required;
	%\STATE Use set $N_{R}$ to perform mutation on assets and pick $\{\min(m,[r_{e}*|N|])\}$th assets from the set each time replacement is required;
	\ENDIF
	\IF {the condition for set $P^{s}$ is satisfied}
	\STATE Perform mutation on them and pick $P^s_{\{\min(|P|,[r_{e}*|P|])\}}$ if replacement is required;
	%\STATE Use set $N_{s}$ to perform mutation on assets and pick $\{\min(m,[r_{e}*|N|])\}$th assets from the set each time replacement is required;
	\ENDIF
	\STATE $w_{i} = r_{u}$ whenever the weight of the $i$th element has to be changed and leave it as $w_{i}$ otherwise;
\ENDIF

\IF {the condition for the third type mutation is satisfied}
\STATE Create $[n_{mi}*|p|]$ completely new portfolios;
\ENDIF
\STATE Normalize weights considering boundaries;

\end{algorithmic}
\end{algorithm}

\subsubsection{Explorer:}
\label{Explorer}

Assume an individual that violates CC. To repair it, additional asset(s) must be added to it
while its previous capital distribution is assigned without the possibility for additional
asset(s). Clearly, adding new stock(s) with no readjustment in fund allocation might lead to 
the worst performance regarding objectives. However, modification of weights might help 
obtain better results when new portfolios or new combinations of assets are added to the 
population. With these considerations in mind, we designed a new operator called 
‘explorer’ to readjust capital allocation of individuals so that better solutions could be 
sought. As it intends to readjust budget allocation in the portfolio, the main purpose behind 
the explorer is to find better weights for repaired individuals while it could be generally used to find better solutions as well. If the latter purpose is intended, it should be activated
only after phase two has started.

The idea is to move a solution (subjected individual) in the search space toward another 
(candidate solution) that is not dominated by the chosen solution and, thus, explore the 
space in quest of better results. A candidate solution is the subjected portfolio with a new 
distribution of wealth that is produced either by a heuristic capital distribution technique or 
randomly. Since it is mandatory to pick out only one portfolio as the candidate one and 
because several budget allocation methods are employed, the non-dominated solutions need 
to be modified in order to pick out the best in case multiple alternatives are available. To 
achieve this goal, improvements in the objectives gained by each heuristic technique, as
compared to the subjected individual, may be calculated via Equation \hyperref[Eq:Improvement]{(\ref{Eq:Improvement})}.

\begin{equation} \label{Eq:Improvement}
I(T,C) = \lambda(\dfrac{C_r}{T_{r}}-1)+(1-\lambda)(1-\dfrac{C_{R}}{T_{R}})
\end{equation}

where, $T_r$ and $T_R$ represent the return and risk associated with the targeted portfolio, 
respectively; $C_r$ and $C_R$ are potential candidate's return and risk; and $\lambda$ is an importance coefficient of 
objectives, where $\lambda=0$, $\lambda=1$ only consider improvements gained in risk and return 
objectives, respectively. Meanwhile, $\lambda=0.5$ means no objective is superior to another.
Finally, it should be noted that altering $\lambda$ values can skew the exploration process in favor of 
certain objectives.

\hyperref[Selecting the candidate individual]{{Algorithm 5}} is the pseudo-code for selecting the candidate individual from a targeted 
portfolio. The $H^1$ technique involves equal distribution of money among all the assets in the 
portfolio while random fund allocation is represented by $H^2$. In the $H^3$ 
technique, an asset 
with a high volatility receives a lower proportion of the budget; this technique is more 
suitable for risk-averse investors. $H^4$ allocates funds to each asset based on their $\frac{\text{return}}{\text{volatility}}$ ratio such that the higher the return of an asset compared to its volatility, the more money 
is invested in that asset in the portfolio. Finally, $H^5$ is similar to $H^4$ but only more sensitive 
to volatility so that assets with higher volatility values are restricted in the portfolio.

\begin{figure}[H]
\centering
\subfloat[$H^j$s and original portfolio are  in a non-dominated set.]{\includegraphics[width=0.49\textwidth]{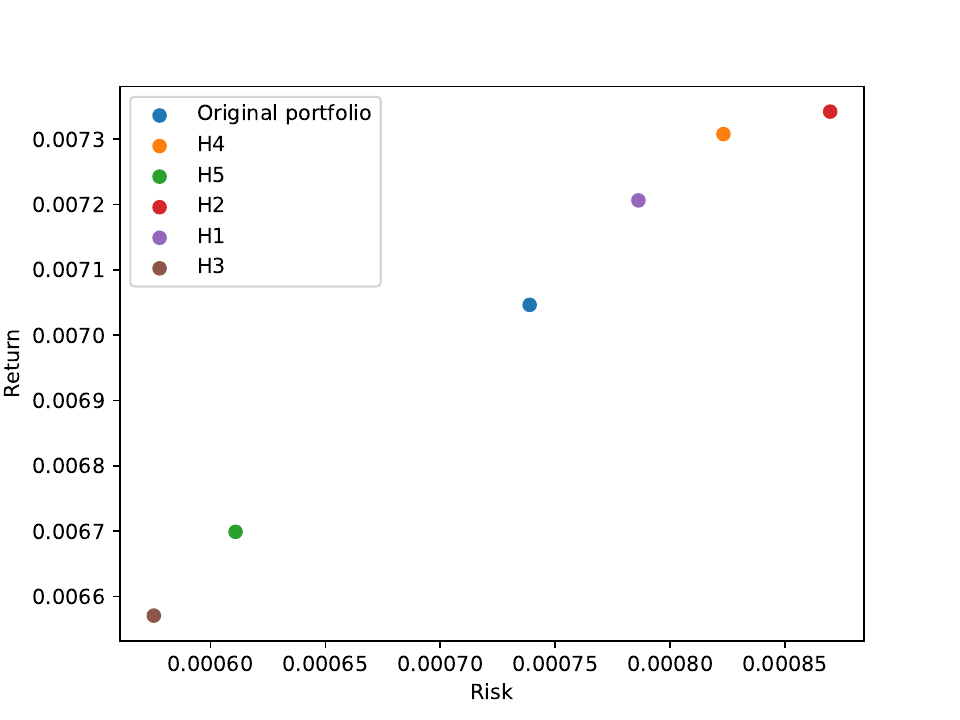}}
\hfill
 \subfloat[$H^1,H^4,H^5$ dominate the original portfolio.]{\includegraphics[width=0.49\textwidth]{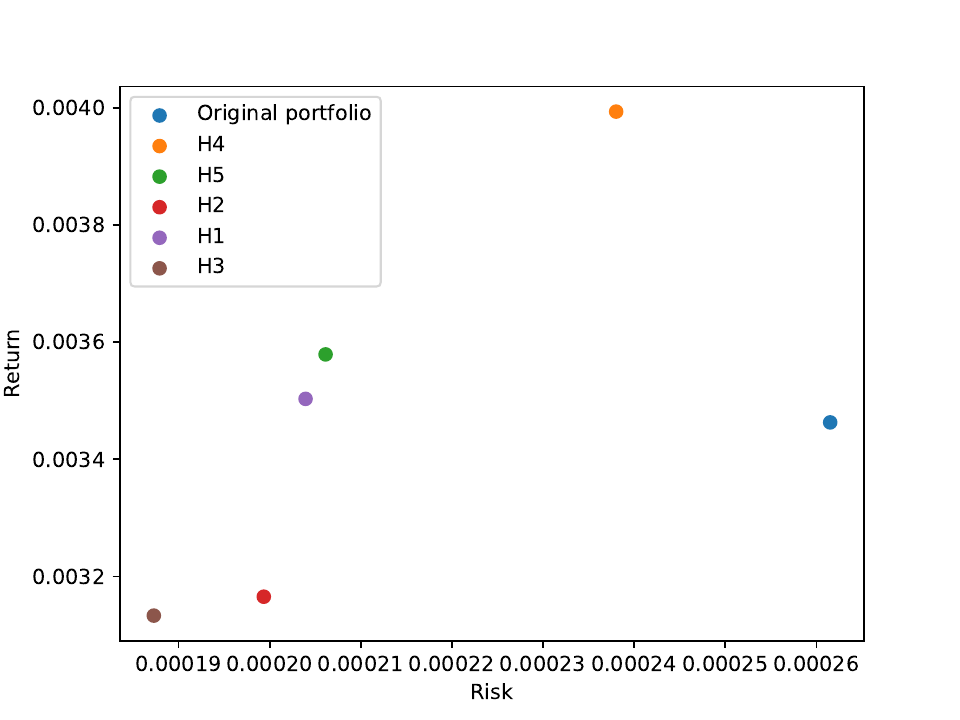}}
\caption{Heuristic fund allocation methods in the explorer operator.}
\label{fig:optimizer_candidates}
\end{figure}

Using this approach to select one individual out of a non-dominated set, we prefer the one 
solution that contributes more to one objective and sacrifices less of another. \hyperref[fig:optimizer_candidates]{{Figure 4}} depicts all the possible outcomes of the operator indicating that the individuals dominated 
by the targeted portfolio are deleted from the selection procedure. \hyperref[fig:optimizer_candidates]{{Figure 4(a)}} shows the 
situation in which none of the heuristic techniques dominates the original solution, nor are 
they dominated by it. In this situation, a solution with a higher value of Equation \hyperref[Eq:Improvement]{(\ref{Eq:Improvement})} is chosen 
for exploration purposes. \hyperref[fig:optimizer_candidates]{{Figure 4(b)}} illustrates the condition that some techniques, namely, $H^1,H^4$ and $H^5$, dominate the targeted portfolio; in this case, only these methods are selected 
for evaluation against Equation \hyperref[Eq:Improvement]{(\ref{Eq:Improvement})}. This signifies that the one solution with higher 
contributions to both objectives is selected since it dominates the original solution. In the 
next step, the space between the targeted portfolio and the candidate is explored using 
Equation \hyperref[Eq:Movement]{(\ref{Eq:Movement})}.

\begin{equation} \label{Eq:Movement}
  w_{i}^1 =
  \begin{cases}
    w_{i}^1 + (r_{u}(0,c)) (w_{i}^2-w_{i}^1)  & \text{} r_{u}(0,c) \geq \displaystyle \frac{-w_{i}^1}{w_{i}^2-w_{i}^1} \\
    r_{u}(w_{i}^1,w_{i}^2) & \text{otherwise }
  \end{cases}
,\hspace{5mm} i=1,...,K_{\text{max}}
\end{equation}
where $w_{i}^1$, $w_{i}^2$ represent allocations to the asset in position $i$ in the subjected individual 
and candidate solutions, respectively. $r_{u}(0,c)$ and $r_{u}(w_{i}^1,w_{i}^2)$ are the uniform random numbers between $(0,c)$ and $( w_{i}^1, w_{i}^2)$ respectively where $c=2$.

\hyperref[Searching]{{Algorithm 6}} explains the process of exploring and selecting the best individual for a targeted 
solution and a candidate. The main goal of this operator is to find a better allocation of wealth 
to the chosen individual that is randomly picked out at no more than $\Omega$ attempts. In this 
situation, three possible scenarios can be envisioned. First, all the heuristic methods fail to 
detect a portfolio that is not dominated by $T$. In this case, we skip \hyperref[Searching]{{Algorithm 6}}. Second, portfolio $C$ dominates $T$. In this case, only the solution that dominates $C$ is accepted as an offspring and is passed on for subjection to the non-dominated sorting procedure since we 
already have $C$. The last possible and more common situation is when $C$ and $T$ are in a non-dominated set. In this scenario, \hyperref[Searching]{{Algorithm 6}} explores their surrounding space for a maximum 
number of $\Omega$ times. If a better solution dominating $T$ is found before $\Omega$ is reached, the search
process will be terminated. This will save computational resources for the obvious reason 
that it is pointless to continue the search process. However, if $T$ remains undominated by the 
solutions found, once the dominated portfolios are eliminated by $T$, the algorithm turns to equation \hyperref[Eq:Improvement]{(\ref{Eq:Improvement})} to select the individual that adds more value to one objective while it sacrifices
less of the other.

\begin{algorithm}[H]
\label{Selecting the candidate individual}
\caption{Selecting the candidate individual}
\begin{algorithmic} 
\STATE Let $S_{i}$ be the asset $i$ in the targeted portfolio;
\STATE Let $H_{i}^{j}$ be $S$ where its corresponding capital allocated to asset $i$ is determined with method $j$;
%\STATE Let $F^{j}$ be the fitness of $P$ where method $j$;
\STATE Let $\eta$ be a small positive number;
%\STATE Let $D$ be an empty array;
\STATE $z = \{\}$;

\FORALL{assets in $S$}
\STATE $H_{i}^1 = k_{\text{max}}^{-1}$;
\STATE $H_{i}^2 = r_u(0,1)$;
\STATE $H_{i}^{3} =  \frac{1}{\sigma_{i}}$;
\IF{$\mu_{i}\geq0$}

\STATE $H_{i}^{4} =  \frac{\mu_{i}}{\sigma_{i}}$;
\STATE $H_{i}^{5} =  \frac{\mu_{i}}{\sigma_{i}^2}$;
\ELSE
\STATE $H_{i}^{4},H_{i}^{5} = \eta$;
%\STATE $H_{i}^{5} = \eta$;
\ENDIF
\ENDFOR
\STATE Normalize each $H^{j}$ considering boundaries if necessary;
%\STATE Delete solutions dominated by $P$;
%\FOR{$j=1:$ to $5$}
%\IF{$W^j$ dominates $P$}
%\STATE Add $W^j$ to $D$
%\ELSE
%\STATE Add $W^j$ to $C$
%\ENDIF
\STATE $H = \{H^j \hspace{2mm}|\hspace{2mm}	H^j \hspace{2mm} \text{is not dominated by S}\} $;
%\STATE Discard each $H^j$ that is dominated by $S$;
\STATE $z = \{z \cup  H^j  \hspace{2mm}|\hspace{2mm} 	H^j \hspace{2mm} \text{dominates} \hspace{2mm} S$	\};
%\STATE Add every $H^j$ that dominates $P$ to $D$;
\IF{$z = \emptyset$}
%\STATE Calculate and Return the $H^j$ with the largest improvement as the candidate based on equation(10);
\STATE Candidate = $\{ H^j \hspace{2mm}|\hspace{2mm} \arg\max_{j \in H} I(T,H^j)	\}$;
\ENDIF
\IF {$|z| = 1$}
\STATE Candidate = $z$;
\ENDIF
\IF{$|z|>1$}
\STATE Candidate =  $\{ H^j \hspace{2mm}|\hspace{2mm} \arg\max_{j \in z} I(T,H^j)	\}$;
\STATE Preserve rest of them;
\ENDIF
%\ENDFOR

\end{algorithmic}
\end{algorithm}

%$\Omega \in \mathbb{N}$
\begin{algorithm}[H]
\label{Searching}
\caption{Searching process}
\begin{algorithmic} 
\STATE Let $T$ and $C$ be the targeted and candidate solution respectively;
\STATE Let $\Omega$ be the maximum number of attempts to find a desirable solution; 
\STATE Let $F$ be a boolean flag that is True if $C$ dominates $T$ and False otherwise;
%\STATE Let $B$ be an empty array;
\STATE $y = \{\}$;

\FOR{$i=1$ to $\Omega$}
\STATE $E_{i} = $ Changed $T$ with equation \hyperref[Eq:Movement]{(\ref{Eq:Movement})};
\STATE Normalize $E_{i}$ considering boundaries if necessary;
\IF{$F$ is True}
	\IF{$E_{i}$ dominates $C$}
		\STATE Return $E_{i}$;
		\STATE Break;
	\ELSE
		\STATE Continue;
	\ENDIF
\ELSE
	\IF{$E_{i}$ dominates $T$}	
		\STATE Return $E_{i}$;
		\STATE Break;
	\ELSE
		%\STATE Calculate $I(E_{i},T)$
		\IF{$I(E_{i},T)>0$}
			\STATE $y = y \cup E_i$;
		\ENDIF
		
	\ENDIF
\ENDIF
\ENDFOR
\IF{$y = \emptyset $}
\STATE Return $C$;
\ELSE
\STATE Return the $E_{i}$ in $y$ with the highest value of $I(E_{i},T)$;
\ENDIF

\end{algorithmic}
\end{algorithm}

\begin{figure}[H]
\centering
\subfloat{\includegraphics[width=0.24\textwidth]{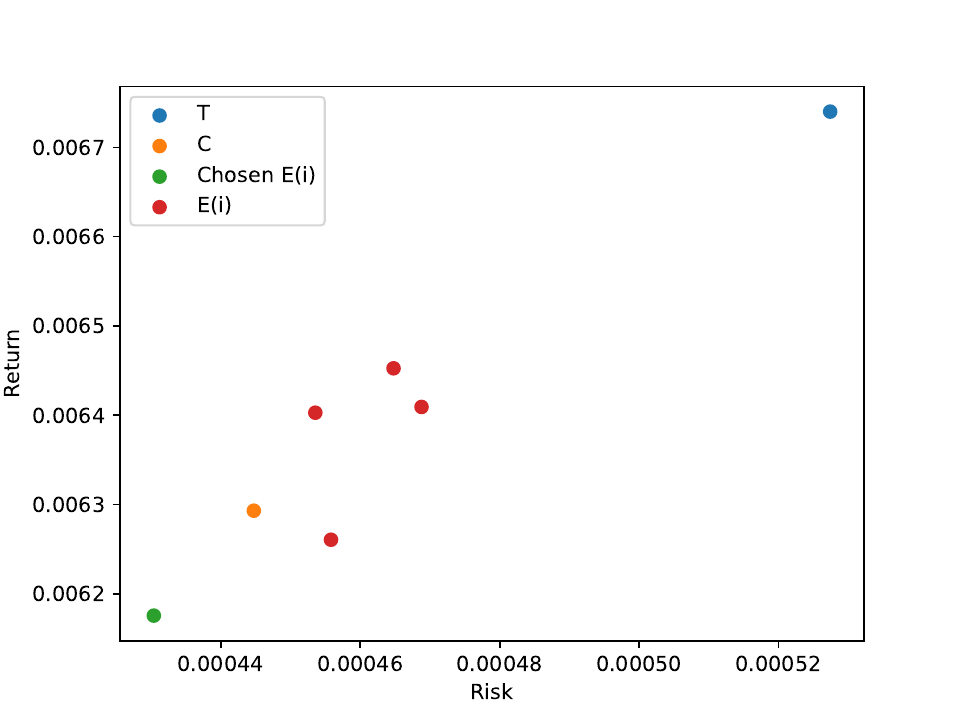}}
\hfill
 \subfloat{\includegraphics[width=0.24\textwidth]{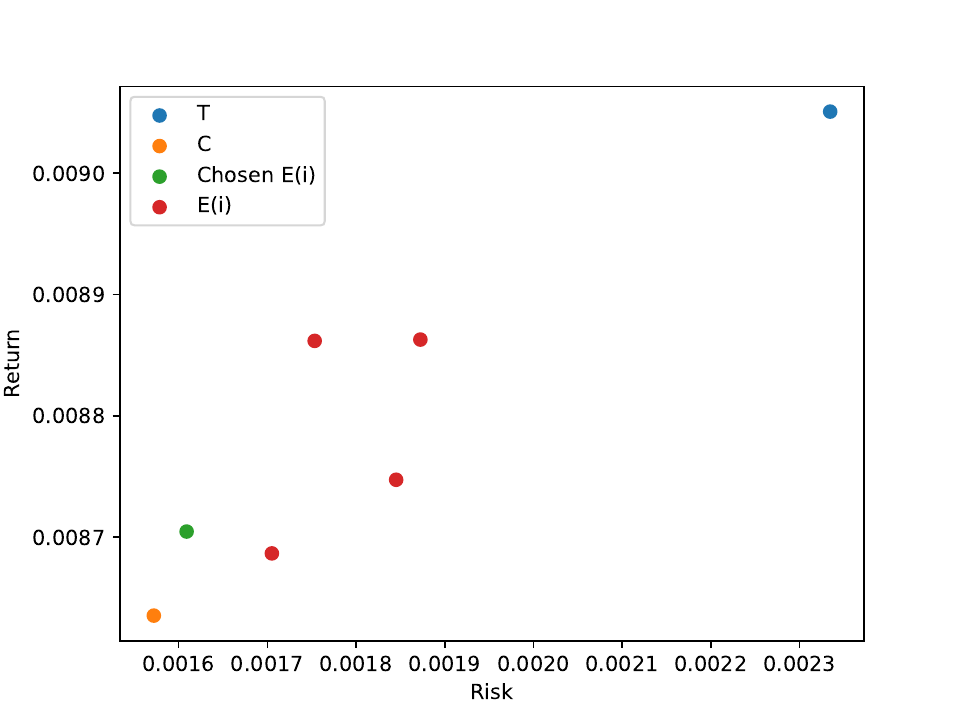}}
\hfill
 \subfloat{\includegraphics[width=0.24\textwidth]{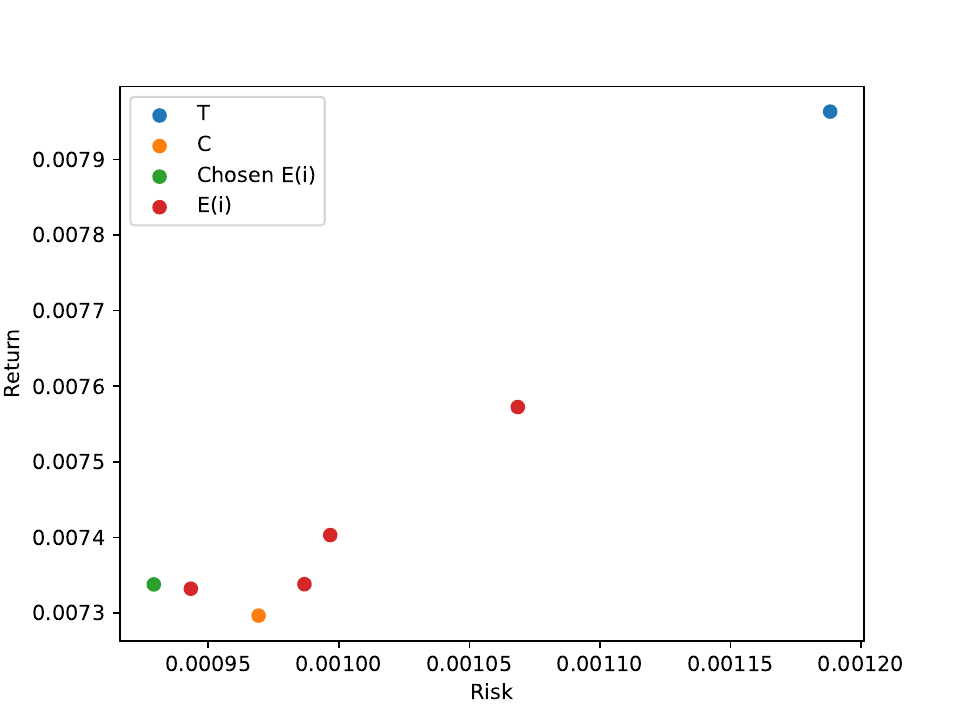}}
\hfill
 \subfloat{\includegraphics[width=0.24\textwidth]{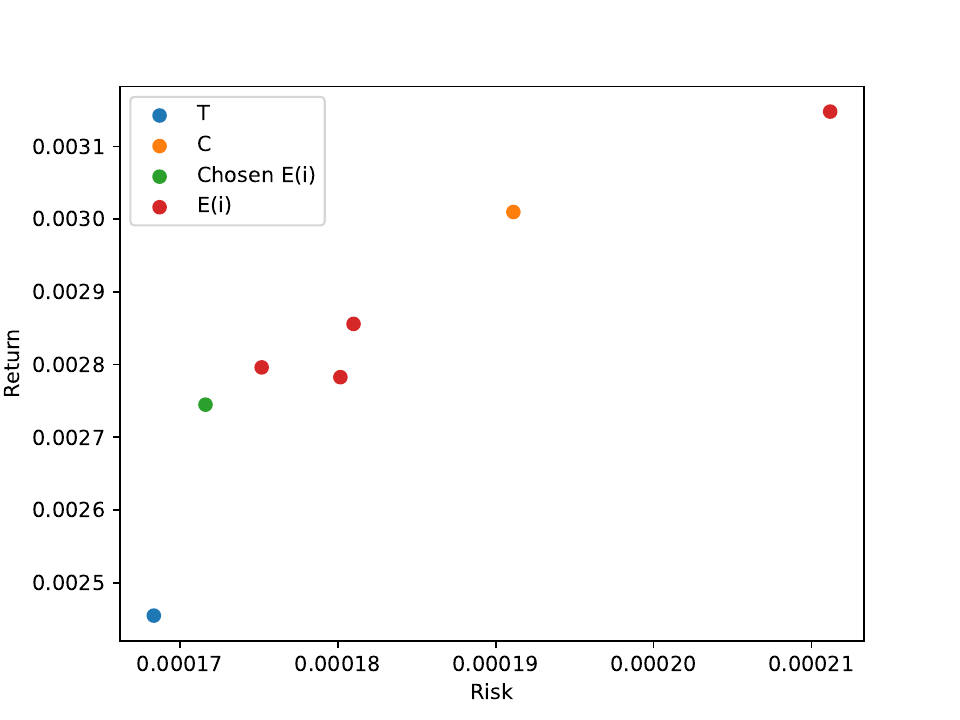}}
\caption{Explorer operator and its output ($\Omega=5$).}

\label{fig:optimizer_search}
\end{figure}

\hyperref[fig:optimizer_search]{{Figure 5}} demonstrates each $E_{i}$ found by the operator along with $T$ and $C$ in the last scenario, where no better solution that dominates $T$ has been found. Green points represent the 
chosen $E_{i}$. Whilst this operator is suitable for such general purposes as generating elite 
offspring, its main objective is to readjust the capital allocation of solutions that are repaired 
via the CC handling algorithms. It is also worth noting that, as opposed to the other operators,
this operator offers the possibility for variable rates of offspring generation. In each iteration, 
the repaired individuals have a higher priority for selection as a targeted portfolio. If the 
repaired solutions are not sufficient to promote a predetermined rate of offspring 
generation, then random individuals will be chosen to fill the gap. Nevertheless, there is a 
subtle contrast between these situations in that if the targeted portfolio is merely a repaired 
individual, the output is then either the individual itself or a portfolio that dominates it, in 
which case the portfolio will directly replace the solution since our purpose is improving the 
solution rather than exploring the space or generating new offspring. However, if the 
targeted portfolio is not a repaired individual, the discovered solution will be substituted for 
the targeted portfolio if it is a superior one.  It will be appended to the pool of 
solutions as an offspring provided it is neither dominated by the targeted portfolio nor does 
it dominate it.

\subsection{Cardinality constraint handling techniques:}
\label{Cardinality constraint handling techniques}

This Section presents an approach to handling CC. Given the solution representation
explained above, it is solely necessary to repair algorithms to append assets to those 
portfolios that violate Equation \hyperref[Eq:cardinality]{(\ref{Eq:cardinality})}  since violation in Equation \hyperref[Eq:portfolio_size]{(\ref{Eq:portfolio_size})} is not possible.

\begin{equation} 
\label{Eq:cardinality}
\sum_{i=1}^{N} z_{i} < K_{\text{min}} ,\hspace{5mm} i=1,...,N
\end{equation}

\begin{equation}
\label{Eq:portfolio_size}
\sum_{i=1}^{N} z_{i} > K_{\text{max}} ,\hspace{5mm} i=1,...,N
\end{equation}

The salient question to ask with regard to appending new assets to a portfolio in the 
repairing stage is which stocks must be added. We propose three techniques to answer this 
question. The first one is called ‘associated-based technique’, which is blind to asset features. 
It repairs portfolios by analyzing the approximation curve and extracting the patterns of the 
current generation to repair solutions for the selection of the next generations. The second,
called ‘clustered-based technique’, evaluates assets to determine a cluster of them suitable 
for the repairing process. Finally, the ‘threshold-based technique’ searches for a combination 
of low-correlated assets to append to the violated portfolio; the assets thus found are
compatible with all in the portfolio.

Once the assets to be included in the portfolio are identified, they will be replaced for the 
repeated stocks. Owing to the fact that the repaired solutions are being marked for weight readjustment purposes via the explorer operator, no specific method is 
required for replacement but it will simply suffice to replace them with the repeated ones. 
In all the methods to be discussed more thoroughly in this article, the chosen assets are 
appended to a set representing unique assets in the relevant solution representation before 
they are replaced in the encoding system with the lowest allocated repeated stocks.

Three cardinality handling techniques are proposed herein to keep the solutions feasible. \hyperref[Associated-based technique]{{Subsection 3.4.1}} describes the associated-based technique while the score-based and the 
threshold-based ones are described in \hyperref[Score-based technique]{{Subsection 3.4.2}} and \hyperref[Threshold-based technique]{{Subsection 3.4.3}} respectively.

\subsubsection{Associated-based technique:}
\label{Associated-based technique}

Individuals in each zone of the PF vary in their composition of assets and their attributes. An 
asset can appear anywhere in the approximation curve accompanied by different stocks.
Since greedy EAs rely on the idea that only the fittest can survive and thrive, solutions that 
survive and reach generation $i$ enjoy more qualified features. We applied a technique to repair individuals based on the frequency with which assets appear in the violated portfolio. The idea behind the associated-based technique is to extract and exploit 
these features in the process of repairing cardinality violations. By adopting this strategy, 
even minute features of groups of solutions that were more qualified than others are
incorporated into violated solutions to handle CC.

To determine how strong the relationships of assets are in each iteration, a comparison set is 
needed that is defined as a set containing the mix of assets selected through the evolution 
process by the algorithm to construct portfolios in the approximation curve for generation $i$. 
Given that the comparison set can be as large as the population size, it is time-consuming to 
check the stock associations in all the portfolios in each generation. Additionally, an asset appears alongside various other assets in each region of the approximated curve. Moreover, it may not 
be appropriate to repair a portfolio placed in a segment where high-risk solutions are located 
in the curve based on the data extracted from risk-averse individuals because it combines 
the features of different kinds of solutions. Finally, each solution in the approximated PF has 
features closer to its neighboring individuals.

To address these issues, we reduce the size of the comparison set by clustering solutions in 
PF. Clustering is the process of grouping individuals with the intention of maximizing 
similarities and minimizing dissimilarities within the clusters. By reducing the size of the 
comparison set in this way, we diminish the size of assets from all those present in the 
current generation to the adjacent stocks, creating solutions close to the violated portfolio,
and picking out those that more regularly co-occur with the elements of the solution that did 
not satisfy CC.

Clustering solutions have a direct impact on the quality of this technique. There are several 
options for clustering. The ideal one is the k-means algorithm employed to determine the 
number of clusters via the elbow method. The obvious drawback associated with this 
method is the additional complexity that k-means will add to our algorithm. Given the fact that the 
individuals in POP have to be clustered based on their returns and risks, clustering them 
based on solely one of the objectives (risk values used in the present case) can be a reliable 
estimate that avoids the use of K-means algorithm with numerous features. Another option 
is to divide the risk objective up into equal intervals based on the highest and lowest risk 
values at intervals defined by Equation \hyperref[Eq:cluster1]{(\ref{Eq:cluster1})}:
\begin{equation} \label{Eq:cluster1}
\text{Interval}_{i} = [R_{\text{min}}+(i-1)x,R_{\text{min}}+(i)x] ,\hspace{5mm} i=1,...,m
\end{equation}
where $R_{\text{min}}$ and $R_{\text{max}}$ , respectively, are the minimum and maximum risks of the present 
portfolios in the iteration; $m$ is number of clusters, and $x$ is the step size calculated as $\frac{R_{\text{max}}-R_{\text{min}}}{\text{m}}$. Each solution whose risk value lies within period $i$ belongs to cluster $i$.

In this clustering technique, compared to the previous ones, more solutions are labeled as 
prime clusters, even in curves where all the solutions are equidistant from their neighbors. 
This is because, at equal intervals, more portfolios can lie on steep lines compared to nearly 
horizontally-shaped individuals in high-risk areas. In a situation with a greater density in 
low-risk areas in the approximation curve, the comparison set in the initial clusters contains 
many more assets while the last clusters suffer from inadequacy of stocks to allow accurate assessment of their relationships. To address this issue, we consider another clustering 
method that groups individuals into equal portfolios. Solutions are sorted in an ascending 
order based on their risk values with members of cluster $i$ specified as follows:

\begin{equation}
\text{Cluster}_i = [\frac{i-1}{m}(P),\frac{i}{m}(P)], \hspace{5mm} i=1,...,\text{m}
\end{equation}
where $P$ is population size.

The range of sorted solutions that belong to cluster $i$ is determined for each value of $i$. Thus,
high values of $m$ yield sparse comparison sets while large comparison sets arise from smaller 
values. It is, therefore, clear that the value of $m$ should be specified in such a way that a
balance is maintained among the various scenarios described above.

The general idea behind this method is to repair each individual based on the data provided 
by the comparison set in its cluster. Nevertheless, if a problem arises with the comparison 
set of portfolio $i$, such as low size or lack of sufficient assets, another set must be constructed 
for comparison. Solutions located in the knee area of the PF are assumed to be more 
desirable to decision-makers due to the balance between the two objectives. Accordingly, 
the following normal random number whose mean is the middle cluster was adopted to 
select a comparison set for these occasions:

\begin{equation}
r_n \sim \mathcal{N}(\frac{m}{2},1)
\end{equation}

It must be mentioned that the underlying idea is to repair portfolios based on good 
individuals, rather than random ones, to reproduce their features. It is, therefore, crucial to 
give some time to the algorithm to build a reasonably reliable Pareto curve for the repairing 
process. To this end, the technique is deactivated until certain iterations have passed. \hyperref[Associated-based handling]{{Algorithm 7}} captures the associated-based handling strategy.

\subsubsection{Score-based technique:}
\label{Score-based technique}

In this method, assets are clustered and portfolios repaired based on the comparative overall 
performance of clusters. The idea is to limit the number of assets from the universe of assets $P$ to a cluster of assets $PR$ so that fewer assets are analyzed more accurately. Having 
clustered assets into $m$ groups, a measurement approach is required to evaluate the 
performance of each cluster to select one. Suppose $O$ be the file containing the correlation 
data of all the assets sorted in an ascending order based on correlation values. The score for clusters $i$ and $j$ will be the numbers of times an asset from clusters $i$ and $j$ has appeared in 
the top $\alpha \%$ of $O$. If $\alpha$ is too small, the data may not be reliable and if it is too large, it takes 
highly correlated stocks into account which are of no interest to our purposes. Once the value 
of $\alpha$ is determined, the value of $m$ may be said to play a crucial role. If the assets do not 
cluster uniformly, the score of cluster $i$ , which contains a larger population of assets 
compared to those of the rest of the clusters, is expected to be higher only due to the 
frequency the assets from cluster $i$ appear in the top $\alpha \%$ of the records, rather than due to 
their lower correlation values.

\begin{algorithm}[H]
\label{Associated-based handling}
\caption{Associated-based handling}
\begin{algorithmic} 
%\STATE Let $P$ be set holding universe of assets;
\STATE Let $G_i$ be required assets to append to the solution $i$ to repair CC;
\STATE Let $PV_{i}$ be the set of assets that is included in the portfolio $i$ that violated CC; 
%\STATE Let $C_i$ be the comparison set for violated portfolio $i$;
\STATE Let $PC_{ij}$ be the portfolio $j$ in comparison set related to the violated portfolio $i$;
%\STATE Let $V$ be a key-value set that keys are all assets in $c_i$ with initial values of 0;
\FORALL{Violated individuals}
\STATE $q_i = \{PV_i\}$;
\STATE Create a key-value array $V$ with keys of $\displaystyle \{\bigcup_{j=1}^{|PC_{ij}|} (PC_{ij})\} \setminus \{PV_i\}$ and initial values of 0;
\FORALL{$PC_{ij}$}
\IF{$PV_{i} \cap PC_{ij} = \emptyset $}
\STATE Continue;
\ENDIF
\IF{$PV_{i} \cap PC_{ij} \neq \emptyset $ \AND $PV_{i} \neq PC_{ij}$}
\STATE Add $|PV_{i} \cap PC_{ij}|$ to values of $PV_i \triangle PC_{ij}$ in $V$ ;
\ENDIF
\ENDFOR
%\STATE Delete $G_i$ repeated assets from the $q_i$ with the lowest corresponding capital allocation;
\IF{$|V|=G_i$}
\STATE $q_i = q_i \cup {V}$;
%\STATE Add all assets in $V$ to the violated individual;
\ENDIF
\IF{$|V|<G_i$}
\STATE $q_i = q_i \cup {V}$;
\STATE $q_i = q_i \cup \{x \hspace{2mm}|\hspace{2mm}	x \in \{P \setminus q_i\} \hspace{1mm},\hspace{1mm} |x| = G_i - |V|\}$;
%\STATE $\displaystyle q_i = q_i \bigcup_{x \in \{P \setminus q_i\},|x| = G_i - |V|} x 	$;
%\STATE Add all assets in $V$ and $G_i-|V|$ random assets from $P\setminus \{PV_i\}$ set to the violated individual;
\ENDIF
\STATE Sort $V$ in descending order based on its values;
\IF{$|V|>G_i$ \AND the value of $G_i$th element of $V$ is greater than $G_i +1$th}
\STATE Add first $G_i$th elements of $V$ to the $q_i$;
\ENDIF
\IF{$|V|>G_i$ \AND the value of $G_i$th element of $V$ is equal to $G_i +1$th}
\STATE $v$ = value of $G_i$th elements of $V$;
\STATE $U=$ assets in $V$ that have a lower value than $v$;
\STATE $q_i = q_i \cup U$;
%\STATE $q_i = q_i \cup \{x \hspace{2mm}|\hspace{2mm}	x \in \{U \setminus q_i\} \hspace{1mm},\hspace{1mm} |x| = G_i - |U|\}$;
\STATE Add $G_i-|U|$ assets with the value of $v$ that have highest $\frac{\mu_i}{\sigma_i}$ ratio to the violated portfolio;
\ENDIF
\STATE Mark the $q_i$ to be optimized with the explorer operator;
\ENDFOR
\STATE Return $q_i$;

\end{algorithmic}
\end{algorithm}

While equally and randomly grouping assets is a solution that immediately comes to mind, 
this approach entails running the risk of having unreliable random data. In this case, the 
assets were clustered using the K-means algorithm to overcome the above problem and the 
elbow method was employed to determine the value of $m$. Subsequently, the score set was
normalized by dividing the score values of clusters $i$ and $j$ by the sum of the sizes of both 
clusters. \hyperref[Score-based handling]{{Algorithm 8}}  is the pseudo code capturing the score-based technique.

To recapitulate, a cluster will be picked out from the set of candidate clusters based on the 
score values of the clusters. Next, either a random stock from the selected cluster will be 
picked out or the assets in the cluster are analyzed and the best one is selected based on the 
normalized sets of $W$ and $Y$, with $\lambda$ being a parameter ranging between 0 and 1 that specifies 
the importance of $W$ and $Y$. In the final stage, an asset from the chosen cluster that has 
achieved a higher rank will be appended to the violated portfolio. It is worth mentioning that,
if CC is to be tackled via this technique, sets like $pv,CR$ and $F$ are generated once for each market data to serve as input to the algorithm and not to be calculated through the process of optimization.

\subsubsection{Threshold-based technique:}
\label{Threshold-based technique}
The goal in this technique is to append assets that have low correlations with already chosen 
assets in the violated portfolios. Even though the process seems straightforward, it is not 
rational to check all the possible assets with all the elements in violated individuals since the 
process is cumbersome. Also, the difficulty increases significantly with growing number of 
assets. To address this issue, a new set was built and fed into the algorithm in order to limit 
the number of assets; the number is limited by a correlation threshold $\gamma$ that can be any value 
in the top percentile and is dependent on the data of each market. Basically, the value of $\gamma$
defines a trade-off point between time and accuracy in this method. \hyperref[Threshold-based handling]{{Algorithm 9}} explains our approach.

Initially, a stock should be picked out from a violated portfolio to analyze the candidate 
assets. Since not all the assets are to be considered for comparison against each selected 
stock, the order in which the assets are selected affects the outcome; this is because the 
candidates will form a new pool of assets as a result each asset being picked out. In this 
situation, either of two approaches could be adopted. Random rearrangement is one possible 
option; however, we opted for the second one that orders them based on their ratios of $\displaystyle \frac{\mu}{\sigma}$ stating that the higher the ratio, the sooner the asset will be selected.

If $t = |PV_i|$ the candidate has a correlation of less than $\gamma$ with all the assets in the violated 
portfolio; this is the ideal scenario for our purposes. What we call ‘the strictness coefficient’ 
is captured by $\displaystyle \frac{|B|}{\varphi}$; it starts with a value of 1, meaning that either a chosen asset has a 
correlation lower than $\gamma$ with all the assets in the portfolio or that candidate is discarded. 
This coefficient is dependent on set $B$ such that the more stocks get eliminated from $B$, the 
less rigid it becomes due to the fact that fewer stocks are available for comparison, and that 
if no stocks are left, the portfolio cannot be repaired. Stocks with a higher correlation than $\gamma$
are gradually permitted to be added during the repair process. Finally, each set $u$ is sorted, 
which helps the low-correlated ones to be found faster.

\begin{algorithm}[H]
\label{Score-based handling}
\caption{Score-based handling}
\begin{algorithmic} 
%\STATE Let $P$ be set holding universe of assets;
\STATE Let $PR_i$ be assets in cluster $i$;
%\STATE Let $G_i$ be required assets to append to the solution $i$ to repair CC;
%\STATE Let $PV_{i}$ be the set of assets that is included in the portfolio $i$ that violated CC; 
%\STATE Let $P_{ij}$ be asset $i$ from set $P$ that is labeled as cluster $j$ based on $\mu_i$ and $\sigma_i$;
\STATE Let $pv_{j}$ be the cluster of asset $j$ in $PV_i$;
\STATE Let $CR_{ij}$ be the correlation score of cluster $i$ and $j$;
\STATE Let $\tau$ be the chance of adding a random asset from the chosen cluster;
%\STATE Let $\Psi= \{x \in \mathbb{N} \mid 0 < x < \text{number of clusters}\}$; 
\STATE Let $F_{i}$ be the top $\Psi$ high scored clusters with $pv_j$, $\Psi \in \mathbb{N}^+ , \Psi \leq \text{number of clusters}$;
%Explaining measurement of scores;
\FORALL{Violated individuals}
%\STATE Delete $G_i$ repeated assets from the $PV_i$ with the lowest corresponding capital allocation;
\WHILE{$G_i \neq 0$}
%\STATE $U= \displaystyle\arg\max_{X'\subseteq X,|X'|=\Psi}\sum_{x\in X'} x$;
\STATE $Q = \displaystyle \{\bigcup_{i=1}^{|PV_i|} F_i\}$;
\STATE $\Theta_i = \displaystyle \sum_{j \in pv}CR_{ij}, \hspace{5mm} \forall i \in Q$;
\STATE  $U=\{PR_{i}:i=\displaystyle\arg\max_{i \in Q} \Theta_{i}\} \setminus \{PV_{i}\} $;
\IF{$rand(0,1) < \tau $}
\STATE Add a random asset from $U$ to $PV_i$;
\STATE $G_i=G_i-1$;
\ELSE
\STATE $W_i=\displaystyle \frac{\mu_i}{\sigma_i}, \hspace{5mm} \forall i \in U$;
\STATE $Y_k = \displaystyle \sum_{j \in PV_i}\sigma_{kj}, \hspace{5mm} \forall k \in U$;
\STATE $W = \displaystyle \{\frac{x}{\max(W)}  : x \in W\}$;
\STATE $Y = \displaystyle \{\frac{\min (Y)}{x}  : x \in Y\}$;
\STATE $\kappa_i = \displaystyle \lambda(W_i)+(1-\lambda)(Y_i),\hspace{5mm} \forall i=1,...,|U|$;
\STATE Add $ U_{\displaystyle\arg\max_{i \in U} \kappa_i}$ to $PV_i$;
\STATE $G_i=G_i-1$;
\ENDIF
\ENDWHILE
\STATE Mark the repaired portfolios to be optimized with the explorer operator;
\ENDFOR
\end{algorithmic}
\end{algorithm}

%numerous methods are possible for clustering assets, here we clustered based oj return and risk to be able to test our results with benchmark data from OR-library.
%Clustering can be applied by almost every feature as long as they capture similarity of movement of 

\begin{algorithm}[H]
\label{Threshold-based handling}
\caption{Threshold-based handling}
\begin{algorithmic} 
%\STATE Let $G_i$ be required assets to append to the solution $i$ to repair CC;
%\STATE Let $PV_{i}$ be the set of assets that is included in the portfolio $i$ that violated CC; 
%\STATE Let $RP_i$ be assets of $PV_i$ but in a random order; 
\STATE Let $I$ be a binary flag that take value 1 if assets $i,j$ have lower correlation of $\gamma$ and 0 otherwise;
\STATE Let $SC_{i}$ be the assets that have correlation less than $\gamma$ with asset $i$ that is sorted in ascending order;
\FORALL{Violated portfolios}
%\STATE Delete $G_i$ repeated assets from the $PV_i$ with the lowest corresponding capital allocation;
%\IF{Condition satisfied }
%\STATE $U = \displaystyle \{\bigcup_{j \in PV_i} SC_{ij}\} \setminus PV_i $;
%\WHILE {$G_i \neq 0$}
%\STATE $PV_i = \displaystyle PV_i \cup \{\text{mode}(U)\}$;
%\STATE $U = U \setminus \{\text{mode}(U)\}$;
%\STATE $G_i = G_i-1$
%\ENDWHILE
%\ENDIF

%\IF{Condition for value-based is satisfied }
%\WHILE{$G_i \neq 0$}
\STATE $\varphi = |PV_i|$;
\STATE $B = \{PV_i\}$;
\FORALL{$PV_i$}
\IF{$G_i = 0$}
\STATE Break;
\ENDIF
\STATE $b = \{x \hspace{2mm} | \hspace{2mm} \displaystyle \arg\max_{x \in B}(\frac{\mu_x}{\sigma_x}) \}$;

\STATE $U = \{SC_{i}\hspace{2mm} |\hspace{2mm} i \in b\} \setminus \{PV_i\}$;
\FORALL{$u \in U$}
\IF{$G_i = 0$}
\STATE Break;
\ENDIF
\STATE $ t = \displaystyle \sum_{x \in PV_i} I_{ux}$;
\IF{$t = |PV_i|$}
\STATE $PV_i = \displaystyle PV_i \cup {u}$;
\STATE $G_i = G_i - 1$;
%\ENDIF
\ELSIF{$t \displaystyle \geq (\frac{|B|}{\varphi})(|PV_i|)$}
\STATE $PV_i = \displaystyle PV_i \cup {u}$;
\STATE $G_i = G_i - 1$;
\ENDIF
\ENDFOR
\STATE $B = \{B\} \setminus \{b\}$;
\ENDFOR
%\ENDWHILE
%\ENDIF
\STATE Mark the repaired portfolios to be optimized with the explorer operator;
\ENDFOR

\end{algorithmic}
\end{algorithm}

\subsection{Boundary constraint handling technique:}
\label{Boundary constraint handling technique}
To handle BC, a modified version due to \cite{chang2000heuristics} was employed to ensure 
boundary feasibility in our solution representation. The constraint guarantees that the 
position of each allocated capital corresponding to asset $i$ in the encoded solution stays 
between $\epsilon_i, \delta_i$ while the summation of all the positions is equal to unity. In our encoded 
solution, this technique exhibits perfect performance in handling BC for $K_{\min} = K_{\max}$ However, the more $K_{\min}$ and $K_{\max}$ diverge from each other, the chances for the 
method to fail increases due to repeating assets. In our problem, if asset $i$ appears more than 
once in the encoded solution, the summation of all those positions should also vary between $\epsilon_i, \delta_i$; this calls for the repair of these types of violations to be duly considered. The BC 
repairing stage is the last step in each run of the algorithm proposed in this study. Hence, all 
the chromosomes follow the cardinality policy, all positions observe BC, and the total 
proportion of money allocated to all is equal to our budget. Thus, there is nothing else to do 
but to check for the boundary violation of repeated assets and, thereby, either to assign the 
violated amount to one asset or to distribute it among others.

Nonetheless, the repair will not be possible if either of two conditions hold. The first involves
the non-uniqueness of assets in a chromosome in the sense that if there are only a few unique 
assets in the encoded solution while the rest of the positions are just duplicates of those 
assets, it is possible that the total budget allocated to asset $i$ will violate $\delta_i$ even if all the
positions have a value of $\epsilon_i$; that is, the lowest possible weight allowed for investment below
which the chromosome will be overexposed to asset $i$. The second situation causing a 
problem is when the summation of the upper bounds of all unique assets falls below unity
while the summation of all the positions equals unity. These two situations indicate that even 
though the solution follows the cardinality policy, assets are noticeably inadequate. Both 
cases will cause problems to the repair process. Although one approach to address this 
problem is to delete such problematic portfolios on grounds of low chances for such 
situations to occur, these solutions are still repaired for the BC repairing process because the 
computational resources, time, and energy are spent on them. \hyperref[Controlling cardinality]{{Algorithm 10}}  captures this approach for 
lifting the problems cited.

In both scenarios, a new asset has to be replaced with another in an encoded solution. 
Because this is the last step in a run, unlike in the CC handling techniques in which weights 
are to be optimized after the process, it does matter which position of the stock to remove 
from the chromosome. New assets are replaced for those in positions with the least 
contribution to the assets’ total allocation so that the original portfolio experiences a 
minimal change. The contribution of each position is defined as the capital proportion 
associated with the asset in that position divided by the weight of that asset in the portfolio. 
When the problem is a lack
of unique assets, assets from the current generation are 
selected to repair them. Moreover, if an encoded solution is over-exposed to an asset, an 
initial attempt is to determine whether it could be repaired using its constituent assets since, 
in this way, the modification the portfolio undergoes is less than that due to adding another 
asset from the generation.

Once all the solutions are converted into the desirable format, the violated chromosomes will 
have to be readjusted. The only violation that can take place is that of the upper bound of 
assets due to duplication. So, it is necessary to figure out the extent of upper bound violation 
by an asset. Moreover, the amount of capital that assets are allowed to absorb in order for 
them to stay feasible needs to be determined. This is a two-stage process. First, the 
violated amount should be subtracted from the specific position that violated the upper 
bound. In the next step, this amount must be allocated to another feasible position of another 
asset in our encoded solution. The question that arises at this juncture is which position of 
an asset can be subtracted from or allocated to. Clearly, it is reasonable to terminate the 
process in a minimum number of iterations. In each step, therefore, we need to determine 
the violated amounts that can be subtracted from or allocated to a position. This is 
advantageous on economic grounds since it would take place in one swoop, rather than in
several steps, without the need for removing that amount in fractions and distributing them 
among multiple positions. \hyperref[Allocation repairing process]{{Algorithm 11}} explicitly illustrates the BC handling technique.

\begin{algorithm}[H]
\label{Controlling cardinality}
\caption{Controlling cardinality}
\begin{algorithmic} 
\STATE Let $\displaystyle p_i^j$ be the asset $j$ from the portfolio $p_i$;
%\STATE Let $SP_i^j$ be the position $j$ in our solution representation of $p_i$;
%\STATE Let $m_i^j$ be number of the appearance of stock $j$ from $p_i$ in our solution representation;
\STATE Let $m_i^{jk}$ be the proportion allocated to asset $j$ in position $k$ from $p_i$ in our solution representation; 
\STATE Let $S_i^{jk}$ be the asset $j$ in position $k$ from $p_i$ in our solution representation; 
\STATE Let $\chi_i^j$ be the lowest contributed position of asset $j$ in $p_i$ in encoded solution;
\STATE Let $PG$ be the set of all assets that participated in the approximation curve in current generation;

\FOR{$i=1$ to $|p|$}
%\IF {$\displaystyle \sum_{j \in p_i} \delta{j}<1$}
\WHILE{$\displaystyle \sum_{j \in p_i} \delta{j}<1$}
\IF{$|p_i| = K_{\text{max}}$}
%\STATE $p_i = \{p^{'}_i \cup x \hspace{2mm}|\hspace{2mm} p_i^{'} \subset p_i,\hspace{2mm} |p_i^{'}| = |p_i|-1,\hspace{2mm} x \in \{PG \setminus p_i\},\hspace{2mm} \delta_x \geq (1-\sum_{j \in p_i} \delta{j}) ,\hspace{2mm} |x|=1 \}$
\STATE $U = \displaystyle \{x\hspace{2mm}|\hspace{2mm}	x \in \{PG \setminus p_i\},\hspace{2mm} \delta_x \geq (1-\sum_{j \in p_i} \delta{j}) ,\hspace{2mm} |x|=1\}$;
\STATE $p_i = p_i \setminus \{p_i^l \hspace{2mm}|\hspace{2mm} l=\min(\chi_i^j) \hspace{2mm} \forall j \in p_i \}$;
\STATE $S_i^{jk} = U,\text{where} \hspace{2mm} k =\{ \min(\chi_i^j) \hspace{2mm} \forall j \in p_i \}	 $;
\ENDIF
\IF{$|p_i| = K_{\text{max}}-1$}
\STATE $U = \displaystyle \{x \hspace{2mm}|\hspace{2mm} x \in \{PG \setminus p_i\},\hspace{2mm} \delta_x \geq (1-\sum_{j \in p_i} \delta{j}),\hspace{2mm} |x| =1		\}$;
%\STATE $p_i = \displaystyle p_i \cup \{x \hspace{2mm}|\hspace{2mm} x \in \{PG \setminus p_i\},\hspace{2mm} \delta_x \geq (1-\sum_{j \in p_i} \delta{j}),\hspace{2mm} |x| =1		\}$;
\STATE $S_i^{jk} = U,\text{where}\hspace{2mm} k =\{ \chi_i^j \hspace{2mm}|\hspace{2mm} |m_i^j|>1 \hspace{2mm}, \forall j \in p_i ,	 \}$;
\ELSE
\STATE $U = \{x \hspace{2mm}|\hspace{2mm} x \in \{PG \setminus p_i\},\hspace{2mm} |x| =1		\}$;
\STATE $S_i^{jk} = U,\text{where} \hspace{2mm} k = \{\min (\chi_i^j), \hspace{2mm}  |m_i^j|>1 , \hspace{2mm} \forall j \in p_i\}$;
\ENDIF
\STATE $p_i = p_i \cup U$;
\ENDWHILE
%\ENDIF
%\FORALL{ Assets in $p_i$}
%\STATE $|m_i^{j}|$;
%\ENDFOR
%\STATE $U_j = \displaystyle	 |m_i^{j}| , \hspace{2mm} \forall j \in p_i $;
\FOR{$j=1$ to $|p_i|$}
\WHILE{$\displaystyle	 \epsilon_j |m_i^{j}| > \delta_j , \hspace{2mm}$}
\STATE $U = \{x \hspace{2mm}|\hspace{2mm} x \in \{p_i^j \hspace{2mm}|\hspace{2mm} \epsilon_j(|m_i^{j}|+1) \leq \delta_j, \hspace{2mm} \forall j \in p_i \} ,\hspace{2mm} |x| = 1\} $ ;
\IF{$U \neq \emptyset$}
\STATE $S_i^{jk} = U ,\text{where} \hspace{2mm} k = \chi_i^j$;
\ELSE
\STATE $U = \{x \hspace{2mm}|\hspace{2mm} x \in \{PG \setminus p_i\},\hspace{2mm} |x| =1		\}$ ;
\STATE $p_i = p_i \cup U$;
%\STATE $S_i^{jk} = U ,\hspace{2mm} \text{where}\hspace{2mm} k=\min(m_i^{jk},\forall k=1,...,|m_i^j|)$;
\STATE $S_i^{jk} = U , \text{where} \hspace{2mm} k = \chi_i^j$;
%\STATE $\chi_i^j = U$;
\ENDIF

\ENDWHILE
\ENDFOR
\ENDFOR
\end{algorithmic}
\end{algorithm}

\begin{algorithm}[H]
\label{Allocation repairing process}
\caption{Allocation repairing process}
\begin{algorithmic} 
%\STATE Let $ p_i^j$ be the asset $j$ from the portfolio $p_i$;
\STATE Let $\phi_i^j$ be the total proportion of budget allocated to asset $j$ in $p_i$;
%\STATE Let $a_i^j$ be amount of violation for asset $j$ in portfolio $i$,$a_i^j = \max(0,\phi_i^j-\delta_j)$;
%\STATE Let $m_i^{jk}$ be the proportion allocated  asset $j$ in position $k$ from $p_i$ in our solution representation; 

\FOR{$i=1$ to $p$}
\STATE $a_j = \phi_i^j-\delta_j, \hspace{2mm} \forall \hspace{2mm} j=1,...,|p_i| $;
\FORALL{Assets in $p_i$ with $a_j > 0$}
%\STATE $U = \max(m_i^j) $;
\IF{ $ \max(m_i^j) - a_j \geq \epsilon_j$}
\STATE $\max(m_i^j) = \max(m_i^j) - a_j$; $//$ Subtract $a_j$ from the largest position

%distribution

\STATE $l,h = \{(l,h) \hspace{2mm} | \hspace{2mm} (\phi_i^l + a_j ) \leq \delta_l,\hspace{2mm} a_l <0,\hspace{2mm} |l|,|h| =1 ,\hspace{2mm} \forall \hspace{2mm} l \in p_i \hspace{1mm},\hspace{1mm} h \in m_i^{l} \}$;
\IF {$h \neq \emptyset$}
\STATE $a_l = a_l + a_j, m_i^{lh} = m_i^{lh} + a_j $;
%\STATE $m_i^{lh} = m_i^{lh} + a_j$;
\STATE $a_j =0$;
\ELSE

\WHILE{$a_j > 0 $}
\STATE $l,h = \{(l,h) \hspace{2mm} | \hspace{2mm} a_l <0,\hspace{2mm} |l|,|h| = 1, \forall ,\hspace{2mm} l \in p_i \hspace{1mm},\hspace{1mm} h \in m_i^l  	 \}$;
\STATE $t = m_i^{lh}$;
\STATE $m_i^{lh} = \min(\delta_l,m_i^{lh} + a_j)$;
\STATE $a_j = a_j - (m_i^{lh} - t), a_l = a_l + (m_i^{lh} - t) $;
%\STATE $a_l = a_l + (m_i^{lh} - t) $;
\ENDWHILE
\ENDIF

\ELSE
\WHILE{$a_j > 0$}
\STATE $U = \max(m_i^j)$;
\STATE $\max(m_i^j) = \max(\epsilon_j,\max(m_i^j)-a_j)$; $//$ Remove maximum allowed weight from the largest position
\STATE $a_j = a_j - (U - \max(m_i^j))$;
%distribution

\STATE $l,h = \{(l,h) \hspace{2mm} | \hspace{2mm} \phi_i^l + (U-\max(m_i^j)) \leq \delta_l,\hspace{2mm} a_l <0,\hspace{2mm} |l|,|h| =1 ,\hspace{2mm} \forall \hspace{2mm} l \in p_i\hspace{1mm},\hspace{1mm} h \in m_i^l \}$;
\IF{$h \neq \emptyset $}
\STATE $m_i^{lh} = m_i^{lh} + (U-\max(m_i^j))$;
\STATE $a_l = a_l + (U-\max(m_i^j))$;
\ELSE
\STATE $t = U - \max(m_i^j) $;
\WHILE{$t>0$}
\STATE $l,h = \{(l,h) \hspace{2mm} | \hspace{2mm} a_l <0 ,\hspace{2mm} |l|,|h| = 1, \forall ,\hspace{2mm} l \in p_i \hspace{1mm} ,\hspace{1mm} h \in m_i^l	 \}$;
\STATE $t^{'} = m_i^{lh} $;
\STATE $m_i^{lh} =  \min(\delta_l,m_i^{lh} + t)$;
\STATE $a_l = a_l + (m_i^{lh} - t^{'}), t = t -(m_i^{lh} - t^{'}) $;
%\STATE $t = t -(m_i^{lh} - t^{'}) $;
\ENDWHILE

\ENDIF
\ENDWHILE

\ENDIF
\ENDFOR

\ENDFOR

\end{algorithmic}
\end{algorithm}

\hyperref[Our proposed algorithm]{{Algorithm 12}}  summarizes the modified NSGA-II, on which our strategies are implemented to illustrate the chronological order of each step, all in a single view. It is seen that a random 
population of portfolios of the upper bound of cardinality is initially generated. Next, the
algorithm starts its search for the mating process using phase one policies. After 
mating solutions and generating offspring, cardinality and boundary constraints are
handled. In this process, use is made of a combination of all the repair techniques described 
above. If the associated-based handling technique is adopted for repairing the portfolios, the 
algorithm checks whether a certain number of iterations has passed or not. If not, then 
solutions in that specific generation are not repaired. The reason behind this strategy is to 
give time to the solution to mate with themselves rather than artificially manipulating them 
from the beginning. The maximum number of iterations for the handling technique to remain
deactivated is less than that at the beginning of stage two. Once CC is repaired, the solutions 
that violated BC are also repaired. As a last step of each iteration, the parents and offspring 
are subjected to the non-dominated sorting process and crowding distance functions 
whereby the best individuals are selected for the next generation. Phase two is accomplished 
in almost the same manner with some minute changes. In addition to changes in mutations and crossover policies, the explorer operator is activated in even iterations to generate 
offspring independently if the number of repaired individuals falls below certain levels.

\begin{algorithm}[H]
\label{Our proposed algorithm}
\caption{Proposed approach implemented on NSGA-II}
\begin{algorithmic} 
%\STATE Let $p$ be a population of portfolios;
\STATE Let $I_M$ be the  maximum iteration;
\STATE Let $\Pi$ be the first iteration of the second phase, $\frac{1}{2}I_M <\Pi < I_M$;
\STATE Let $n_\text{mark}^i$ be the number of repaired solutions in iteration $i$; 
\STATE Let $n_{\text{opt}}$ be rate of producing offspring with explorer operator;
\STATE Let $\Phi$ be the last iteration that \hyperref[Associated-based handling]{{Algorithm 7}} stays deactivated;
\STATE Initiate a random population p;
%\STATE Let $\Lambda \in \mathbb{N^+},\Lambda > 1 $ ;
\FOR{$i=:1$ to $I_M$}
\IF{$i < \Pi$}
\STATE $C = $ Crossover($p $, Phase one policies for mating individuals in iteration $i$);
\STATE $M = $ Mutation($p$, Phase one policies for mutating individuals, Exclude first-type);
\STATE $A = M \cup C $;
\IF{\hyperref[Associated-based handling]{{Algorithm 7}} is a candidate for handling CC \AND $i \leq \Phi$}
\STATE Do not repair CC in violated solutions in $A$;
\ELSE
\STATE Repair CC in violated solutions in $A$;
\ENDIF
\ELSE
\STATE $C = $ Crossover($p $, Phase two policies for mating individuals in iteration $i$);
\STATE $M = $ Mutation($p$, Phase two policies for mutating individuals,Exclude third-type);
\STATE $A = M \cup C $;

\STATE Repair CC in violated solutions in $A$;
\IF{$n_\text{mark}^i < [(p) n_{\text{opt}}]$ \AND  $i\%2 =0$ }
\STATE  $O = ([(p) n_{\text{opt}}] -n_\text{mark}^i)$ offspring produced by explorer($p$);
\STATE $A = A \cup  O$;
\ENDIF

\ENDIF

\STATE repair BC in violated solutions in $A$ ;
\STATE $p = $ individuals of size $|p|$ selected with Non-dominated sorting and crowding distance($A$);
\ENDFOR

\end{algorithmic}
\end{algorithm}

\section{Experiment results and discussion}
\label{Experiment results}
This section compares the performance of the proposed strategy against the conventional 
solution-encoded NSGA-II. Comparisons are also made between the proposed approach and 
the conventional NSGA-II using our solution representation but only with normal mating 
strategies in order to verify the effectiveness of the proposed strategies and their impacts on 
convergence rate. For this purpose, the approach proposed herein was implemented on 
three markets in the OR-library benchmark dataset, namely, DAX 100, S\text{\&}P 100, and Nikkei 225
containing 85, 98, and 225 assets, respectively. Moreover, weekly prices of the assets 
obtained from TSE over the period from 29 March 2017 to 3 May 2023 were exploited while 
the stocks that had not been traded more than several weeks during the study period were 
excluded; this resulted in a pool of 418 assets.

\begin{figure}[H]
\centering
\subfloat{\includegraphics[width=0.24\textwidth]{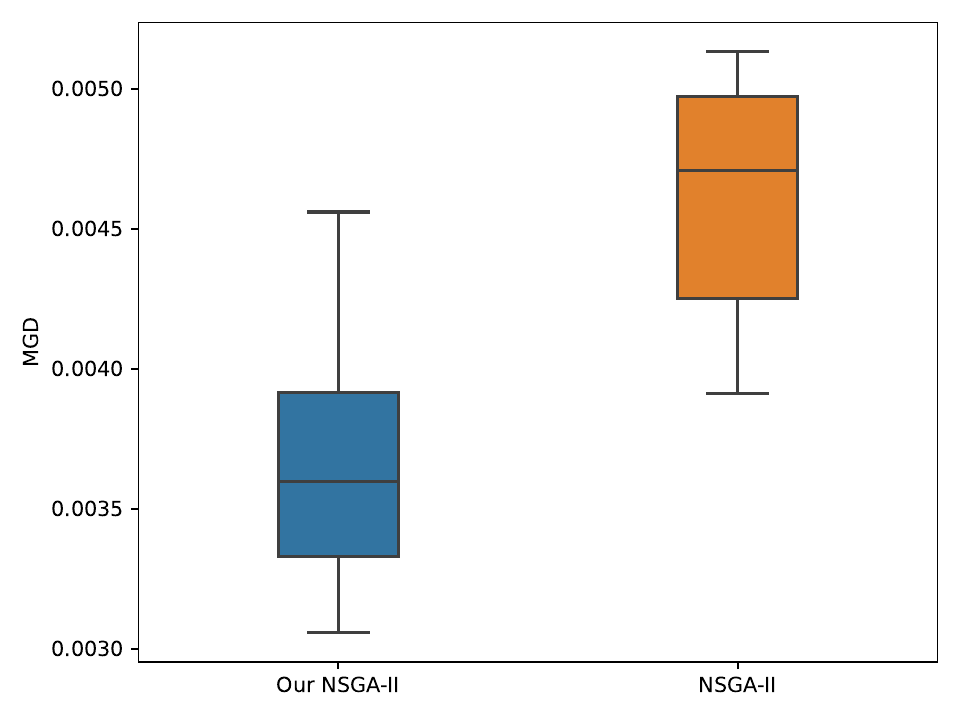}}
\hfill
 \subfloat{\includegraphics[width=0.24\textwidth]{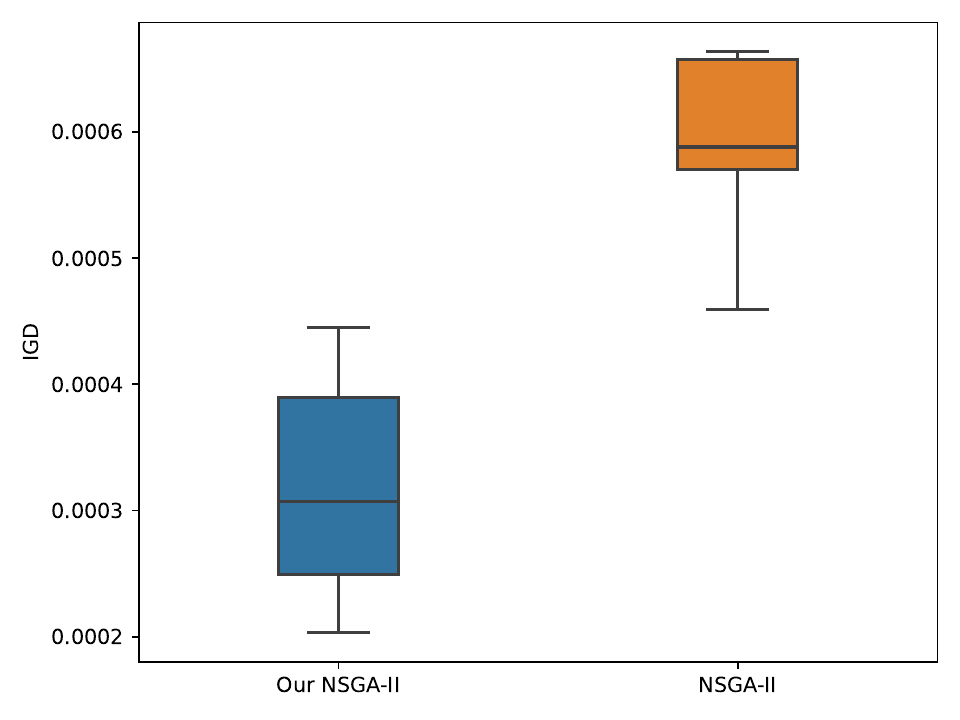}}
\hfill
 \subfloat{\includegraphics[width=0.24\textwidth]{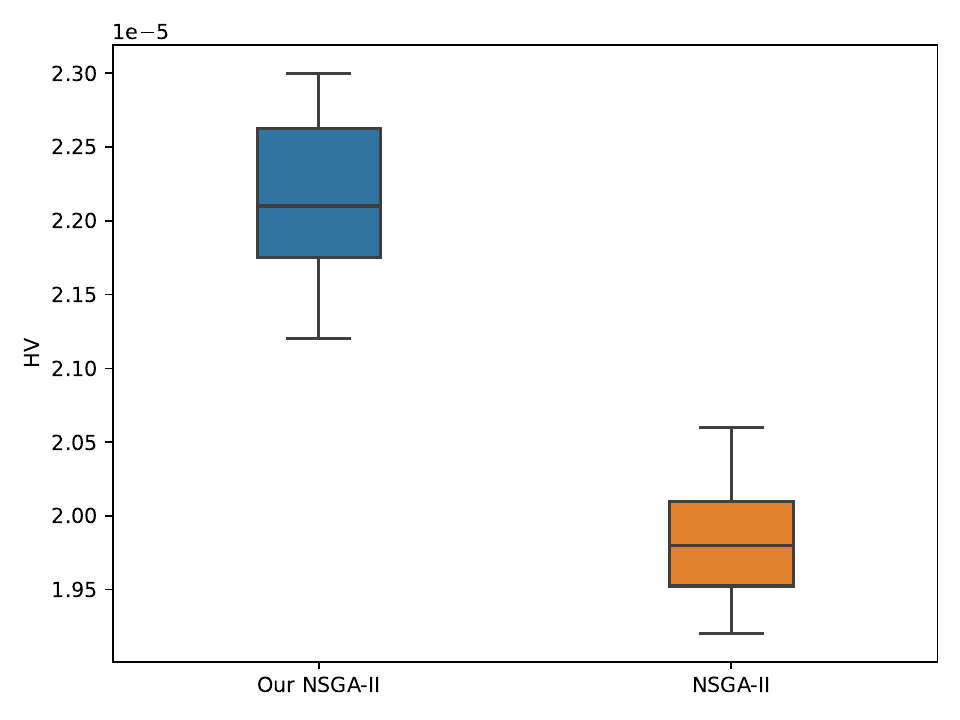}}
\hfill
 \subfloat{\includegraphics[width=0.24\textwidth]{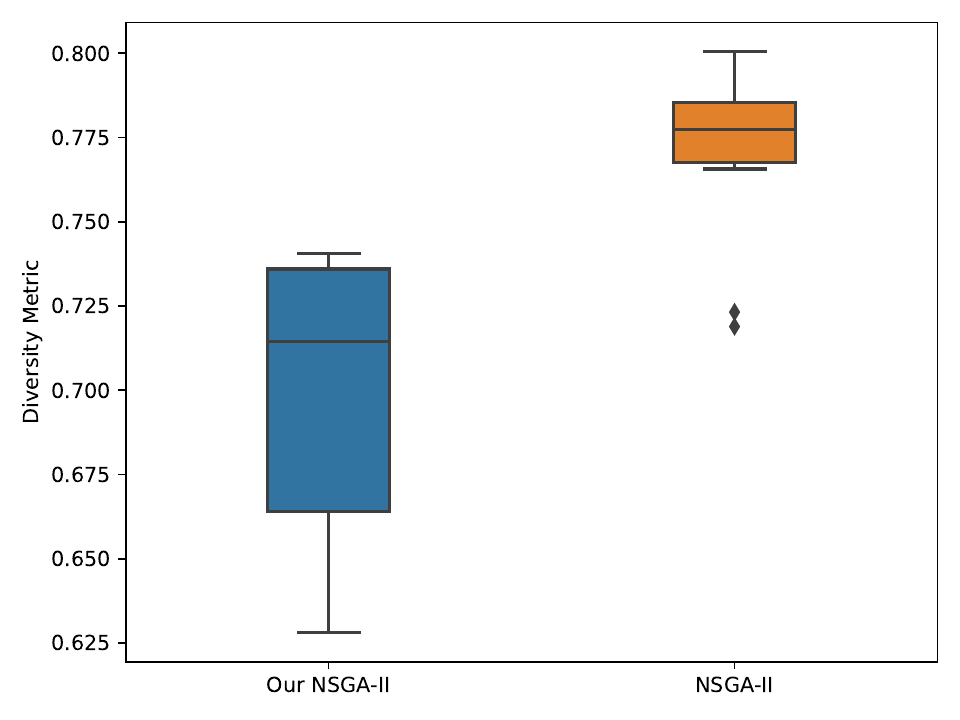}}
\caption{Comparison of performance indicators for DAX 100.}
\label{fig:DAX}
\end{figure}

\begin{figure}[H]
\centering
\subfloat{\includegraphics[width=0.24\textwidth]{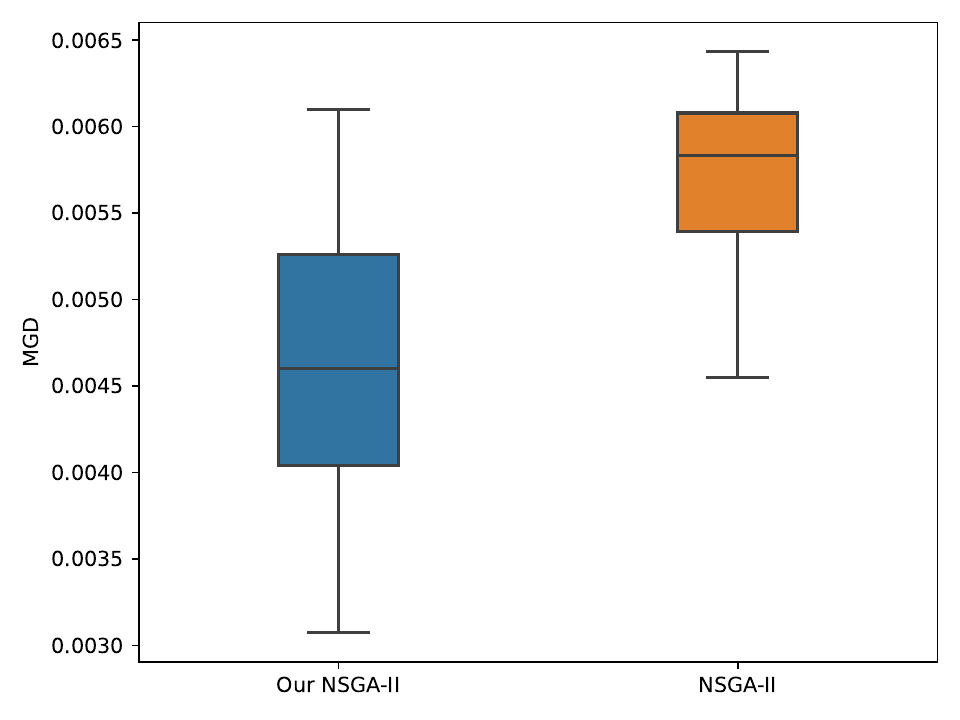}}
\hfill
 \subfloat{\includegraphics[width=0.24\textwidth]{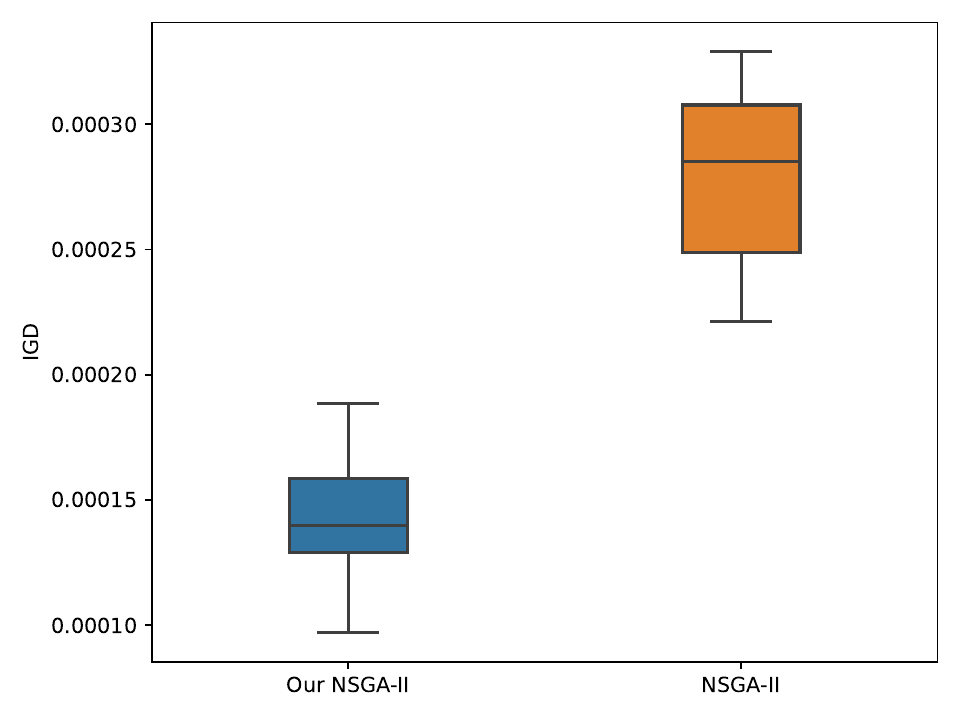}}
\hfill
 \subfloat{\includegraphics[width=0.24\textwidth]{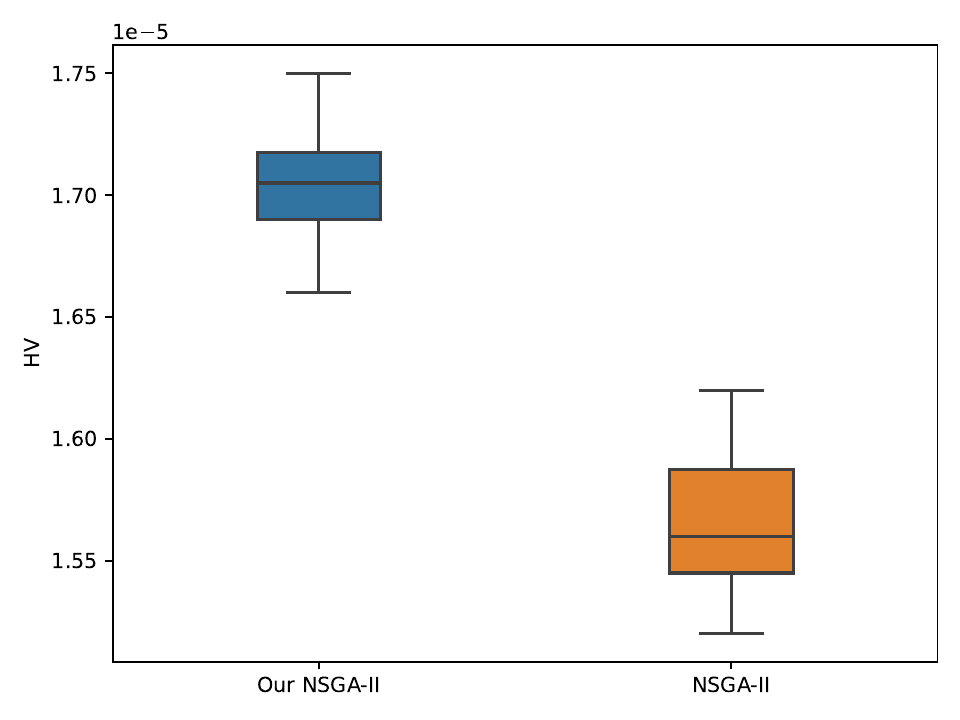}}
\hfill
 \subfloat{\includegraphics[width=0.24\textwidth]{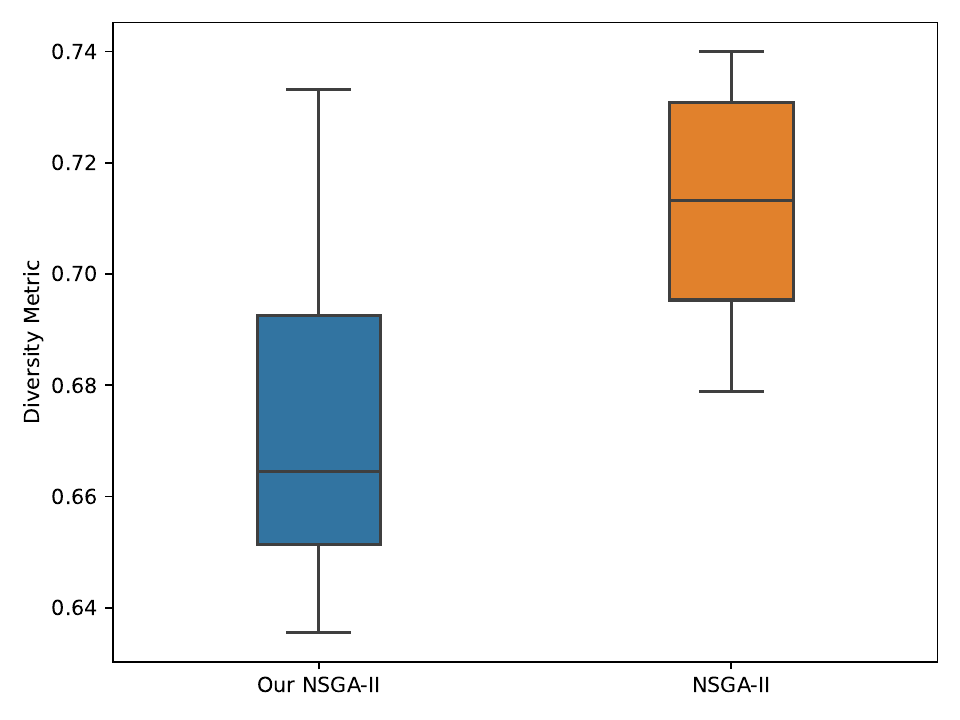}}
\caption{Comparison of performance indicators for S\text{\&}P 100.}
\label{fig:SP100}
\end{figure}

\begin{figure}[H]
\centering
\subfloat{\includegraphics[width=0.24\textwidth]{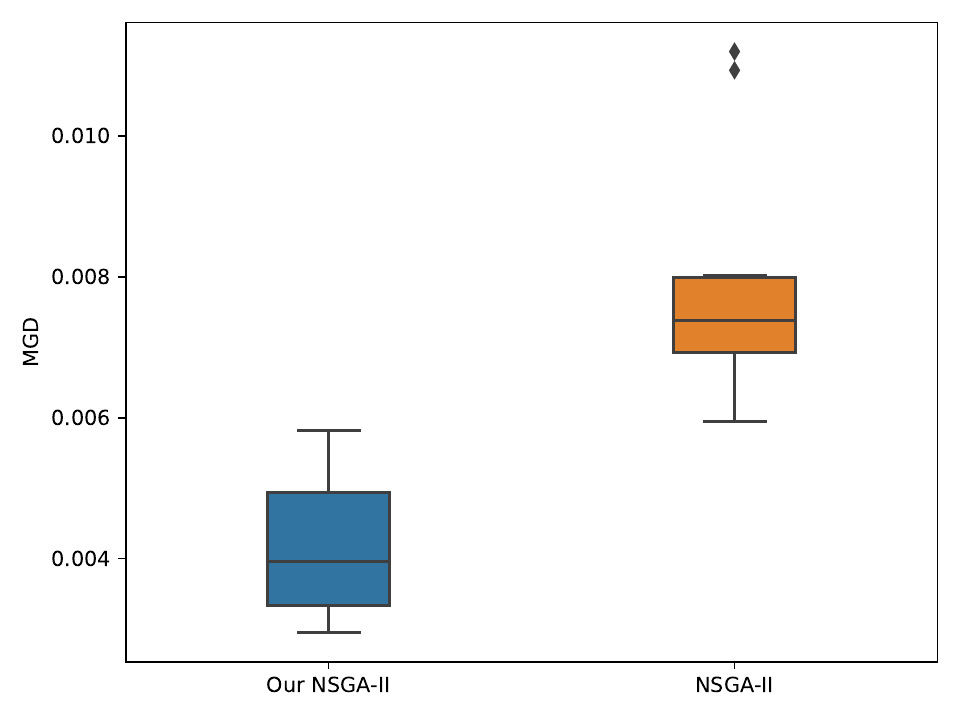}}
\hfill
 \subfloat{\includegraphics[width=0.24\textwidth]{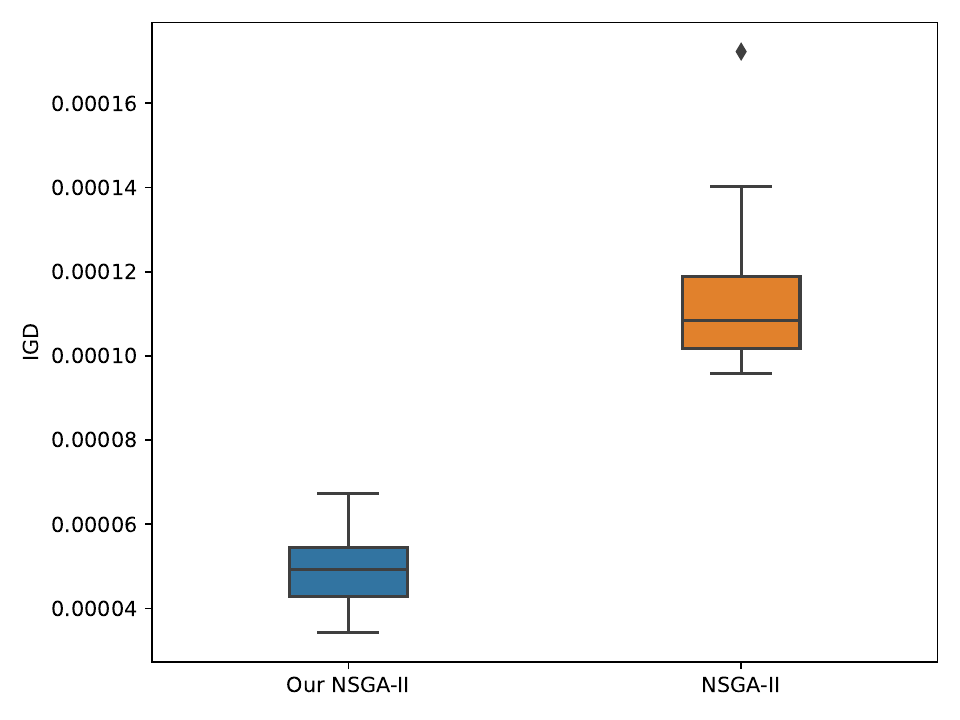}}
\hfill
 \subfloat{\includegraphics[width=0.24\textwidth]{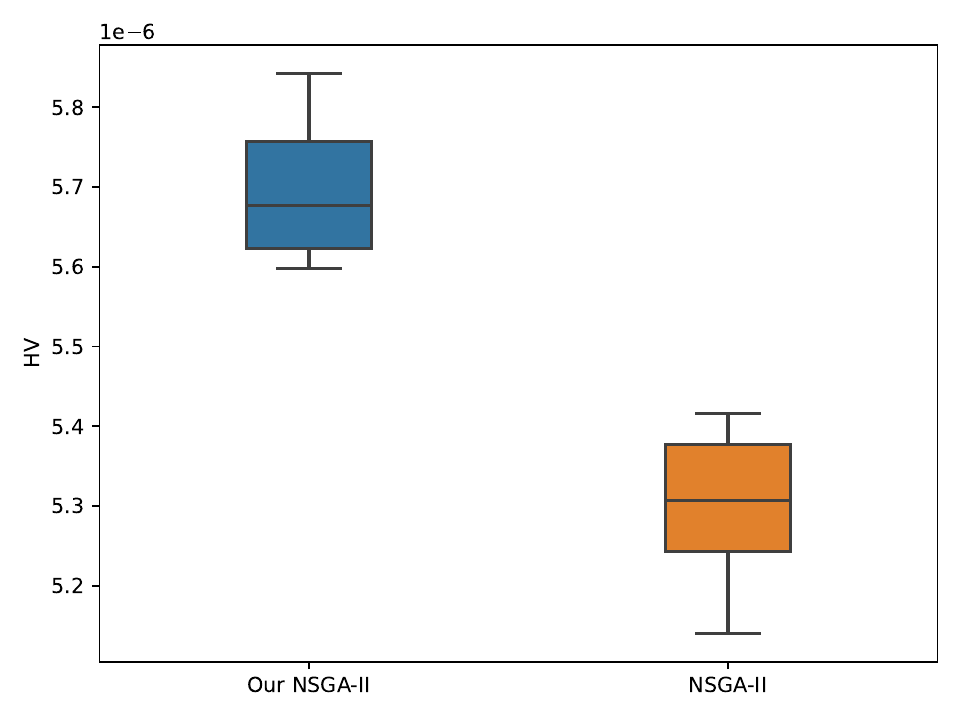}}
\hfill
 \subfloat{\includegraphics[width=0.24\textwidth]{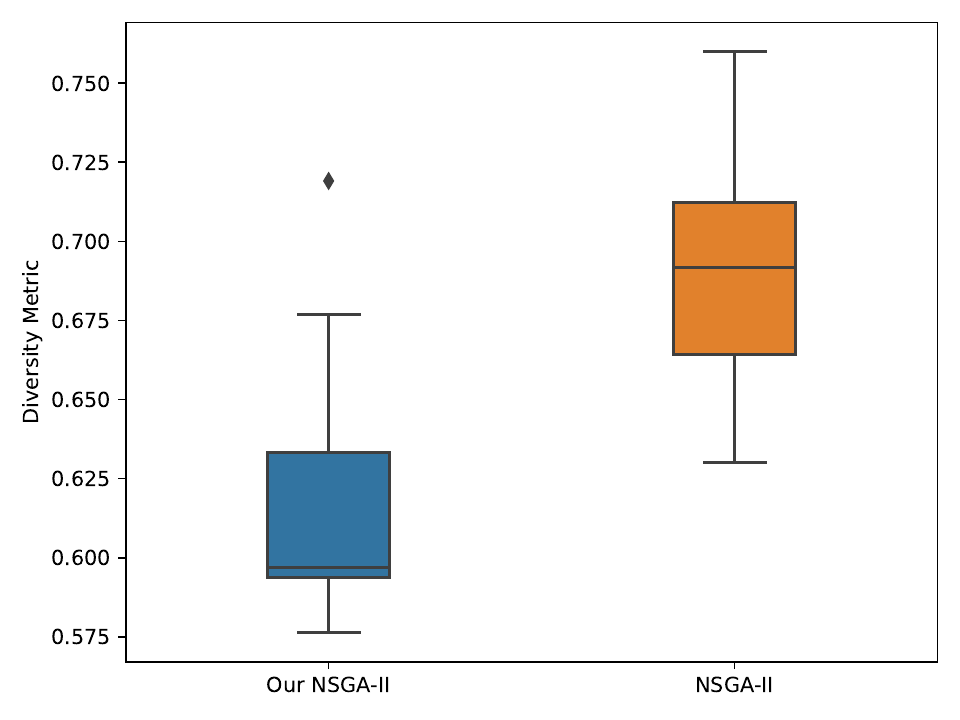}}
\caption{Comparison of performance indicators for Nikkei 225.}
\label{fig:Nikkie}
\end{figure}

\begin{figure}[H]
\centering
\subfloat{\includegraphics[width=0.24\textwidth]{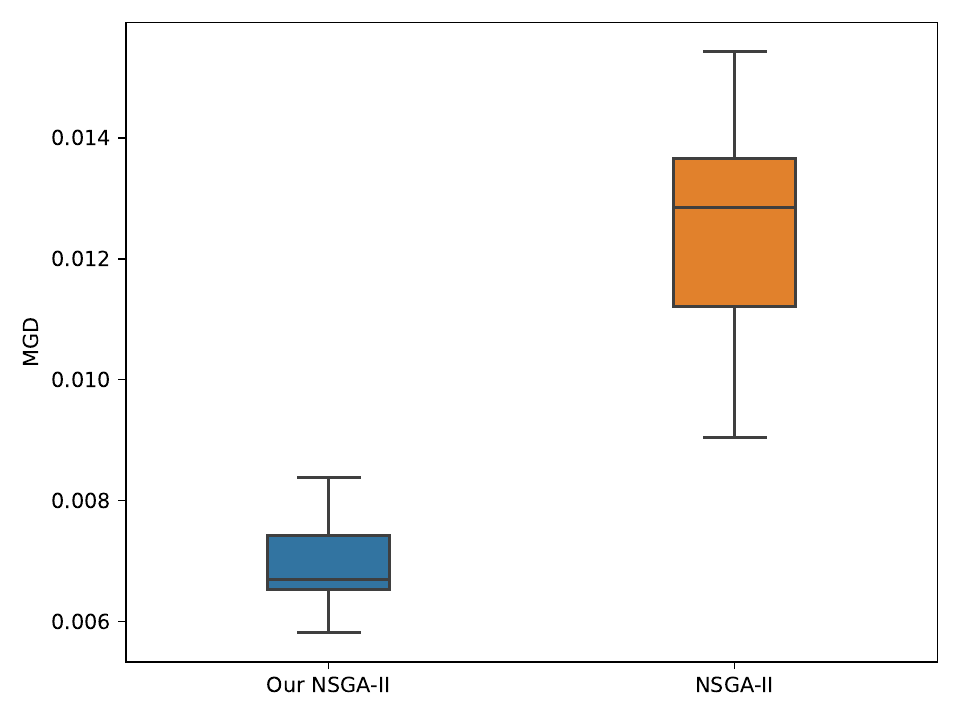}}
\hfill
 \subfloat{\includegraphics[width=0.24\textwidth]{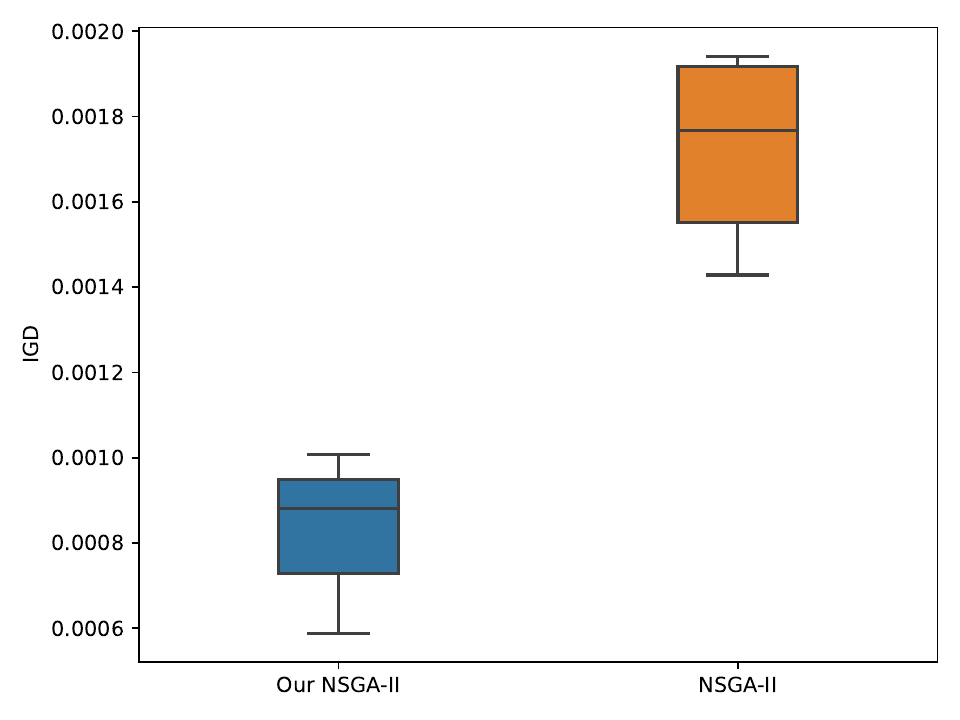}}
\hfill
 \subfloat{\includegraphics[width=0.24\textwidth]{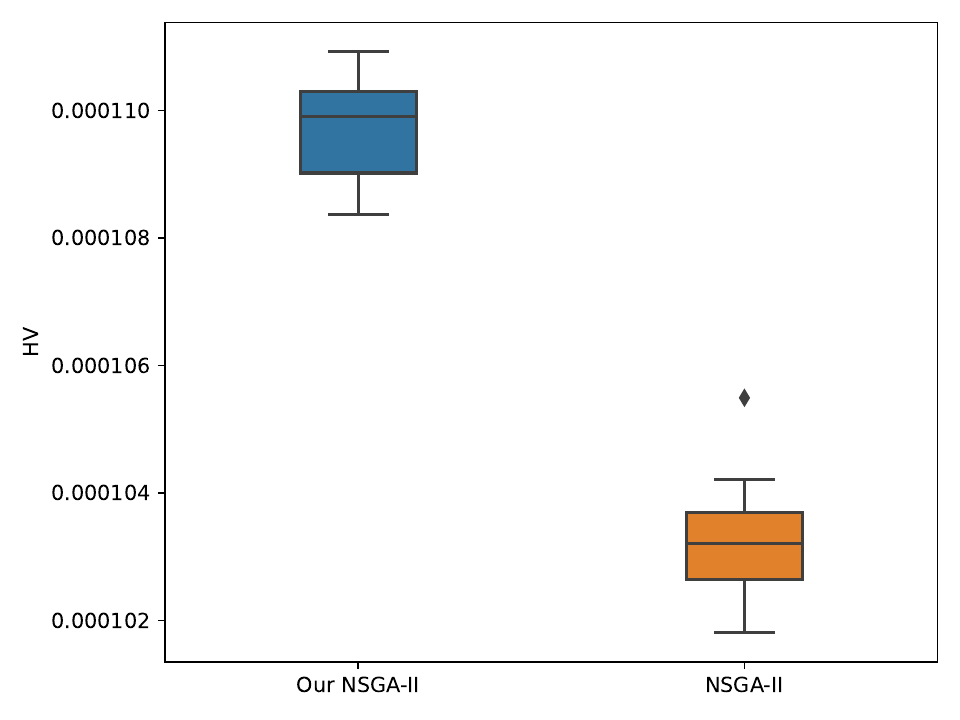}}
\hfill
 \subfloat{\includegraphics[width=0.24\textwidth]{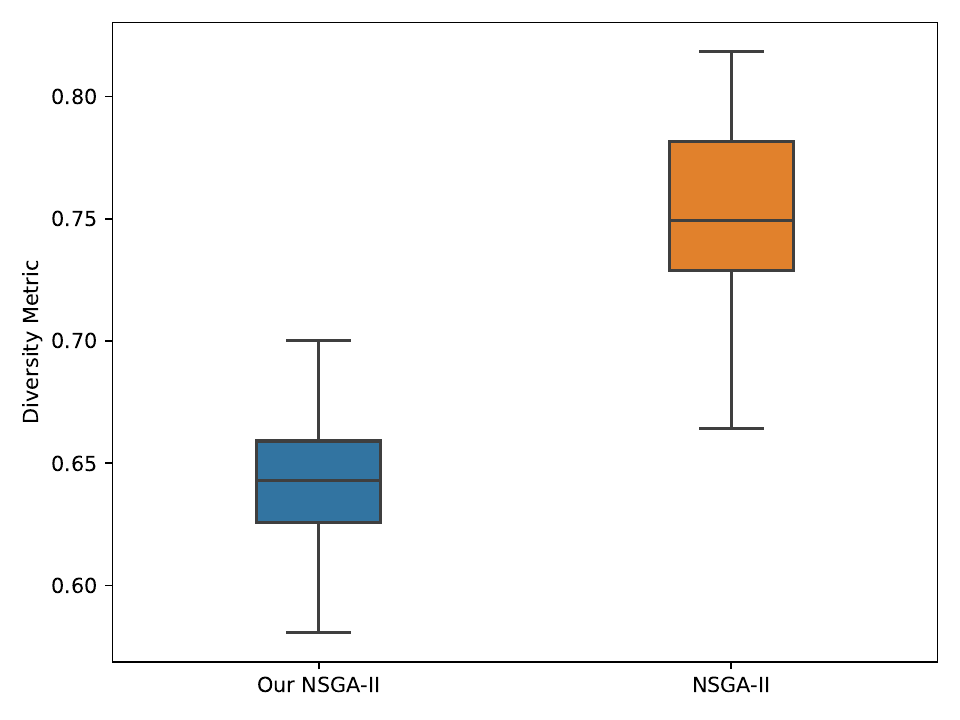}}
\caption{Comparison of performance indicators for TSE.}
\label{fig:TSE}
\end{figure}

For evaluation and comparison of the algorithms, the following four performance indicators 
were employed. 1) Hypervolume (HV) (\cite{zitzler2003performance}) measures the area under the 
approximation curve with respect to a reference point which is a point dominated by all the
solutions; 2) Generational Distance (GD) (\cite{hemici2023multi}) measures the average distance of points of the approximation curve to their closest point in the unconstrained  PF; 3) Inverted 
Generational Distance (IGD)(\cite{chen2021utilizing}) is similar to GD but calculates the distance of 
each point in unconstrained  PF to its closest point in the approximated curve; and, 4) Diversity metric $\Delta$ (\cite{deb2002fast}) measures the spread in solutions by calculating the distance of 
consecutive solutions while considering the distance of the approximated curve from 
extreme solutions. While lower values of GD, IGD, and diversity metric are desirable, higher 
HV values represent better solutions. However, GD can cause confusion in the comparison of 
two curves when they are not distributed within the same range. Imagine two approximation 
curves A and B where A dominates all points in B with few additional points of a greater 
range compared to B located in the tail of A. In this case, the GD of A could be even higher 
than that of B due to those points. To deal with this, we normalized the generational distance
metric by dividing it by the distance of boundary points found by each algorithm. To prevent 
confusion with GD, this new metric was named ‘Modified Generational Distance’ (MGD). \hyperref[fig:DAX]{{Figures 6}},\hyperref[fig:SP100]{{7}},\hyperref[fig:Nikkie]{{8}},\hyperref[fig:TSE]{{9}} demonstrate box plots of the performance indicators for NSGA-II with 
standard solution representations (real-value coded chromosomes with the population size of universe of assets) and the modified NSGA-II proposed herein. The reported results are 
obtained from 100 iterations for both algorithms with population size of 100, $\epsilon_i = 0.01, \delta_i = 0.99 \hspace{2mm} \forall i$, and $\sum_{i=1}^{N} z_i = 10$. In the standard encoded NSGA-II, the BC handling follows \cite{chang2000heuristics} and whenever the number of assets in the portfolio violates the upper or 
lower bounds of CC, a random stock is removed from or added to the portfolio, respectively. 
All the algorithms are implemented in Python and ran on a personal computer Intel(R) Core(TM) i7-4500U CPU @ 1.80GHz and the results are recorded in 10 runs. Evidently, the
approach proposed in this study is superior in all performance indicators for all the selected 
markets to the commonly encoded NSGA-II. Furthermore, the gap between the mean values of each indicator become more obvious as the number of assets increases. These results indicate the effectiveness of our approach regarding searching ability.

\begin{figure}[H]
\centering
\subfloat[DAX 100]{\includegraphics[width=0.49\textwidth]{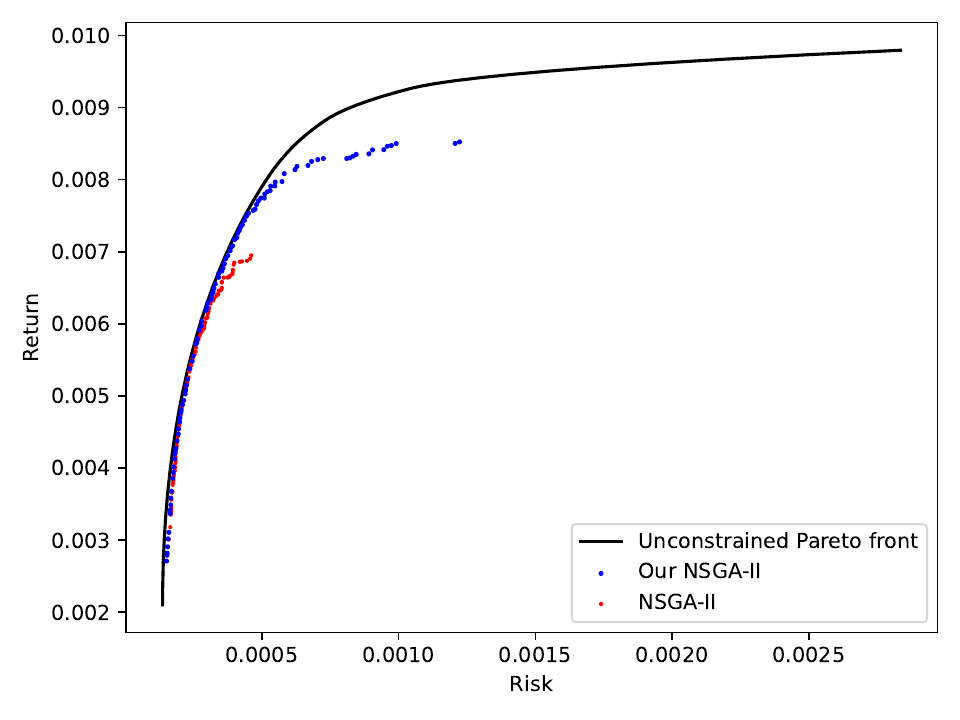}}
%\hfill
\subfloat[S\text{\&}P 100]{\includegraphics[width=0.49\textwidth]{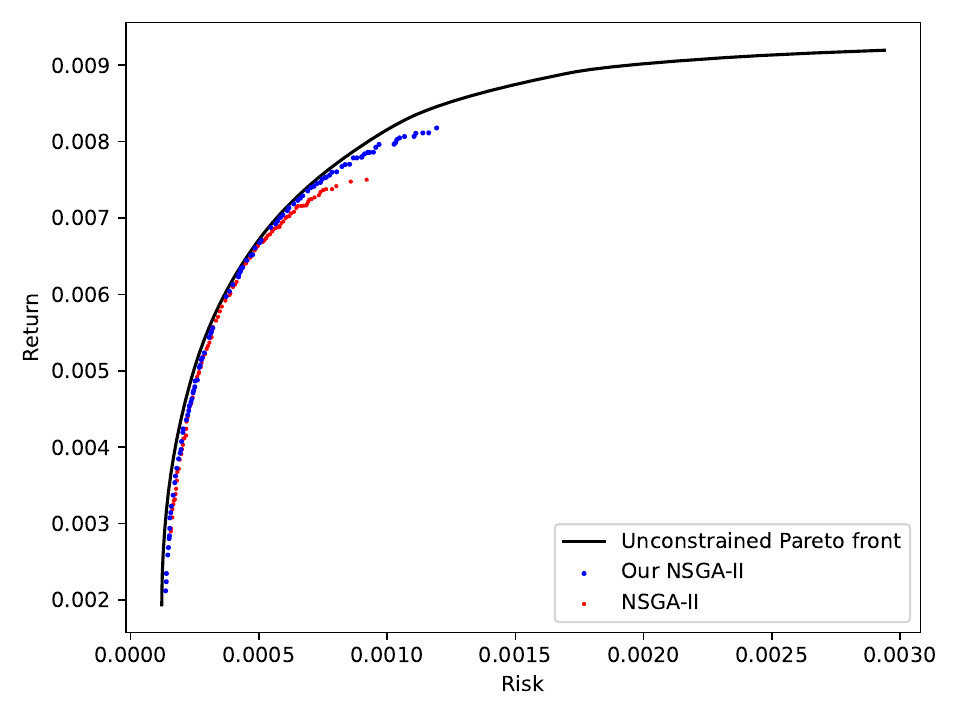}}
\hfill
\subfloat[Nikkei 225]{\includegraphics[width=0.49\textwidth]{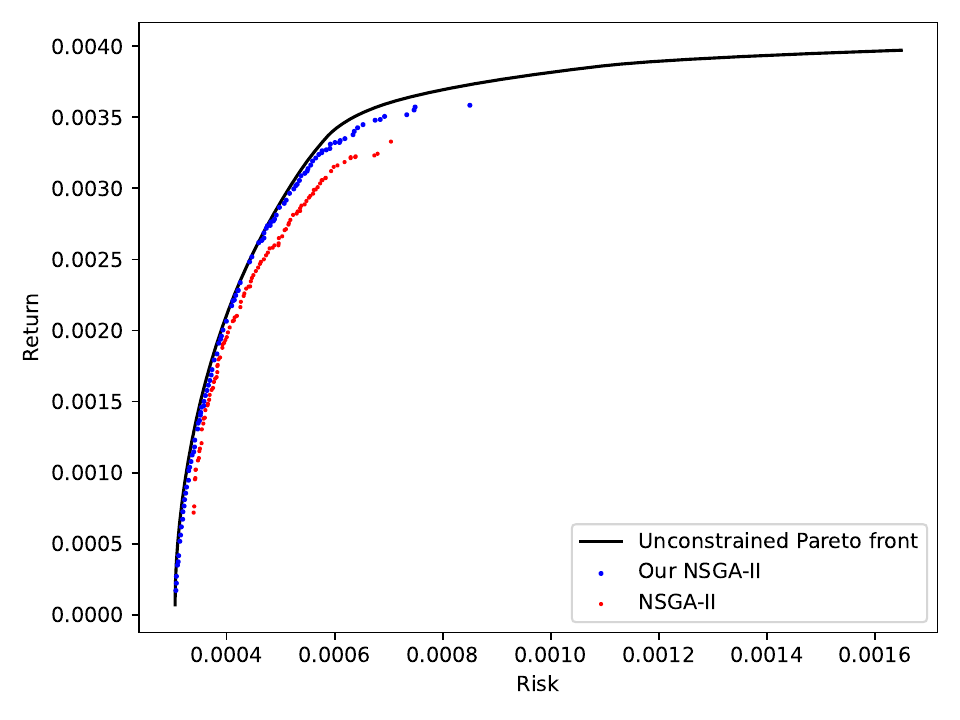}}
%\hfill
\subfloat[TSE]{\includegraphics[width=0.49\textwidth]{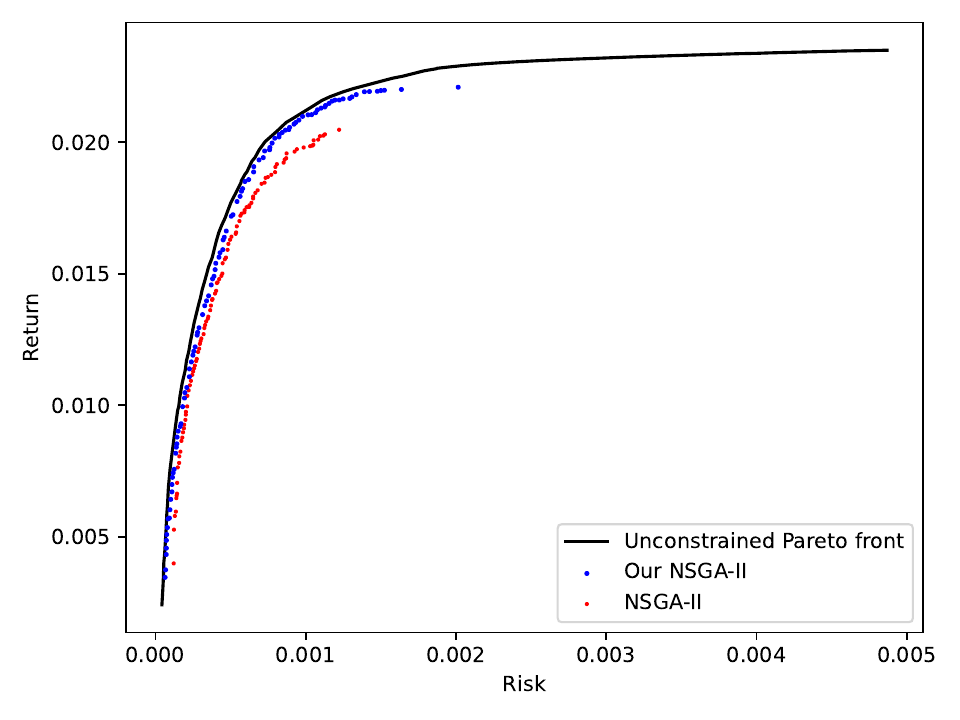}}
\caption{Comparison of approximated efficient frontiers ($\sum_{i=1}^{N} z_i = 10$).}
\label{fig:Results}
\end{figure}

\hyperref[fig:Results]{{Figure 10}} indicates the approximation curves obtained from both NGSA-II and the proposed 
algorithms for each market along with their EF marked by black line. The results show the 
superiority of the proposed approach to the conventional NSGA-II. It is obvious from the 
plots that both algorithms yield close results for relatively small universes of assets in the 
markets. While the differences become more conspicuous with growing numbers of assets. 
One drawback, however, is that the data for the unconstrained Pareto front for the TSE market was not 
available to us and we, therefore, not only executed our algorithm with somewhat relaxed 
constraints for longer iterations but multiplied each return and risk of each data point from 
the approximated curve by a factor with equal positive and negative deviations from one, 
respectively, and extrapolated the tails of the curve along the slope of the last change in the 
data points only to ensure no feasible solution capable of dominating that artificial Pareto 
front is left out. This is also the reason why algorithms are more distant away from the black 
line in the TSE market compared to other markets.

\begin{figure}[H]
\centering
\subfloat[DAX 100]{\includegraphics[width=0.24\textwidth]{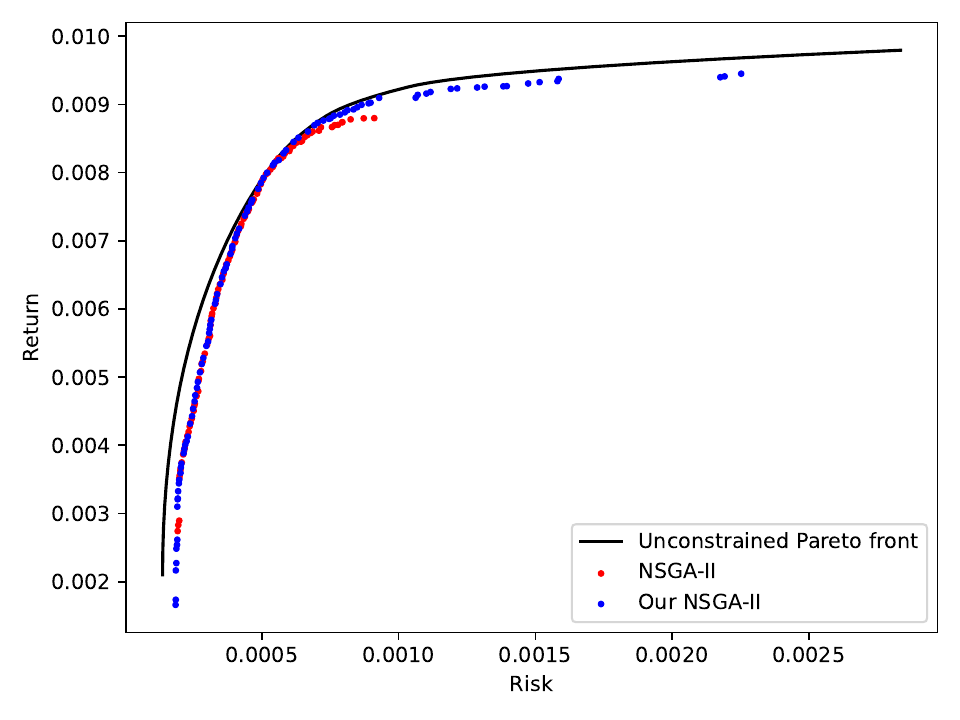}}
\hfill
\subfloat[S\text{\&}P 100]{\includegraphics[width=0.24\textwidth]{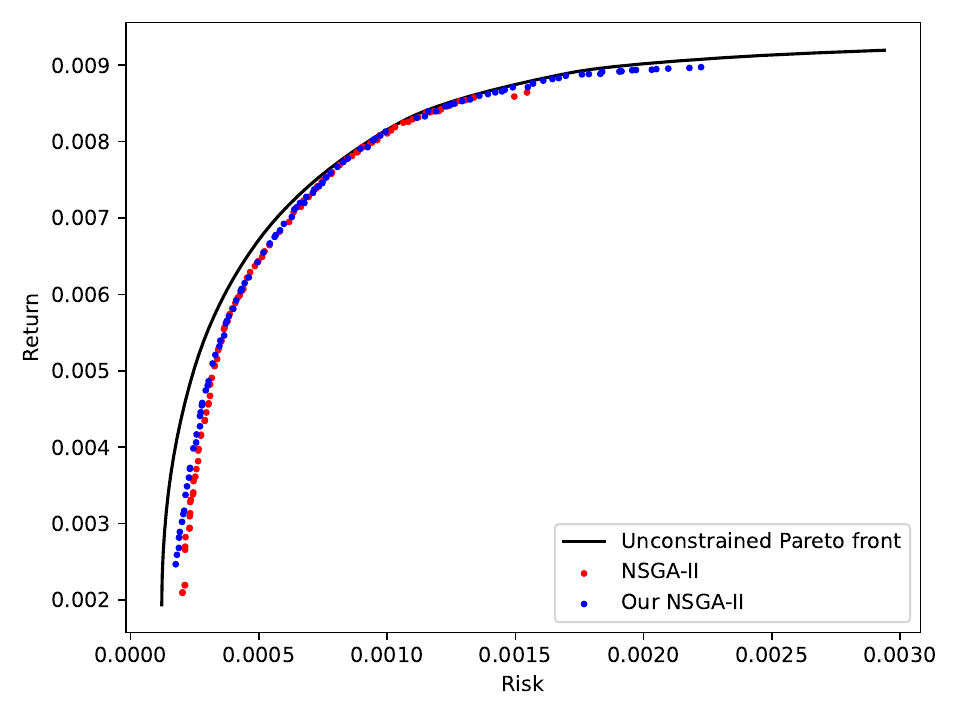}}
\hfill
\subfloat[Nikkei 225]{\includegraphics[width=0.24\textwidth]{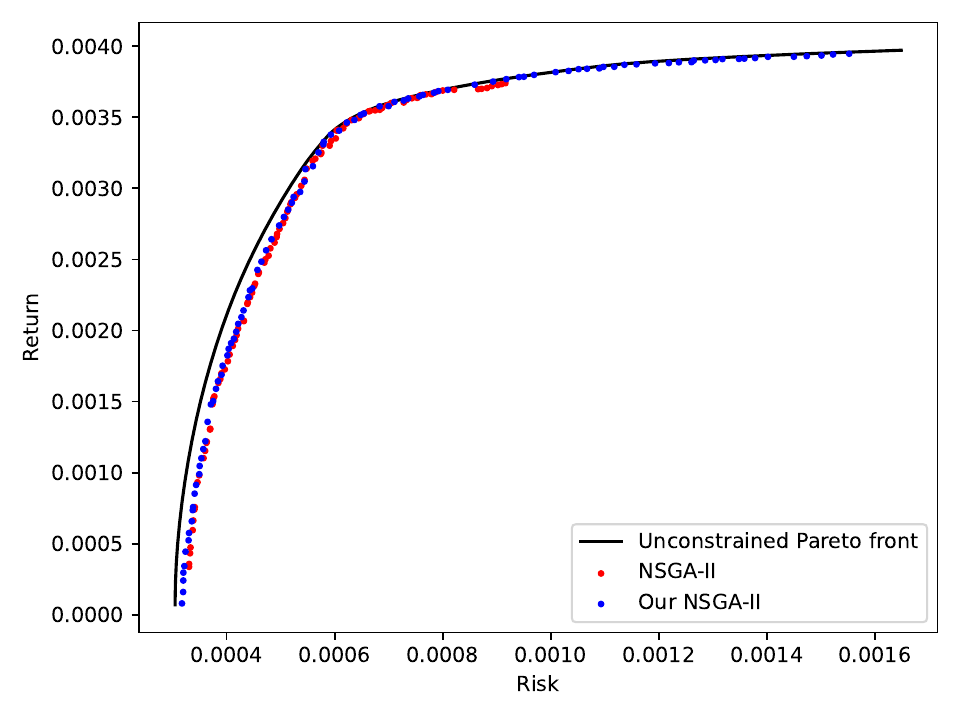}}
\hfill
\subfloat[TSE]{\includegraphics[width=0.24\textwidth]{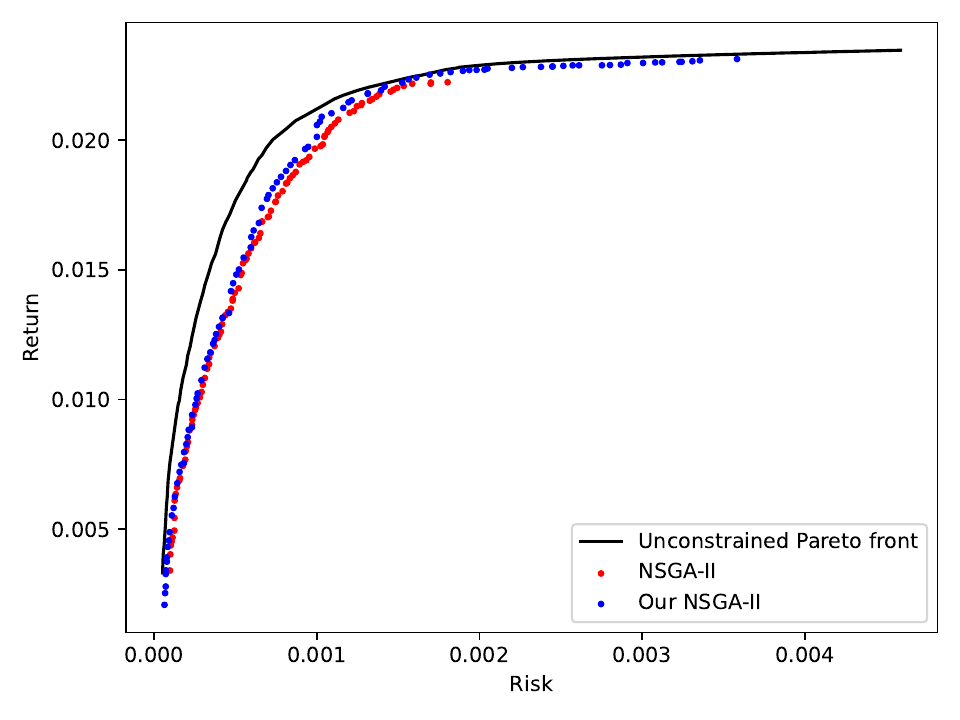}}
\caption{Comparison of approximated efficient frontiers ($\sum_{i=1}^{N} z_i = 5$).}
\label{fig:Results5}
\end{figure}

\begin{figure}[H]
\centering
\subfloat[DAX 100]{\includegraphics[width=0.24\textwidth]{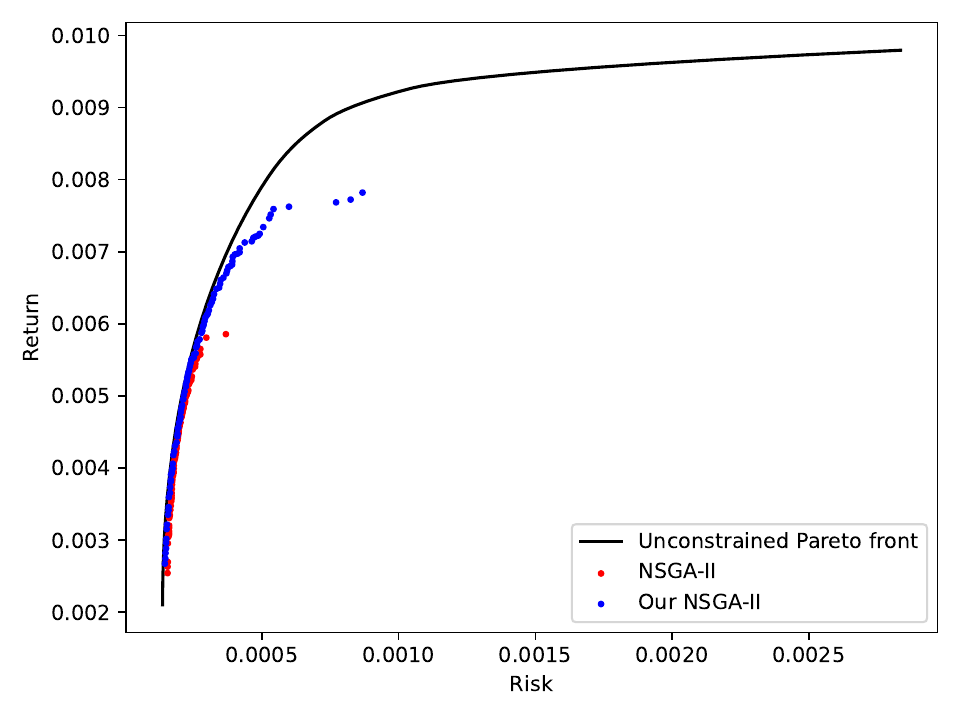}}
\hfill
\subfloat[S\text{\&}P 100]{\includegraphics[width=0.24\textwidth]{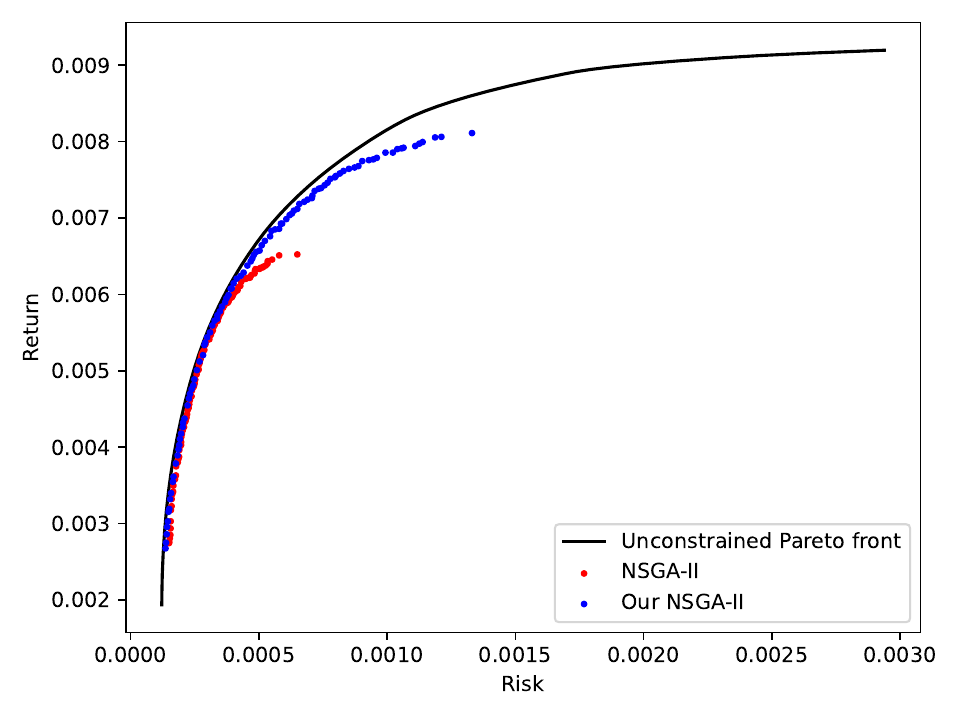}}
\hfill
\subfloat[Nikkei 225]{\includegraphics[width=0.24\textwidth]{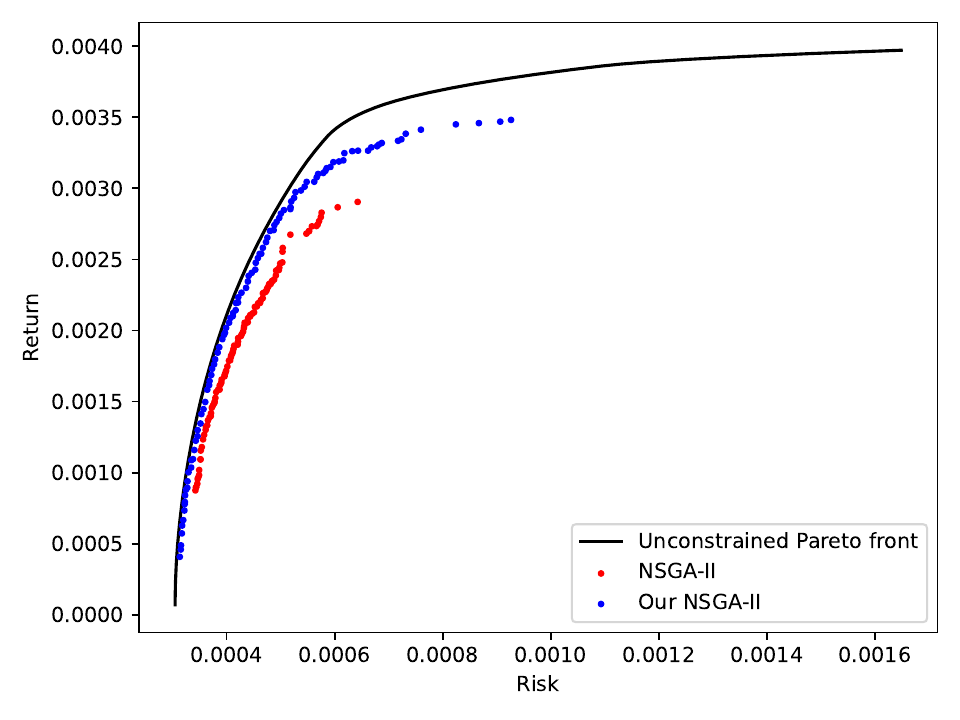}}
\hfill
\subfloat[TSE]{\includegraphics[width=0.24\textwidth]{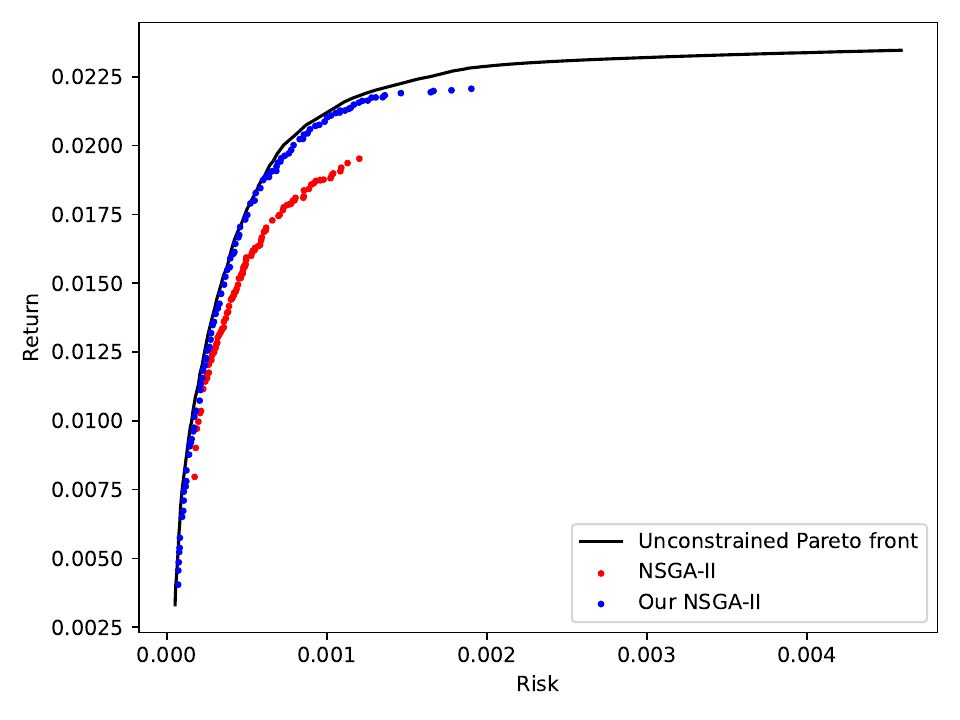}}
\caption{Comparison of approximated efficient frontiers ($\sum_{i=1}^{N} z_i = 15$).}
\label{fig:Results15}
\end{figure}

\hyperref[fig:Results5]{{Figure 11}} and \hyperref[fig:Results15]{{Figure 12}} show comparisons of the two algorithms for different values of $K_{\min}$ and $K_{\max}$. Obviously, the proposed approach was capable of discovering solutions superior to 
those found by NSGA-II with different CC setups, especially with increasing pools of assets. 
Furthermore, the difference between the conventional NSGA-II and the version proposed in 
this article increased with increasing numbers of assets in the portfolio. Aforementioned results further approve our overall strategies since customized NSGA-II are able to find both better and wider ranges of portfolios. Further results for more sparse cardinality configurations are provided in  \hyperref[Appendix A]{{Appendix A}}. Additionally, \hyperref[Appendix B]{Appendix B} presents the results of implementing our mechanisms on another multi-objective algorithm.

\begin{figure}[H]
\centering
\subfloat[DAX 100]{\includegraphics[width=0.49\textwidth]{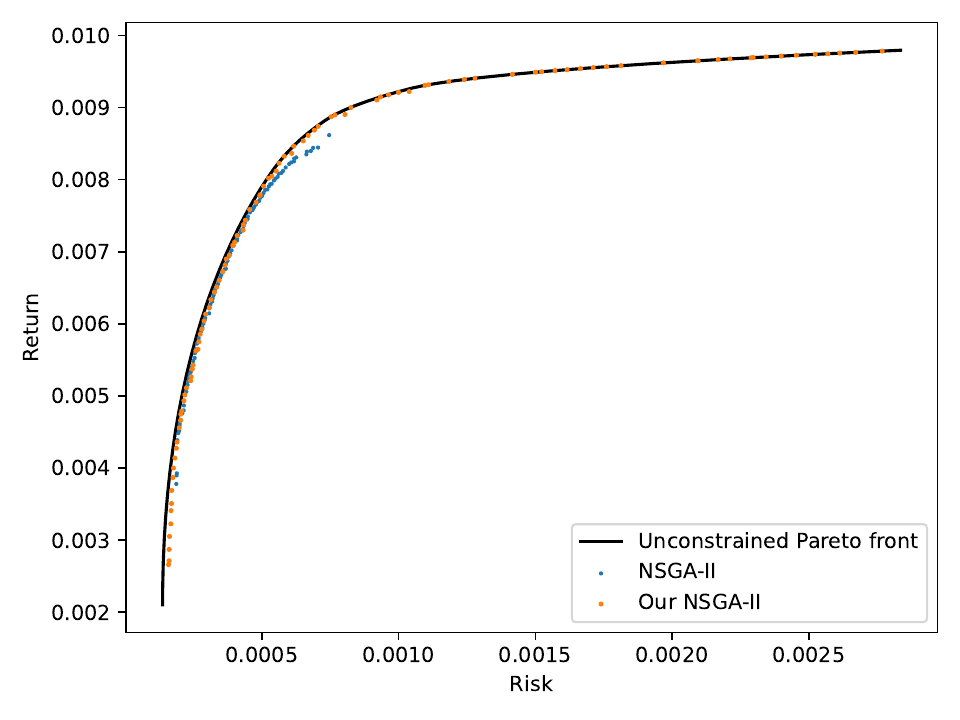}}
%\hfill
\subfloat[S\text{\&}P 100]{\includegraphics[width=0.49\textwidth]{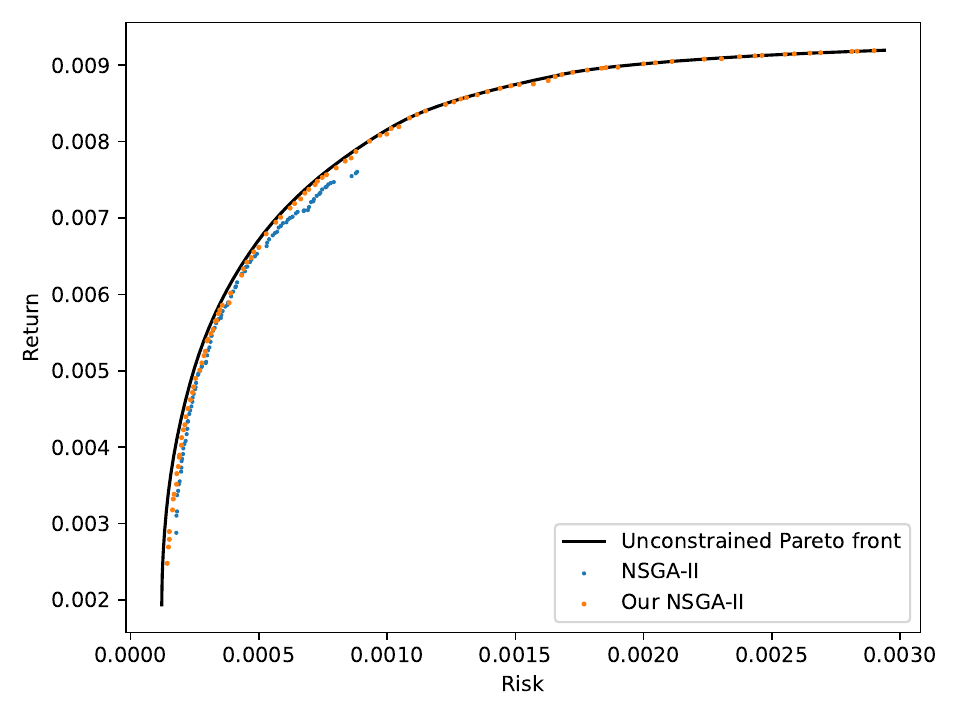}}
\hfill
\subfloat[Nikkei 225]{\includegraphics[width=0.49\textwidth]{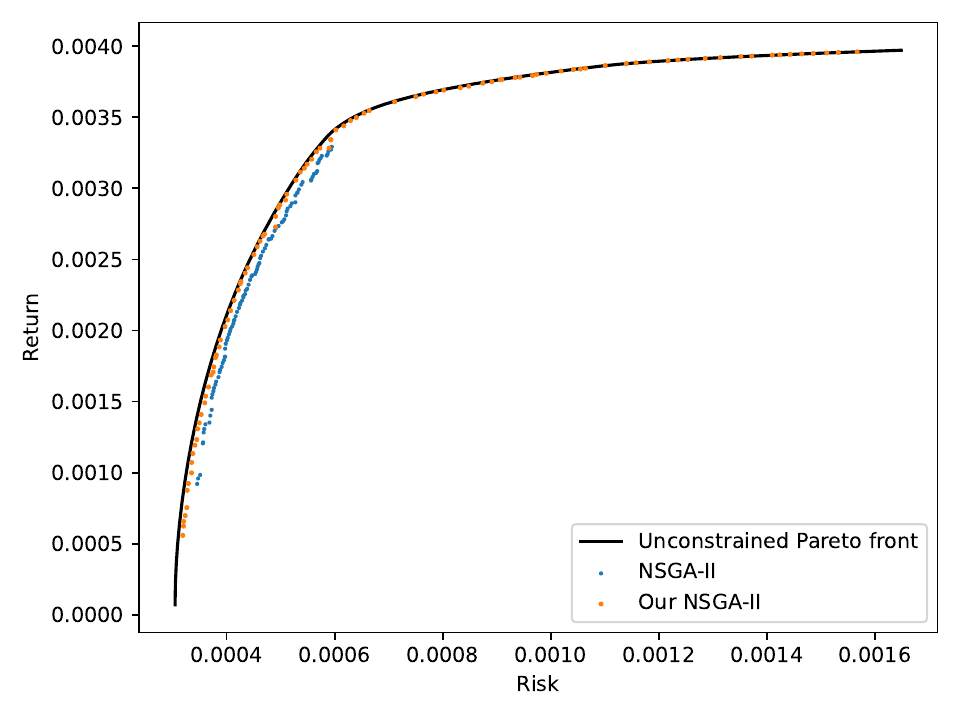}}
%\hfill
\subfloat[TSE]{\includegraphics[width=0.49\textwidth]{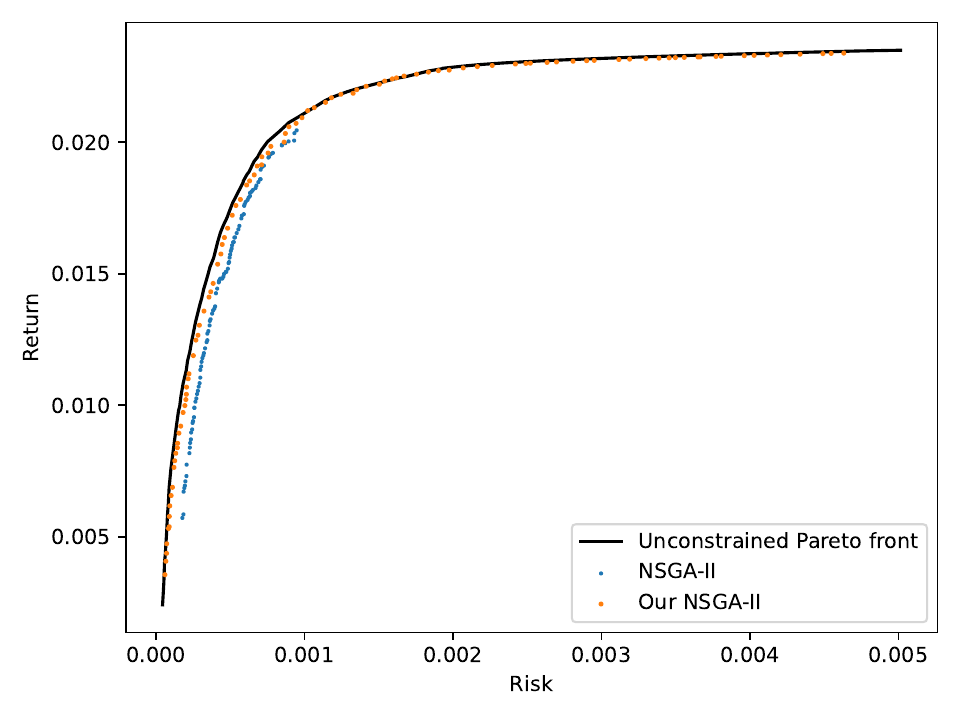}}
\caption{Comparison of approximated efficient frontiers ($K_{\min} = 2 , K_{\max} = 10$).}
\label{fig:Results2-10}
\end{figure}

\hyperref[fig:Results2-10]{{Figure 13}} compares the two algorithms with the setup as previously described except for CC
whose parameters are set to $K_{\min} = 2 $, $K_{\max} = 10 $. Comparison of the approximated curves
reveals that the proposed package found better portfolios and covered a wider range of 
unconstrained  PF even with increasing pools of assets. Similar to previous plots, the results for the 
relatively small market data obtained from the two algorithms shown in this Figure are 
competitively close. While the conventional NSGA-II failed to keep the same performance,
approximated smaller ranges of unconstrained  PF, and yielded less efficient solutions with a growing
universe of assets, the approach proposed herein not only found better dominant portfolios
but also kept the same performance and approximated a wider range of unconstrained  PF with 
increasing numbers of stocks.

\begin{figure}[H]
\centering
\subfloat[Fix $K_{\min}$.]{\includegraphics[width=0.49\textwidth]{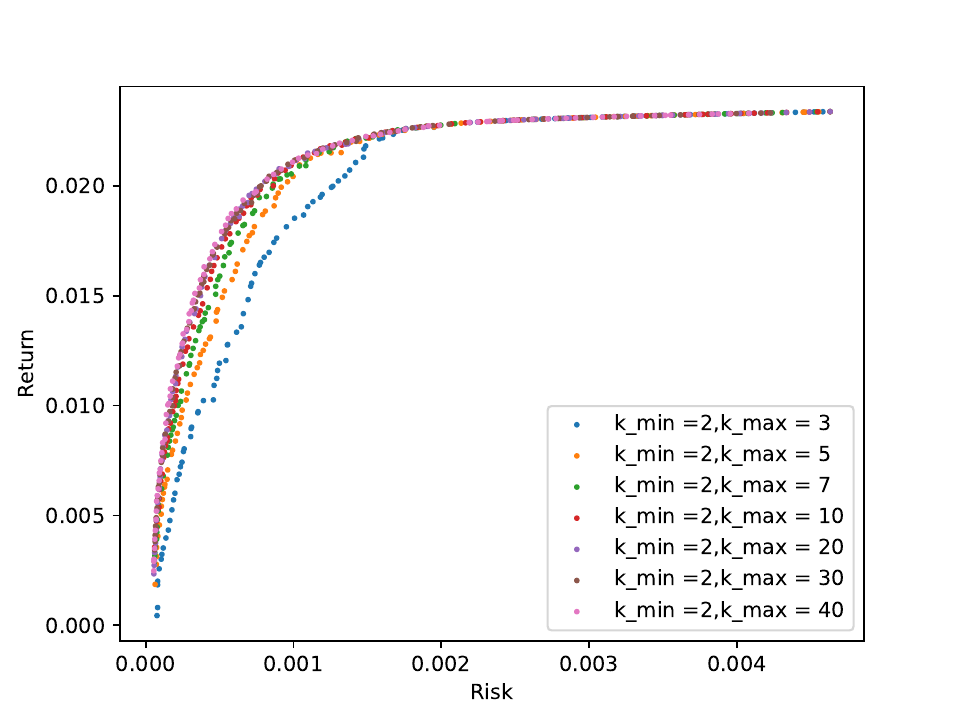}}
\hfill
 \subfloat[Fix $K_{\max}$.]{\includegraphics[width=0.49\textwidth]{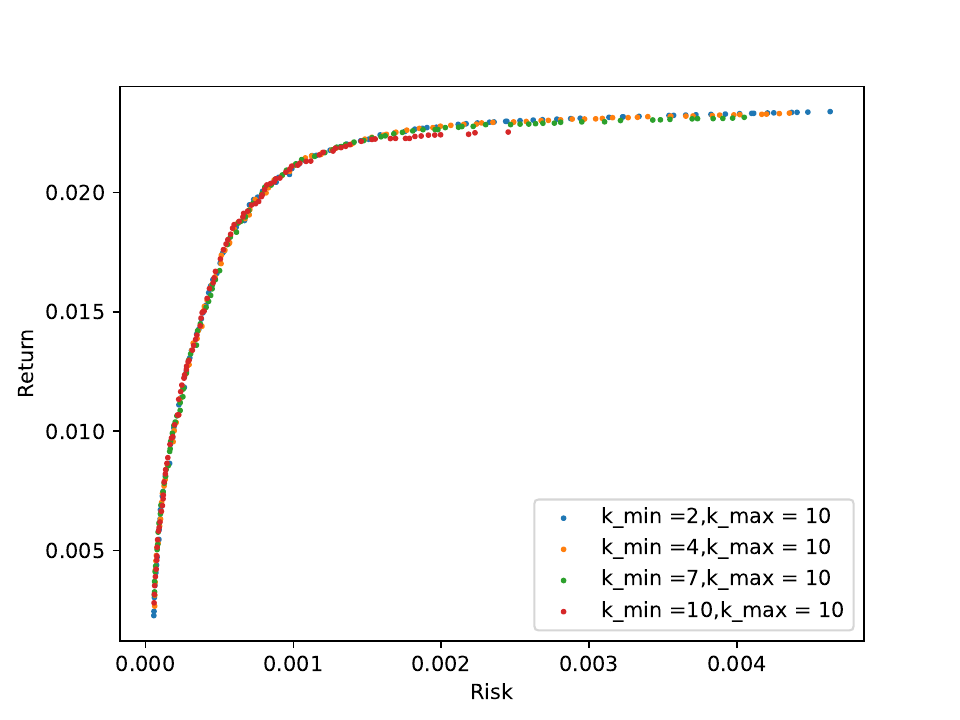}}
\caption{Pareto front with the different cardinality settings.}
\label{fig:cardinality effect}
\end{figure}

\hyperref[fig:cardinality effect]{{Figure 14}} presents the effects of upper and lower bounds of CC on PF generated by the 
modified NSGA-II in the TSE market for longer iterations. In situations with the lower 
bound of CC equal to 2 but a fluctuating upper bound, it is clear that the best choice for an 
investor who seeks a profitable portfolio of low risk (i.e., portfolios in the knee area of the 
curve) would be diversified assets. This is evidenced by the major risk reduction in portfolios 
(namely, a clear shift toward the left side of PF) with increasing maximum number of assets 
in the portfolio from three to five. Even though the risk of the portfolio decreases with a 
greater overall diversification in assets, the amount of risk reduction (shifts) grows smaller 
with increasing numbers of assets in the portfolio, proving that diversification can be achieved with a portfolio of fewer assets. Beyond a maximum number of ten stocks, 
the difference becomes almost negligible and the investor gains not much benefit from 
adding more assets to the portfolio. Another important factor that comes into play with 
growing numbers of assets in the portfolio is the rise in holding costs, including transaction 
and monitoring costs. Meanwhile, holding a portfolio of less than three or five assets does 
not seem a rational decision for a risk-averse investor since the investor can achieve a safer investment by adding just a few assets while avoiding high transaction and monitoring costs. 
Hence, it is costly to hold more than ten to twenty stocks but perilous at the same time for a 
risk-averse investor to hold a portfolio with assets fewer than the number of fingers on one 
hand.

Although imposing a limit on the maximum number of assets in the portfolio changes the left side of the approximation curves, at some point, curves with different $K_{\max}$ values coincide at some point with each other for identical $K_{\min}$ values. In cases with fixed values of $K_{\max}$ but varying $K_{\min}$ the only part that varies is the right side of the curves. The lower 
bound of CC could be considered by fund managers and regulators to ensure a minimal diversification since
it reduces the number of extremely risky portfolios or eliminates them altogether from the 
search space. Over-diversified portfolios could end in inefficient solutions for risk-taking 
investors since they can choose less risky investments with more returns that would entail 
a portfolio with fewer assets.

\begin{figure}[H]
\centering
\subfloat[DAX 100]{\includegraphics[width=0.24\textwidth]{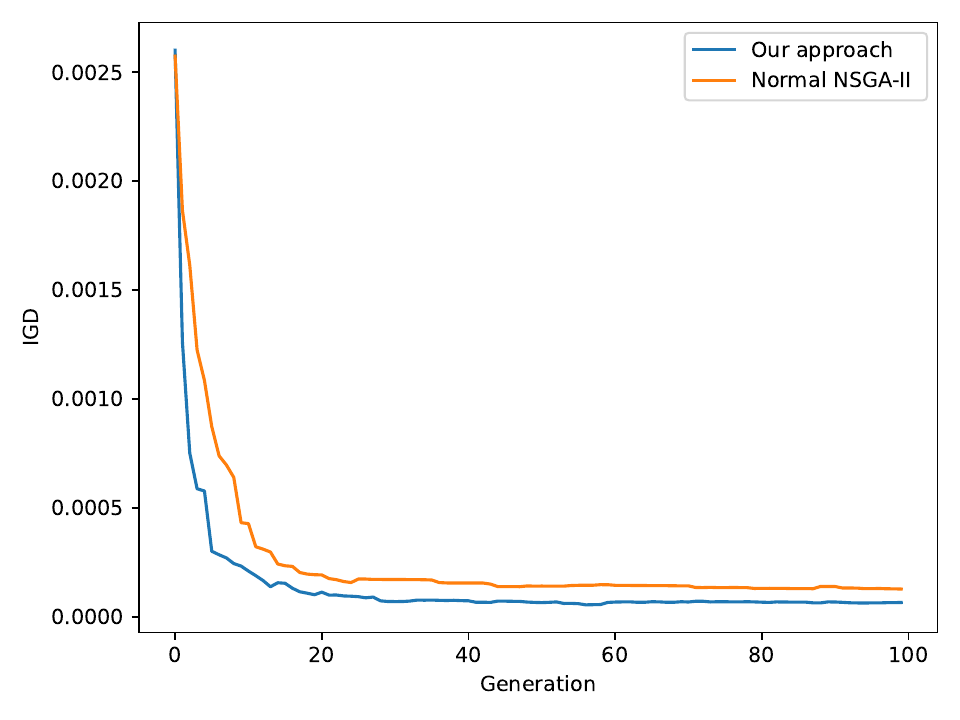}}
\hfill
\subfloat[S\text{\&}P 100]{\includegraphics[width=0.24\textwidth]{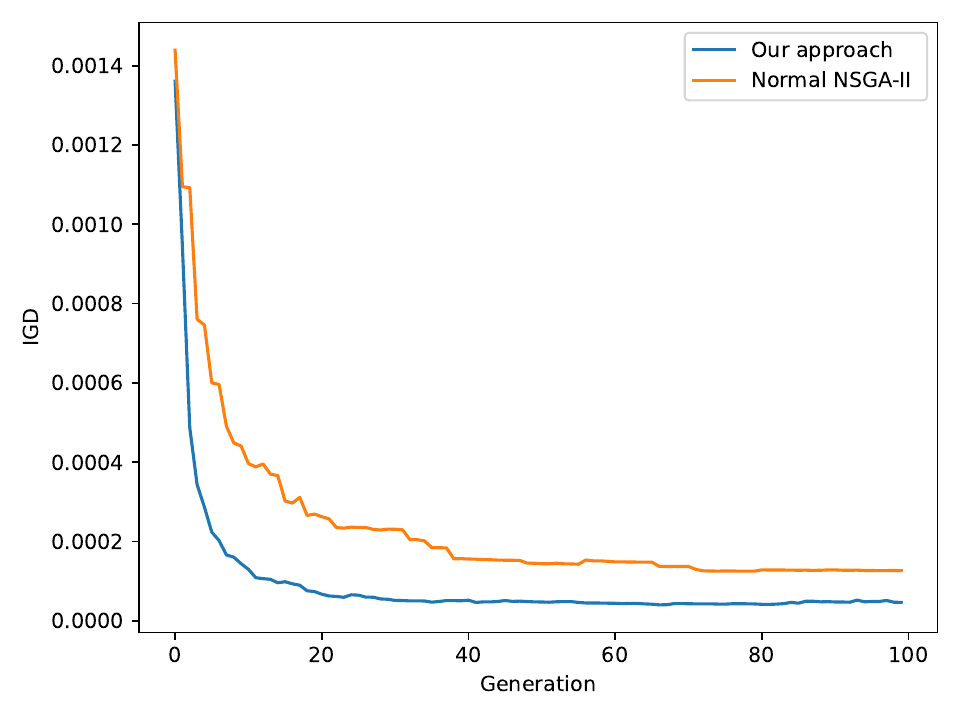}}
\hfill
\subfloat[Nikkei 225]{\includegraphics[width=0.24\textwidth]{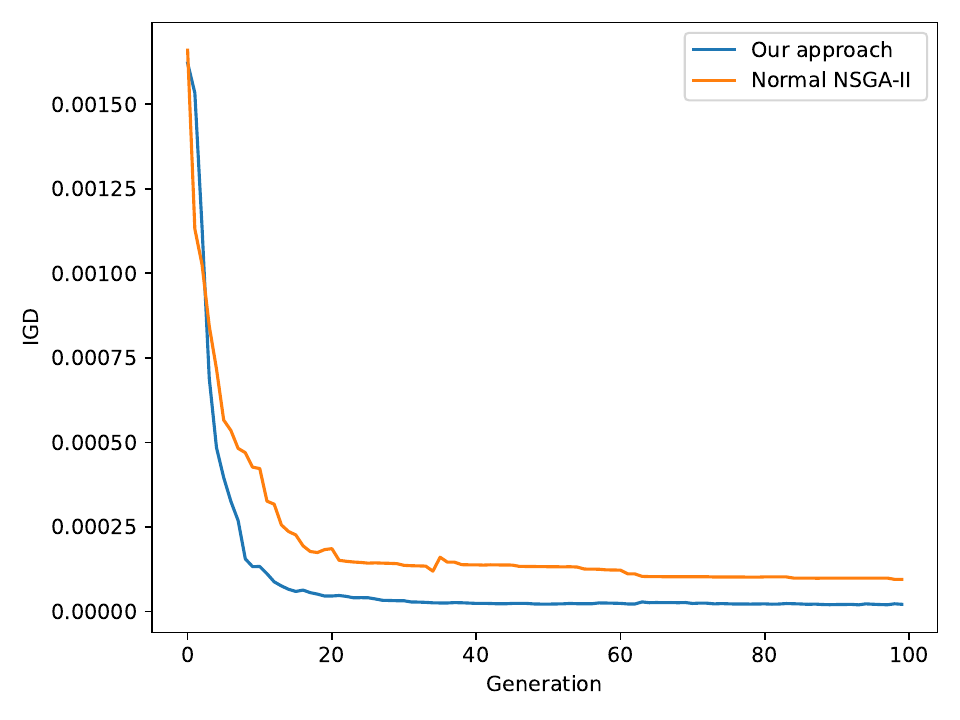}}
\hfill
\subfloat[TSE]{\includegraphics[width=0.24\textwidth]{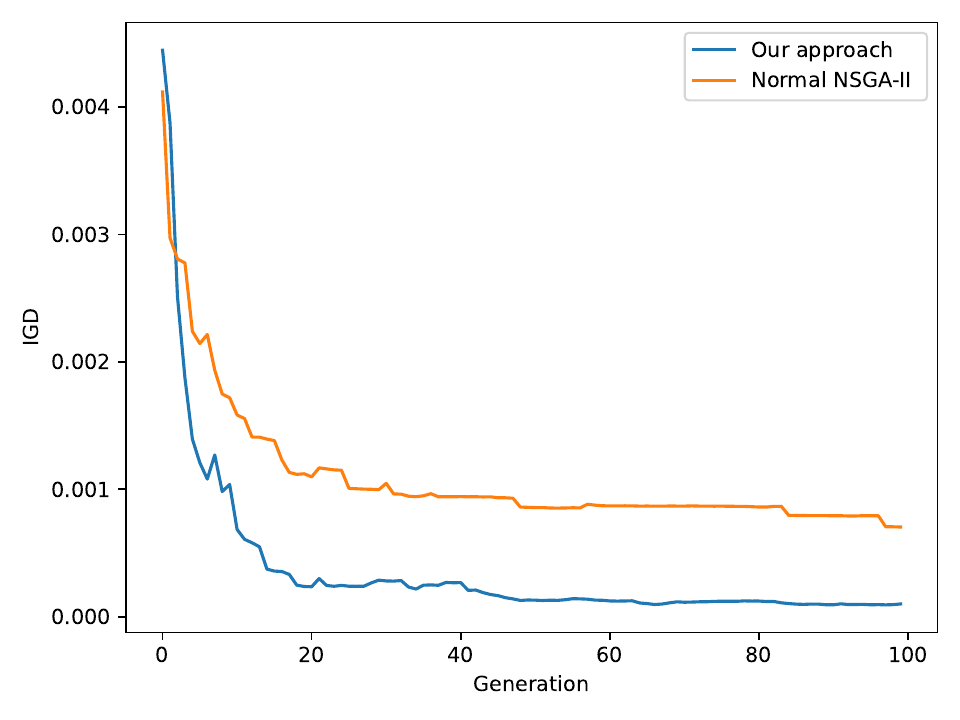}}
\caption{Comparison of IGD of our approach and normal NSGA-II with our encoding system.}
\label{fig:ResultsIGD}
\end{figure}

\hyperref[fig:ResultsIGD]{{Figure 15}} compares the IGDs of both the proposed approach and the conventional NSGA-II 
that have the same solutions encoded with previously explained parameters and iterations 
for different markets, illustrating the effectiveness of our mating strategies due to the faster 
convergence of the proposed strategy against the same encoding system in NSGA-II. The 
plots in this Figure also indicate that our approach is capable of yielding closer 
approximations for the same number of generations. Furthermore, the distance between the 
two lines grows larger and more conspicuous as the universe of assets grows.

\begin{figure}[H]
\centering
\begin{minipage}{0.49\textwidth}
    \centering
    \includegraphics[width=\textwidth]{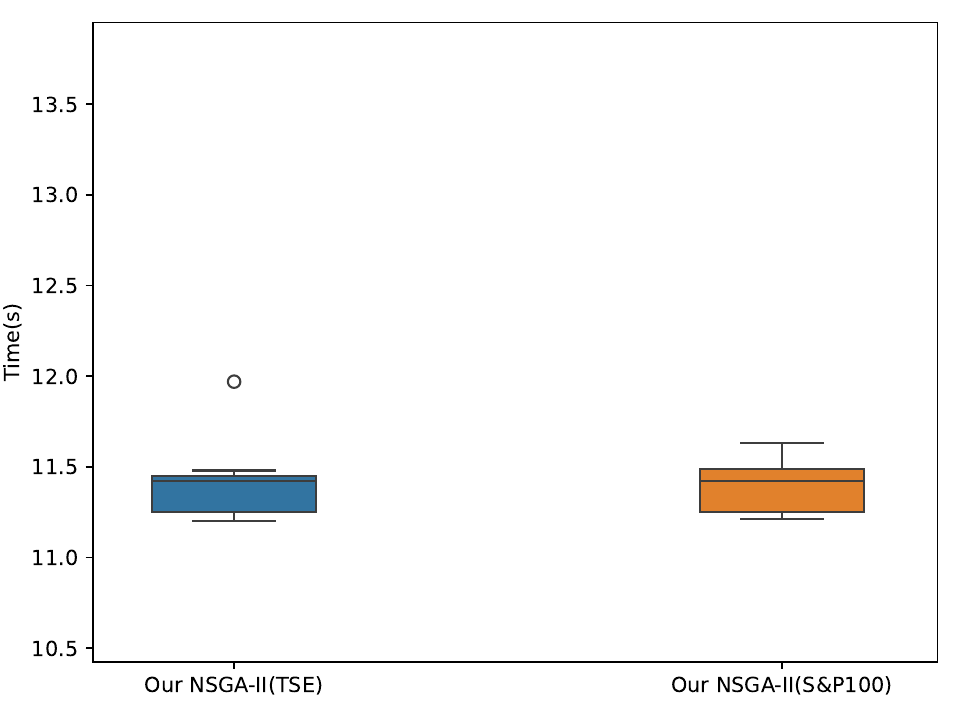}
    \caption*{(a) Magnified view} % Adds (a) as a subtitle
\end{minipage}%
\begin{minipage}{0.49\textwidth}
    \centering
    \includegraphics[width=\textwidth]{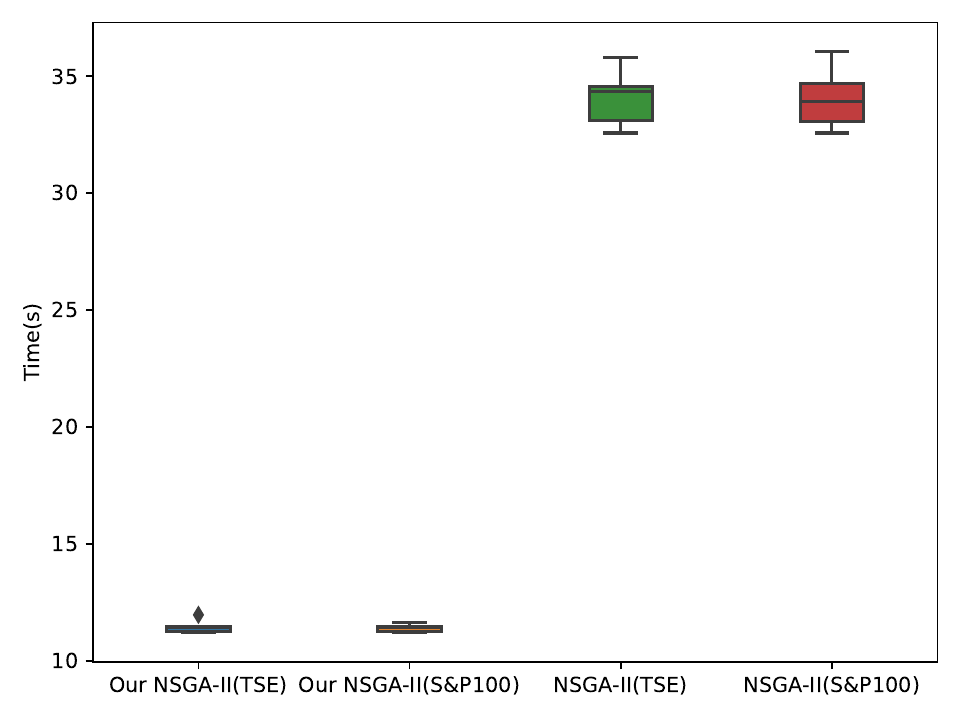}
    \caption*{(b) Original view} % Adds (b) as a subtitle
\end{minipage}
\caption{Comparison of computation times of the conventional and modified NSGA-II involving the proposed solution encoding.}
\label{fig:Time-magnify}

\begin{tikzpicture}[overlay, remember picture]
    \draw[blue, thick] (4,3.5) rectangle (1.2,2.9);
    \draw[blue, thick] (-0.5, 5.5) rectangle (-7,3.75);
    \draw[<-, ultra thick, blue] (-0.5, 3.75) -- (1.2,2.9);
\end{tikzpicture}

\end{figure}

Regarding computation time, running codes in Python is slower than that in such other 
programming languages as C/C++ or machine codes. Indeed, we identified functions that 
required more computation power and acted as bottlenecks in the algorithm. It was found
that multiprocessing or translating them into machine code using the Just In Time (JIT) complier provided 
by Numba library (\cite{lam2015numba}) would reduce the overall execution time. Multiprocessing, which involves distributing computation across various CPU cores, allows code to execute faster because more computational power is allocated by the CPU. 
Moreover, optimized functions translated into machine code via Numba are executed faster
after each iteration. This is because the compiled code is reused after its initial compilation during the first execution, and it is cached to eliminate the need for recompilation in subsequent calls, making it ideal for use in repeated calls, such as loops. \hyperref[fig:Time-magnify]{{Figure 16}}  presents a comparison of the computation times for TSE and S\text{\&}P100 markets by both the conventional and modified NSGA-II algorithms (modified with
the above boosting execution techniques); both follow our solution encoding system. Results 
indicate a significantly lower computation time for the proposed approach and almost 
identical execution times for both markets despite the larger size of one market by four 
times.

\begin{comment}
\begin{figure}[H]
\centering
{\includegraphics[width=0.50\textwidth]{time}}
\caption{Comparison of computation times of the conventional and modified NSGA-II 
involving the proposed solution encoding.}
\label{fig:Time}
\end{figure}
\end{comment}

Additionally, two non-parametric tests are performed to analyze the effectiveness of our approach. All four previously-discussed performance metrics are used for evaluation in these two tests.

The first test is the sign test, as demonstrated in \hyperref[tab:markets_wins_losses]{Table 1}. The critical value for this test is 9 wins out of 10 for $\alpha = 0.05$ and 8 out of 10 for $\alpha = 0.1$. In this study, we consider $\alpha = 0.05$. As shown in the table, although there was no significant improvement in the MGD metric for the DAX 100 and S\&P 100 markets at $\alpha = 0.05$, our strategies proved effective in all other metrics across all other markets.

\begin{table}[H]
    \centering
    \caption{Sign Test. Wins and Losses per Market}
    \label{tab:markets_wins_losses}
    \begin{tabular}{lccccccccc}
    \toprule
        Market & \multicolumn{2}{c}{MGD} & \multicolumn{2}{c}{IGD} & \multicolumn{2}{c}{HV} & \multicolumn{2}{c}{Diversity Metric} \\
        & Wins & Losses & Wins & Losses & Wins & Losses & Wins & Losses \\
    \midrule
        DAX 100       & 7 & 3  & 10 & 0 & 10 & 0 & 10 & 0 \\
        S\&P 100      & 8 & 2  & 10  & 0 & 10 & 0 & 9 & 1 \\
        Nikkei 225    & 10 & 0  & 10  & 0 & 10 & 0 & 9 & 1 \\
        TSE       & 10 & 0  & 10  & 0 & 10 & 0 & 10 & 0 \\
    \bottomrule
    \end{tabular}
\end{table}

However, there is a downside to the sign test, as it does not consider the extent to which a performance metric performed better for each market. To address this issue, we performed the Wilcoxon signed-rank test, which accounts for this limitation. \hyperref[tab:wilcoxon-results]{{Table 2}} presents the results of the Wilcoxon signed-rank test for the markets used in this study. It includes the values of $R_+$, $R_-$, and the P-value for each metric. The symbol "$\sim$" indicates no significant difference, while an asterisk (*) denotes a significant improvement of our approach over the traditional NSGA-II. As shown in the table, our approach was effective across all metrics and markets with a significance level of $\alpha = 0.05$, except for the MGD metric in the DAX 100 market.

\begin{table}[H]
    \centering
    \caption{Wilcoxon Signed-Rank Test Results (Two-Tailed)}
    \label{tab:wilcoxon-results}
    \begin{tabular}{llcccc}
        \toprule
        Performance Metric & Market & $R_+$ & $R_-$ & P-value & Result \\
\midrule

        % MGD results
        \multirow{4}{*}{MGD} & DAX 100     & 11  & 44 & 0.1054 & $\sim$ \\
                            & S\&P 100    & 3  & 52 & 0.0097 & * \\
                            & Nikkei 225  & 0 & 55 & 0.0019 & * \\
                            & TSE      & 0 & 55 & 0.0019 & *\\

        % IGD results
        \multirow{4}{*}{IGD} & DAX 100    & 0 & 55 & 0.0019 & *\\
                             & S\&P 100   & 0 & 55 & 0.0019 & *\\
                             & Nikkei 225 & 0 & 55 & 0.0019 & *\\
                             & TSE     & 0 & 55 & 0.0019 & *\\

        % HV results
        \multirow{4}{*}{HV} & DAX 100     & 55 & 0 & 0.0019 & *\\
                            & S\&P 100    & 55 & 0 & 0.0019 & *\\
                            & Nikkei 225  & 55 & 0 & 0.0019 & *\\
                            & TSE      & 55 & 0 & 0.0019 & *\\

        % Diversity Metric results
        \multirow{4}{*}{Diversity Metric} & DAX 100    & 0 & 55 & 0.0019 & *\\
                             & S\&P 100   & 1 & 54 & 0.0039 & *\\
                             & Nikkei 225 & 5 & 50 & 0.0195 & *\\
                             & TSE     & 0 & 55 & 0.0019 & *\\

        \bottomrule
    \end{tabular}
\end{table}

\section{Conclusion and future research}
\label{Conclusion and future research}
In this study, promising strategies are presented to improve the performance of multi-objective evolutionary algorithms, aiming to achieve a better convergence rate, strengthen search power, and find higher-quality portfolios in less time for large-scale portfolio optimization problems under cardinality constraints, a well-known issue in finance with numerous real-world applications. The aforementioned strategies are applied to a multi-objective evolutionary algorithm and tested against a standard one, using four stock market datasets, three of which are benchmark data from well-established stock markets around the world. In order to maintain the feasibility of the solutions, a novel solution 
representation was used in lieu of the standard real-value vector of size $N$ along with a novel 
operator and specially designed repair mechanisms. Moreover, new mating strategies were 
used for producing high-quality solutions. Results revealed the faster convergence rate of the 
proposed approach as well as its capability to explore a broader range of the search space in 
comparison with the original version of the algorithm. Furthermore, the proposed handling techniques proved effective for different CC settings, and our approach was also shown to be effective through the use of two non-parametric statistical tests, demonstrating superior performance across four common performance metrics in all markets. Finally, it was observed that the proposed 
approach was more appropriate for application to larger markets when CC is present since, 
unlike the the original algorithm, our approach did not decline in performance with 
increasing number of assets.

While our approach has shown promising results, several opportunities for future research remain. Although the formulation used in this paper is widely studied in the literature (\cite{kalayci2019comprehensive}), there are additional constraints that investors may be concerned about, such as transaction costs or liquidity constraints. Since we did not use penalty functions to ensure the feasibility of these constraints, due to their limitations (\cite{yeniay2005penalty}), incorporating such constraints into our model would require modifications to the proposed approach. Future work could explore these adaptations.

\section*{Appendix A: Comparison with fewer assets}
\label{Appendix A}

As mentioned in the text, it is not a wise choice to opt for portfolios with few assets, such as two or three, as portfolios of this size are inefficient and one can significantly reduce investment risk by adding more assets. Nevertheless, to demonstrate that our approach is not limited to the tested cardinality settings and to illustrate the discontinuity in the efficient frontiers, we also provide a comparison between these portfolios for our improved NSGA-II and the standard version.

\begin{figure}[H]
\centering
\subfloat[DAX 100]{\includegraphics[width=0.24\textwidth]{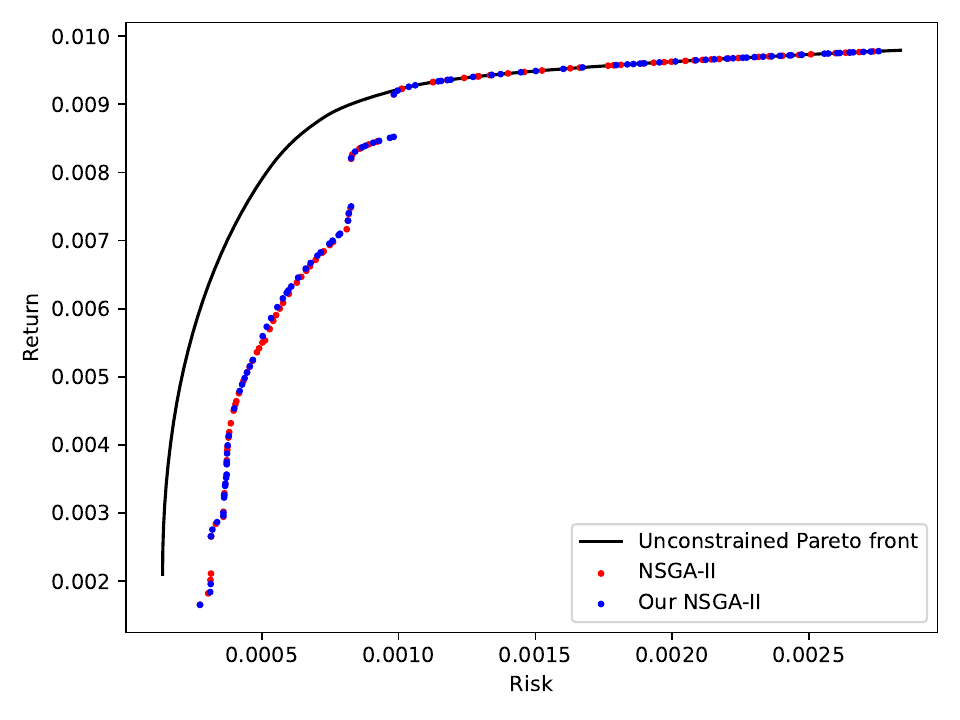}}
\hfill
 \subfloat[S\text{\&}P 100]{\includegraphics[width=0.24\textwidth]{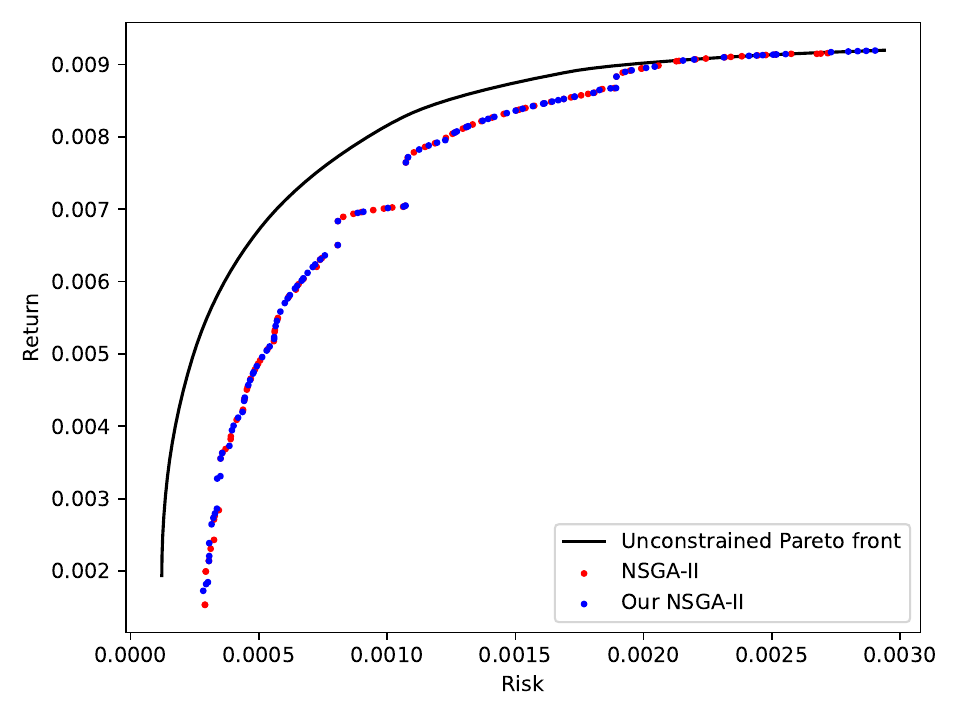}}
\hfill
 \subfloat[Nikkei 225]{\includegraphics[width=0.24\textwidth]{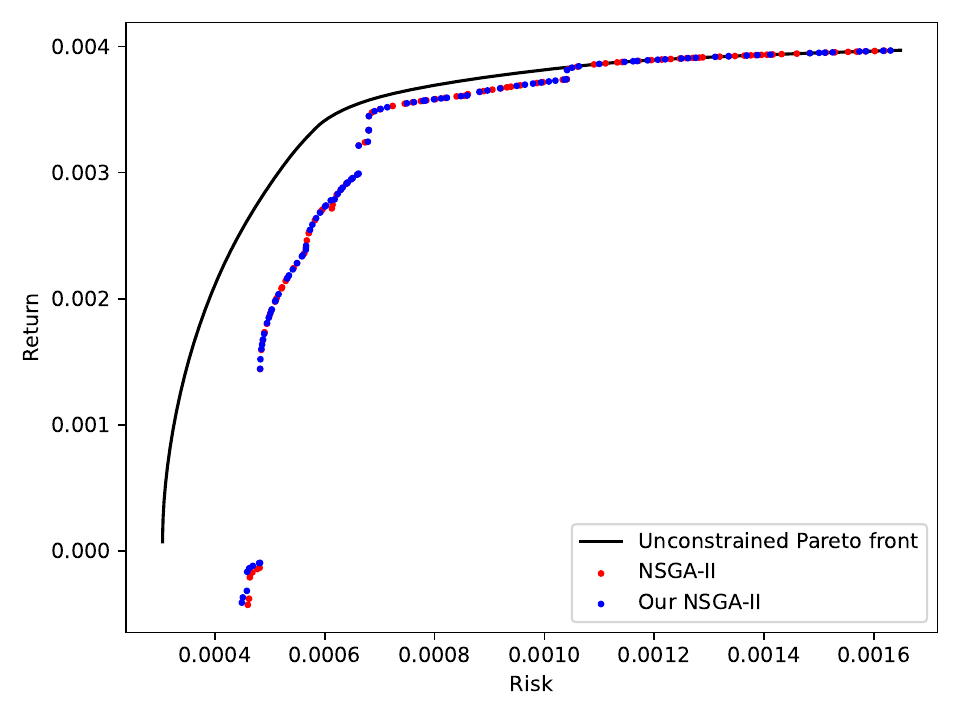}}
\hfill
 \subfloat[TSE]{\includegraphics[width=0.24\textwidth]{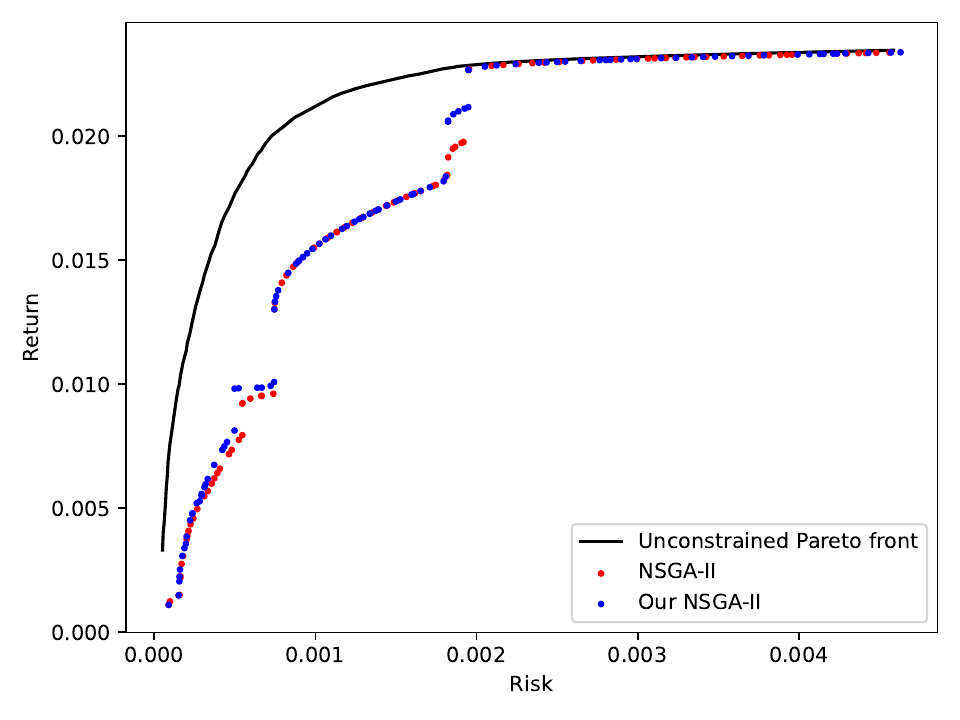}}
\caption{Comparing NSGA-II algorithms with $\sum_{i=1}^{N} z_i = 2$.}
\label{fig:NSGA-II-2}
\end{figure}

\begin{figure}[H]
\centering
\subfloat[DAX 100]{\includegraphics[width=0.24\textwidth]{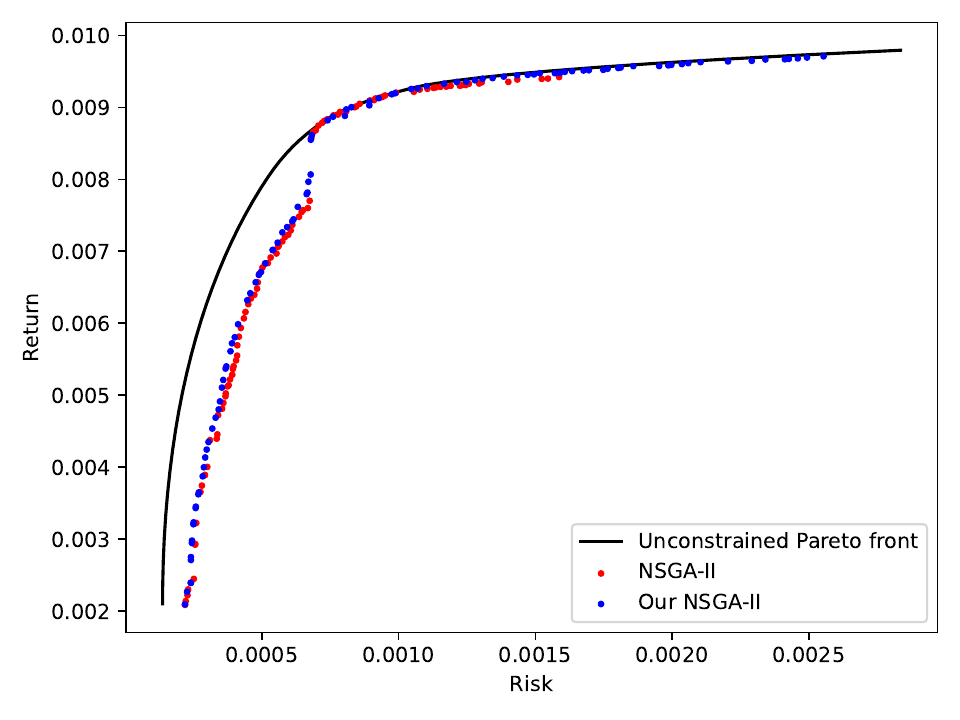}}
\hfill
 \subfloat[S\text{\&}P 100]{\includegraphics[width=0.24\textwidth]{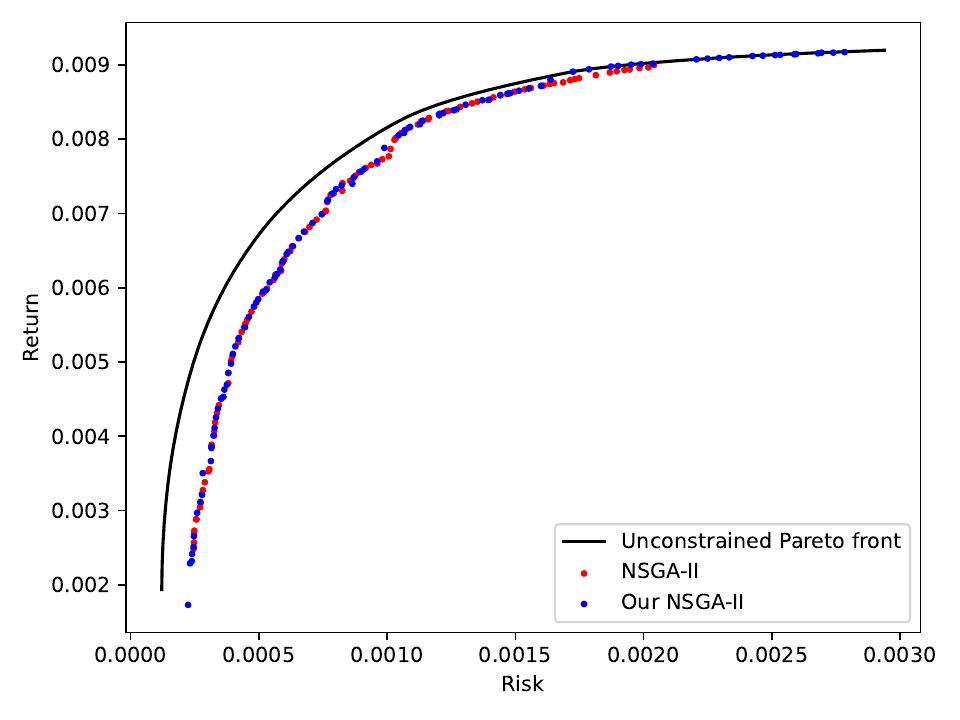}}
\hfill
 \subfloat[Nikkei 225]{\includegraphics[width=0.24\textwidth]{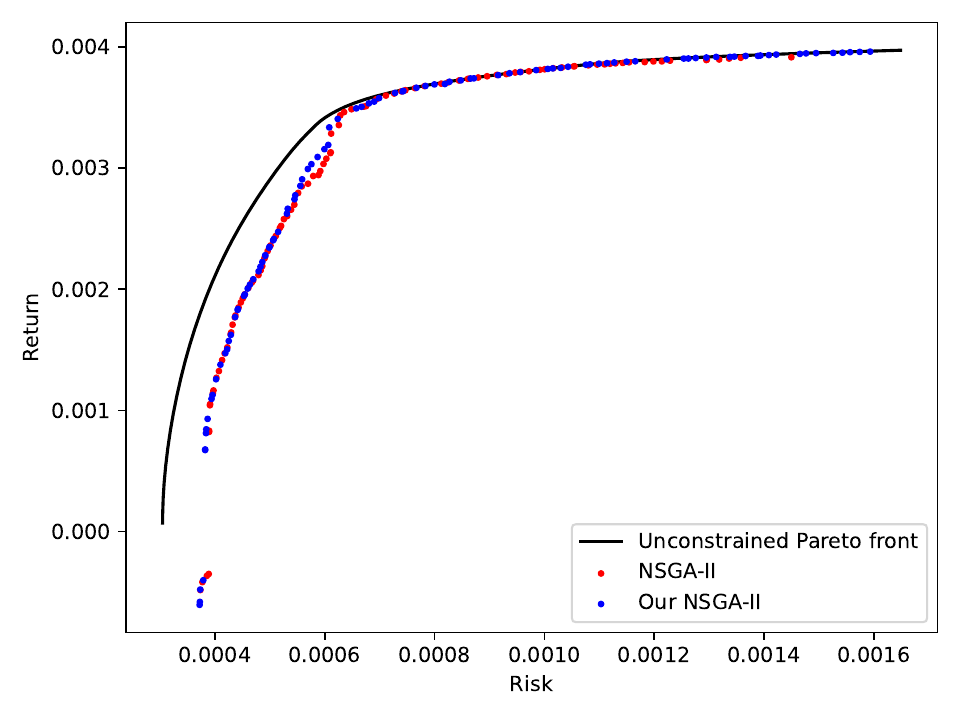}}
\hfill
 \subfloat[TSE]{\includegraphics[width=0.24\textwidth]{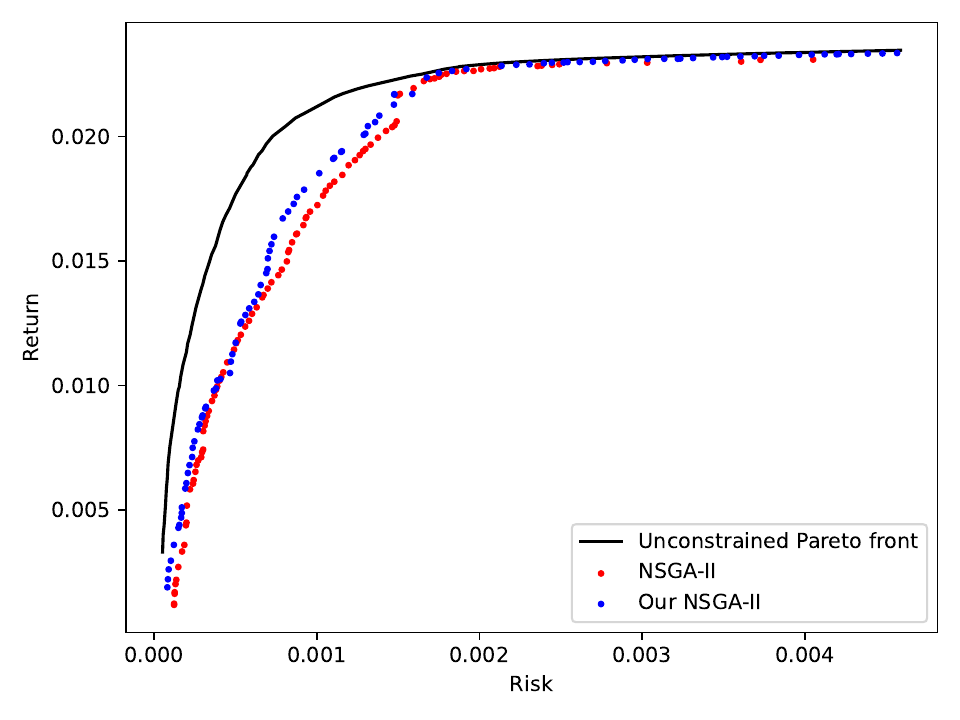}}
\caption{Comparing NSGA-II algorithms with $\sum_{i=1}^{N} z_i = 3$.}
\label{fig:NSGA-II-3}
\end{figure}

\section*{Appendix B: SPEA-II algorithm}
\label{Appendix B}
\begin{figure}[H]
\centering
\subfloat{\includegraphics[width=0.24\textwidth]{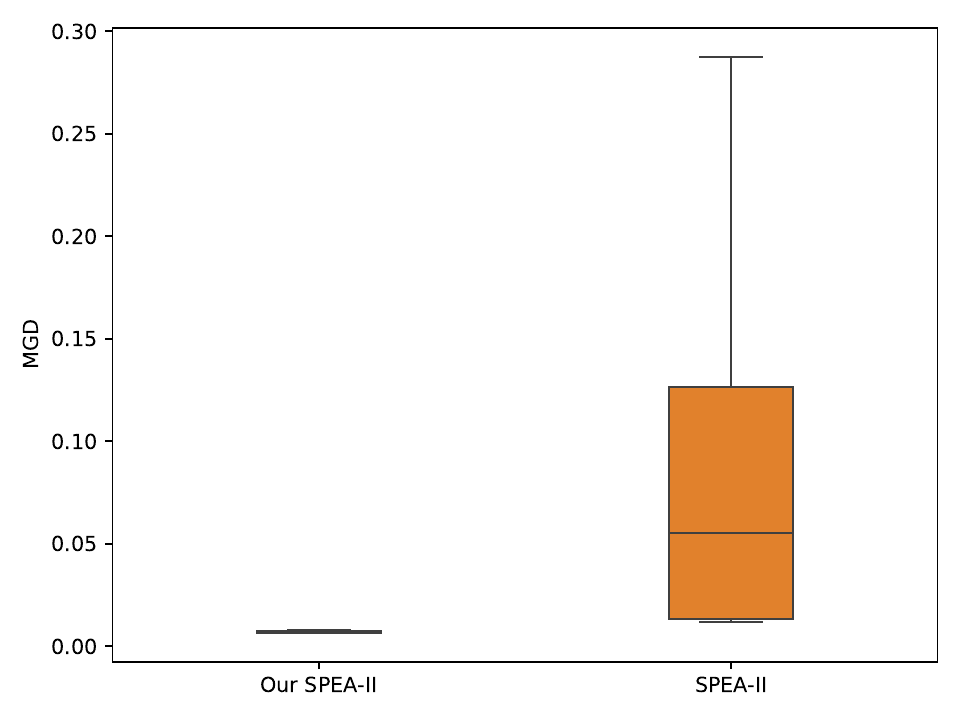}}
\hfill
 \subfloat{\includegraphics[width=0.24\textwidth]{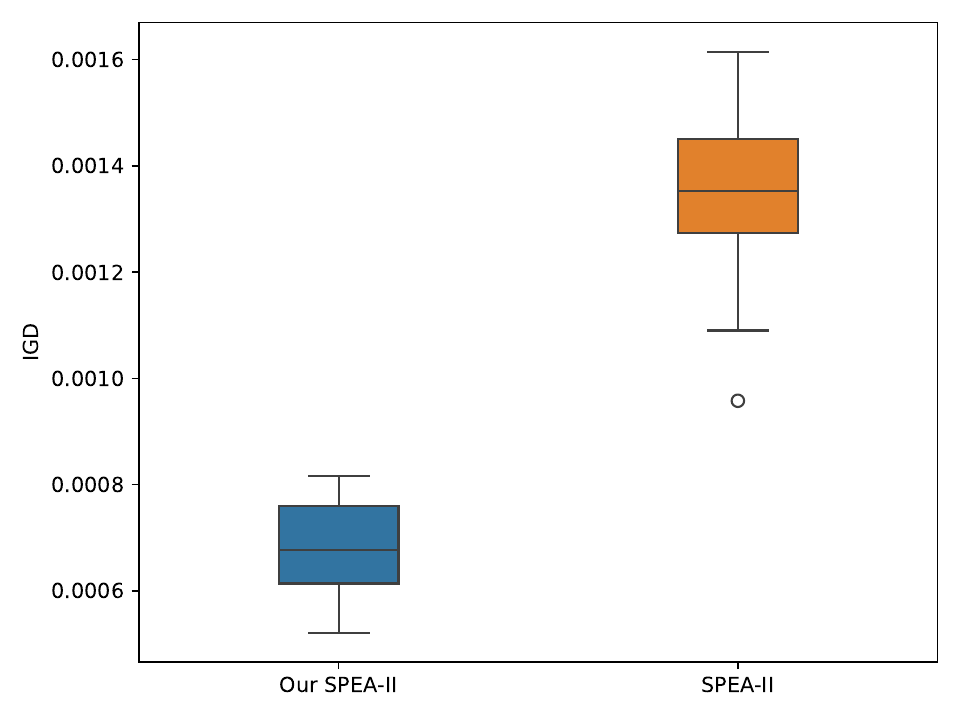}}
\hfill
 \subfloat{\includegraphics[width=0.24\textwidth]{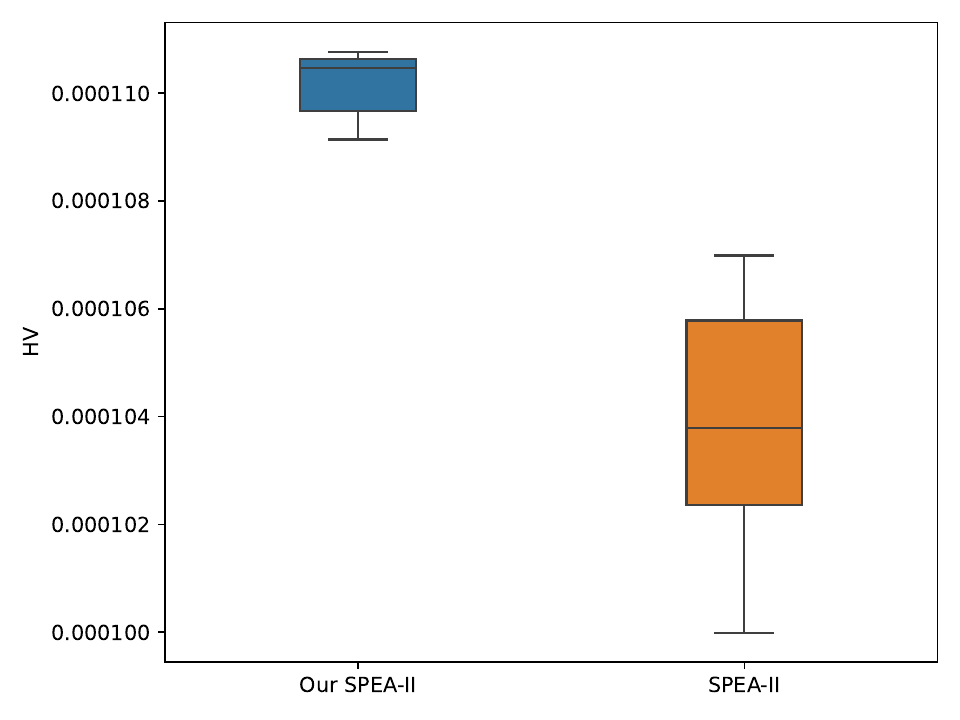}}
\hfill
 \subfloat{\includegraphics[width=0.24\textwidth]{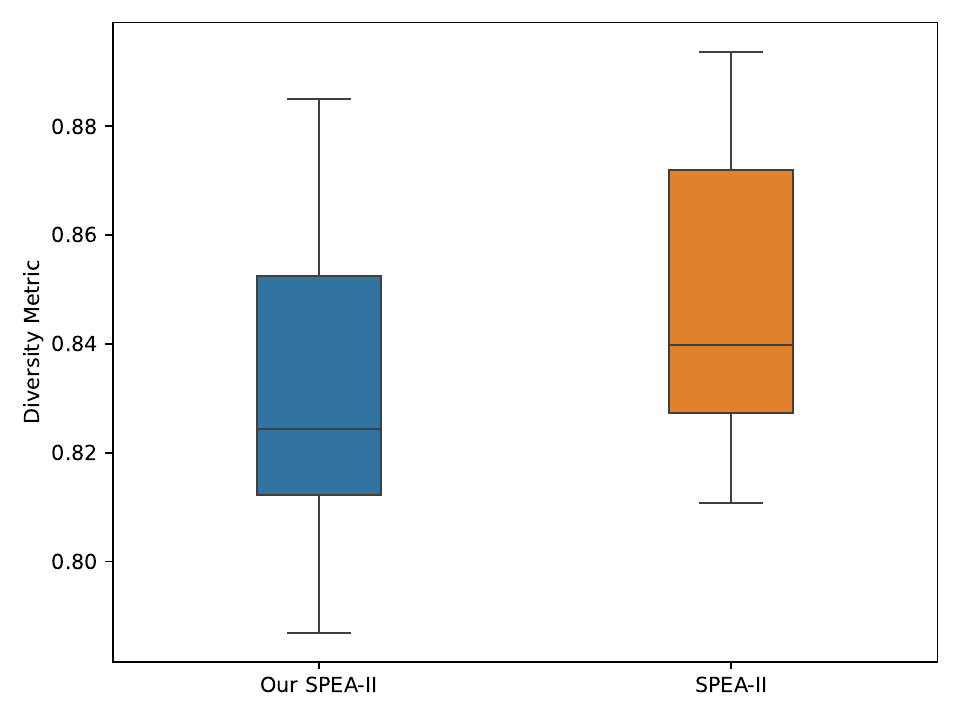}}
\caption{Comparison of performance indicators for TSE using SPEA-II.}
\label{fig:TSE-SPEA-metrics}
\end{figure}

The SPEA-II algorithm is also used for comparison to demonstrate that our strategies are applicable and generalizable to other multi-objective algorithms, and not specific to NSGA-II. To avoid redundancy and manage the length of the paper, and given that the improved NSGA-II consistently outperformed its standard version, we did not repeat all evaluations for every market. Instead, due to the consistent results, further validation with SPEA-II was conducted using only the largest and most complex market dataset, TSE, under identical configurations. \hyperref[fig:TSE-SPEA-metrics]{Figure 19} illustrates the performance of the improved SPEA-II and its standard version. As evident from the indicators, the algorithm enhanced with our mechanisms outperformed the regular SPEA-II across all metrics.

\begin{figure}[H]
\centering
\subfloat[$\sum_{i=1}^{N} z_i = 5$]{\includegraphics[width=0.32\textwidth]{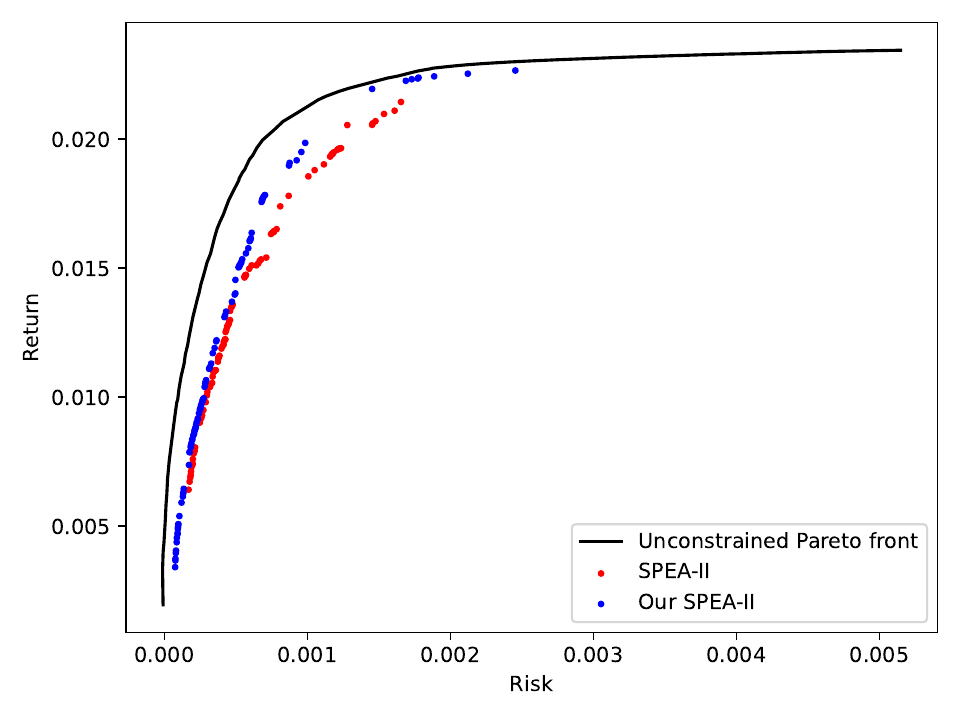}}
\hfill
\subfloat[$\sum_{i=1}^{N} z_i = 10$]{\includegraphics[width=0.32\textwidth]{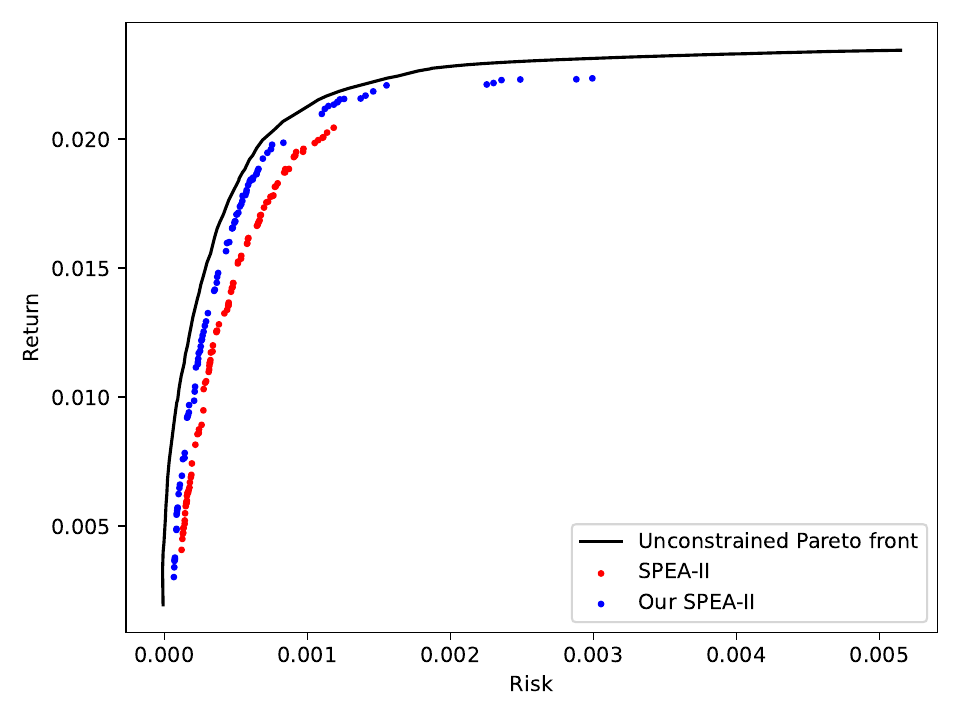}}
\hfill
\subfloat[$\sum_{i=1}^{N} z_i = 15$]{\includegraphics[width=0.32\textwidth]{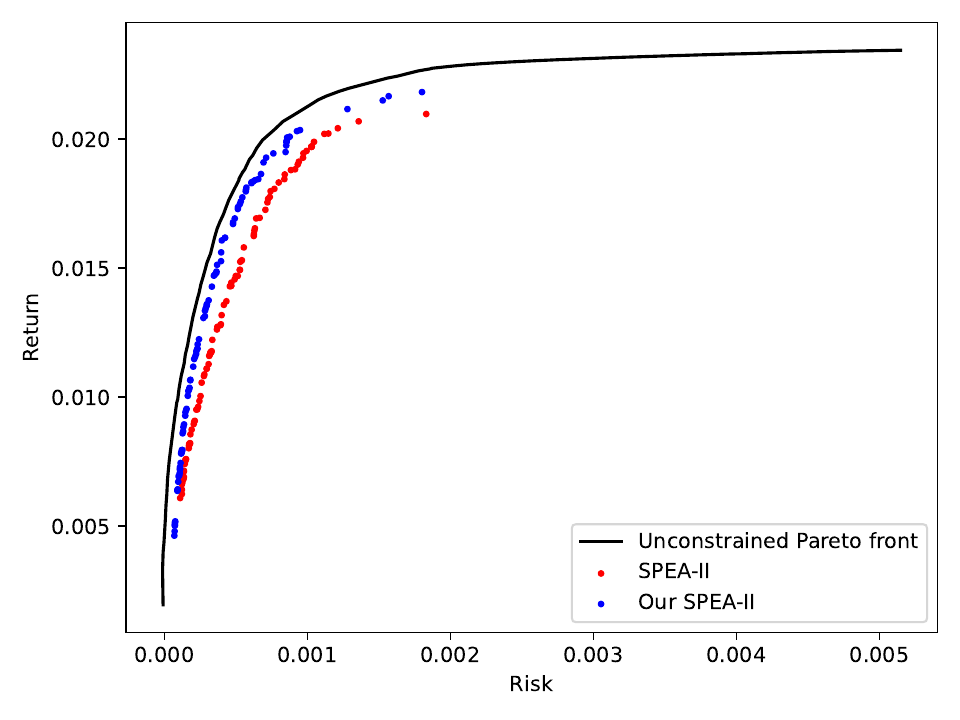}}
\caption{Comparison of approximated efficient frontiers for TSE using SPEA-II.}
\label{fig:TSE-SPEA-PF}
\end{figure}

Additionally, \hyperref[fig:TSE-SPEA-PF]{Figure 20} presents the efficient frontiers under three cardinality settings. Consistent with previous results, our improved SPEA-II dominated the standard version, consistently generating superior portfolios in terms of risk and return.

\section*{Declarations}

\subsection*{Ethics approval and consent to participate:}
Not applicable.

\subsection*{Consent for publication:}
Not applicable.

\subsection*{Funding:}
This research received no specific grant from any funding agency in the public, commercial, or not-for-profit sectors.

\subsection*{Availability of data and materials:} 

The code for our approach, along with the data we used in addition to the publicly available OR-library benchmark, will be available upon request.

\subsection*{Competing interests:}
The authors declare that they have no competing interests.

\subsection*{Authors' contributions:} 
Danial Ramezani contributed to conceptualization, methodology, software, and writing - original draft preparation.\\
Mostafa Abouei Ardakan contributed to supervision, conceptualization, methodology, and writing - reviewing and editing.\\
All authors read and approved the final manuscript.

\subsection*{Acknowledgements:}
No acknowledgements to declare.

\vspace{0.5cm}

\noindent\textbf{Corresponding author:} \\
Danial Ramezani\\
\texttt{danialramezani@khu.ac.ir}

%\pagebreak
\renewcommand{\bibsection}{\section*{References}}
\bibliographystyle{plainnat}
\bibliography{markowitz1952}

\end{document}